\documentclass[12pt,preprint]{aastex}

\usepackage[below]{placeins}

\shorttitle{NEW INTERSTELLAR DUST MODELS}
\shortauthors{ZUBKO, DWEK \& ARENDT}

\begin{document}

\title{INTERSTELLAR DUST MODELS CONSISTENT WITH EXTINCTION, EMISSION,
       AND ABUNDANCE CONSTRAINTS}
\author{Viktor Zubko\altaffilmark{1,2}, Eli Dwek, and Richard G. Arendt\altaffilmark{2}}
  \affil{NASA Goddard Space Flight Center, Code 685,
         Greenbelt, MD 20771}
\altaffiltext{1}{Corresponding author: {\tt zubko@stars.gsfc.nasa.gov}}
\altaffiltext{2}{Science Systems and Applications, Inc.}

\begin{abstract}
We present new interstellar dust models which have been derived by
simultaneously fitting the far--ultraviolet to near--infrared extinction,
the diffuse infrared (IR) emission and, unlike previous models,
the elemental abundance constraints on the dust for different
interstellar medium abundances, including solar, F and G star, and
B star abundances. The fitting problem is a typical ill-posed inversion
problem, in which the grain size distribution is the unknown, which
we solve by using the method of regularization.
The dust model contains various components: PAHs, bare silicate, graphite,
and amorphous carbon particles, as well as composite particles containing
silicate, organic refractory material, water ice, and voids.
The optical properties of these components were calculated using physical
optical constants. As a special case, we reproduce the Li \& Draine (2001)
results, however their model requires an excessive amount of silicon,
magnesium, and iron to be locked up in dust: about 50 ppm (atoms per
million of H atoms), significantly more than the upper limit imposed by
solar abundances of these elements,  about 34, 35, and 28 ppm,
respectively. A major conclusion of this paper is that there is
no unique interstellar dust model that simultaneously fits the observed
extinction, diffuse IR emission, and abundances constraints.
We find several classes of acceptable interstellar dust models, that
comply with these constraints.  The first class is identical in
composition to the Li \& Draine model, consisting of PAHs, bare graphite
and silicate grains, but with a different size distribution that
is optimized to comply with the abundances constraints. The second class
of models contains in addition to PAHs bare graphite and silicate grains
also composite particles. Other classes contain amorphous carbon instead
of graphite particles, or no carbon at all, except for that in PAHs.
All classes are consistent with solar and F \& G star abundances,
but have greater difficulty fitting the B star carbon abundance,
which is better fit with the latter (no carbon) models. Additional
observational constraints, such as the interstellar polarization,
or x--ray scattering may be able to discriminate between the various
interstellar dust models.
\end{abstract}


\section{INTRODUCTION}

Interstellar dust is completely characterized by the composition,
morphology, and size distribution of its various particles, and by
the abundance, relative to hydrogen, of its elemental constituents.
A viable dust model should be able to explain the various astrophysical
phenomena associated with the presence of dust in the interstellar medium
(ISM): the wavelength dependence of the interstellar extinction,
albedo, and polarization, the infrared emission, and the observed
elemental depletion pattern as primary constraints, and the extended
red emission (ERE) as a secondary constraint. Dust characteristics may vary in
the different ISM phases and Galactic locations. Here we concentrate
on characterizing the dust in the local diffuse ISM. 

Interstellar dust models have evolved with the advance of observational data.
Until recently, the most popular dust model was the Mathis, Rumple,
\& Nordsieck (1977; hereafter MRN), which, using the dust optical
constants of \citet{dl84} provided an excellent fit to the average
interstellar extinction curve. The model consisted of a population
of spherical graphite and silicate dust particles with a $a^{-3.5}$
power--law distribution in grain radii in
the \{$a_{\rm{min}},\ a_{\rm{max}}\}$ = \{0.005 $\mu$m, 0.25 $\mu$m\}
size interval. Assuming solar composition for the ISM, the model required
essentially all the interstellar carbon, C/H = 370~ppm (parts per
million), and all the magnesium, silicon, and iron,
Mg/H, Si/H, Fe/H = \{34, 35, 28\}, to be locked up in dust.

The first observational evidence for the incompleteness of the MRN dust
model was provided by the {\it Infrared Astronomical Satellite}
({\it IRAS}) all sky survey, which provided the average infrared
(IR) emission spectrum  at 12, 25, 60, and 100~$\mu$m  from
the diffuse ISM. The observations showed an excess of 12 and 25~$\mu$m
emission over that expected from dust heated by the local
interstellar radiation field (ISRF) and radiating at the equilibrium
dust temperature. \citet{da85} suggested that the MRN grain size
distribution should be extended to very small grains (VSG) with
radii of $\sim$ 5 {\AA} which undergo temperature fluctuations
when heated by the ISRF. Allamandola, Tielens, \& Barker (1985),
and \citet{lp84} identified these VSG with polycyclic aromatic
hydrocarbon (PAH) molecules whose presence in the ISM was inferred
from the ubiquitous solid state emission features at 3.3, 6.7, 7.6,
8.6, and 11.3~$\mu$m. We point out that the identification of PAHs
with these IR features is not universally accepted
[see \citet{tokunaga97} for a review]. 

The {\it IRAS} observations pointed out the importance of the IR
emission as a constraint on interstellar dust models. The first model
that attempted to fit in a self-consistent manner the interstellar
extinction as well as the diffuse IR emission using PAHs as
an interstellar dust component was the model of D\'esert, Boulanger,
\& Puget (1990). However, their model did not use physical optical
constants for the various dust components, and was primarily
empirical in nature. 

Interstellar polarization provides addtional constraints on interstellar
dust particles \citep{km95,km96,lg97}. In particular, the latter
authors presented a interstellar dust model consisting of PAHs,
and cylindrical silicates coated with an organic refractory mantle.
Their model satisfies the interstellar extinction, polarization and
solar abundances constraints. It did not attempt to fit the diffuse
IR emission, and used particles with hypothetical optical constants
to represent the far-UV extinction and the 2200~{\AA} extinction hump.

The Diffuse IR Background Experiment (DIRBE) and Far Infrared Absolute
Spectrophotometer (FIRAS) instrument on board the {\it Cosmic
Background Explorer} ({\it COBE}) satellite provided the most extensive
wavelength coverage (3.5 to 1000~$\mu$m) of the IR emission from
the diffuse ISM, and indirect evidence for the emission from PAHs
from this phase of the medium \citep{dwek97}. \citet{dwek97} attempted
to fit the interstellar extinction and diffuse IR emission using a mix
of bare silicate and graphite particles with \citet{dl84} optical
constants, and PAHs with the optical properties of \citet{dbp90}.
The model failed to reproduce the observed interstellar extinction,
primarily due to the non-physical nature of the UV-optical
properties adopted for the PAHs. \citet{dl01} and \citet{ld01b}
improved on this model by using a more realistic
characterization of the optical properties of the PAHs, based on
laboratory measurements. However, their model requires an excessive
amount of Mg, Si, and Fe to be locked up in dust, almost twice
the available iron for an ISM with solar abundances. 

In this paper we explore possible dust models that simultaneously 
comply with the three major observational constraints: the average
interstellar extinction, the thermal IR emission form the diffuse ISM,
and the interstellar abundances constraints. The model explores
a variety of potential dust compositions, including bare silicate
(MgFeSiO$_4$), graphite, amorphous carbon particles, PAHs, and
composite particles consisting of a mixture of silicate, refractory
organic material (C:H:O:N $\approx$ 1:1:0.2:0.04; we will adopt
C$_{25}$H$_{25}$O$_5$N for simplicity), water ice (H$_2$O), and voids.
The model is physical in the sense that it uses measured optical
constants or observed radiative properties to characterize
the optical properties of the various dust constituents. 
Given the composition and physical properties of the dust particles,
the problem of simultaneously fitting the model to a set of
observational constraints is a typical ill-posed inversion problem,
in which the grain size distribution is the unknown. We solve this
problem using the method of regularization.

In \S\ref{section:theoretical_model} we present the integral equation
that we invert in order to solve for the grain size distribution.
The general equations for the interstellar extinction, IR emission,
and elemental dust abundances are cast as a Fredholm integral equation
of the first kind. The left hand side (LHS) in the equation comprises
of the observational input data, and the RHS is an integral over grain
sizes of a known function, the kernel, multiplied by an unknown fuction,
the grain size distribution. The section summarizes the observational
input data used in our analysis: the average interstellar extinction,
thermal IR emission per H~atom, the average interstellar radiation
field (ISRF) to which the dust is exposed, and the allowable amount
of refractory elements that can be locked up in the dust, and
the physical properties of the dust that are used in the kernel.  
\S\ref{section:regularization} describes the Regularization method
that we used to invert the Fredholm equation to derive
the distribution of grain sizes that satisfies the observational
constraints for a given kernel. Modeling details are presented
in \S\ref{section:mod_details}. The results of our calculations are
presented in \S\ref{section:results}. We show that the Li \& Draine
PAH, bare silicate, and graphite model can be optimized, and
we produce an equally good fit to the observational constraints
without violating the interstellar abundance constraints for
a different, more general, grain size distribution. We also show
that the PAH, bare silicate and graphite grain model is not unique.
A more complex dust model comprising of PAHs, bare silicate and 
graphite grains, as well as organic refractory and icy silicates,
provides a somewhat improved fit to the observational constraints,
including the interstellar abundances. A summary of the paper,
its astrophysical implications, and directions for future
improvements and tests for interstellar dust models are
presented in \S\S\ref{section:implications} and
\ref{section:conclusions}.

\section{THEORETICAL MODEL}
     \label{section:theoretical_model}

\subsection{Constructing the Main Integral Equation}
     \label{subsection:main_equations}

The goal of our studies is to find dust models that
simultaneously fit the observed interstellar extinction, infrared emission,
and elemental abundance constraints. A dust model consists of a set of
dust components ($i=1..n$) each of which is characterized by a chemical
composition and a size distribution function $f_i(a) \> \rm{d}$$a$,
defined as the number of grains per
hydrogen (H) atom in the radius interval $a$ to $a+{\rm{d}}a$.
Thus, if $\rho_i$ is the mass density of $i$-th component,
the expressions $\sum_{i=1}^{n} \int \! f_i(a) \>{\rm{d}}a$ and
$\sum_{i=1}^{n} \int \! \frac{4}{3} \pi a^3 \rho_i f_i(a) \>{\rm{d}}a$,
give, respectively, the total number of dust particles per H atom and
their total mass per H atom,  summed up over all dust components.

The observational constraints we use to characterize the nature of
the dust in the diffuse ISM are:
\begin{enumerate}
\item
the average wavelength-dependent extinction $\tau(\lambda)$ per H column
density:
\begin{equation}
  {\tau(\lambda) \over N_{\rm{H}}} =
    \sum_{i=1}^{n} \int \!
      \Big[ \pi a^2 \> Q^{[i]}_{\rm{ext}}(\lambda, a) \Big]
        \> f_i(a) \> {\rm{d}}a
    \equiv
    \sum_{i=1}^{n} \int \! K^{[i]}_{\tau}(\lambda, a) \> f_i(a) \> {\rm{d}}a
  \label{eq:extinction}
\end{equation}
where $N_{\rm{H}}$ is the line-of-sight hydrogen column density, and
$Q^{[i]}_{\rm{ext}}(\lambda, a)$ is the extinction efficiency of
the $i$-th component of radius $a$.
\item
the spectrum and intensity of the infrared emission from the diffuse ISM:
\begin{equation}
  {I_{\lambda}(\lambda) \over N_{\rm{H}}} =
    \sum_{i=1}^{n} \int \! \Big[ \pi a^2 \> Q^{[i]}_{\rm{abs}}(\lambda, a) \>
       E^{[i]}_{\lambda}(\lambda, a) \Big] \> f_i(a) \> {\rm{d}}a
    \equiv
    \sum_{i=1}^{n} \int \! K^{[i]}_I(\lambda, a) \> f_i(a) \> {\rm{d}}a
  \label{eq:emission}
\end{equation}
where $I_{\lambda}(\lambda)$ is the specific intensity per unit solid angle,
$Q^{[i]}_{\rm{abs}}(\lambda, a)$ is the absorption efficiency factor at
$\lambda$ of the $i$-th dust component of radius $a$, and
$E^{[i]}_{\lambda}(\lambda, a)$ is its emissivity. 
For all dust particles $E^{[i]}_{\lambda}(\lambda, a)$ is given by:
\begin{eqnarray}
E_{\lambda}(\lambda, a) =
  \int_{\frac{hc}{\lambda}}^{\infty}
  {\rm{d}}U \> P(a, U) \> B_{\lambda}[\lambda, T(U, a)],
    \label{eq:emiss1}
\end{eqnarray}
where $T[U, a]$ is the vibrational temperature of the dust with
an internal energy $U$, and $P(a,U)$ is the probability distribution
of internal energy $U$ for grains of radius~$a$. Very small dust particles
will be stochastically heated by the interstellar radiation field and
undergo temperature, or internal energy fluctuations which were first predicted
by \citet{greenberg68}. For sufficiently large grains with the thermal
content significantly exceeding the mean energy of the colliding photons,
$P(a,U)$ can be approximated by a delta function at $U_{\rm{eq}}$
corresponding to an equilibrium temperature $T_{\rm{eq}}(a)$.
In this case, $E_{\lambda}(\lambda, a)$ is simply the Planck function:
\begin{eqnarray}
E_{\lambda}(\lambda, a) \simeq B_{\lambda}[\lambda, T_{\rm{eq}}(a)].
    \label{eq:emiss2}
\end{eqnarray}
The grain equilibrium temperature is defined by the balance of absorbed
and emitted radiation:
\begin{eqnarray}
\int_0^{\infty} \!\! \pi a^2 \> Q_{\rm{abs}}(\lambda, a) \>
  J_{\lambda}^{\rm{ISRF}}(\lambda) \> {\rm{d}}{\lambda} =
\int_0^{\infty} \!\! \pi a^2 \> Q_{\rm{abs}}(\lambda, a) \>
  B_{\lambda}[\lambda, T_{\rm{eq}}(a)] \> {\rm{d}}{\lambda}.
    \label{eq:t_eq}
\end{eqnarray}
where $ J_{\lambda}^{\rm{ISRF}}(\lambda)$ is the intensity of the local
interstellar radiation field (ISRF).
\item
the abundance of elements locked up in the solid phase of the diffuse ISM.
If $k$ is the number of chemical elements constituing the dust,
then the column density $N_j$ of the $j$-th element ($j=1..k$)
locked up in the dust is given by:
\begin{equation}
    {N_j \over N_{\rm{H}}} =
    \sum_{i=1}^{n} \int \! \Big[ \frac{4}{3} \pi a^3 \> \rho_i
    { \alpha_{j,i} \over m_j } \Big] \> f_i(a) \> {\rm{d}}a
    \equiv
    \sum_{i=1}^{n} \int \! K^{[i]}_j(a) \> f_i(a) \> {\rm{d}}a
  \label{eq:abundance}
\end{equation}
where $m_j$ is the atomic mass of $j$-th element, and $\alpha_{j,i}$
is the mass fraction of $j$-th element in the $i$-th constituent.
For example, for silicon locked up in olivine [(Mg,Fe)$_2$SiO$_4$ or
else MgFeSiO$_4$ to imply that $N_{\rm{Mg}}$=$N_{\rm{Fe}}$ in the
grain composition], $m_j$~=~28$\>m_{\rm{H}}$ and $\alpha_{j,i}$~=~28/172.
\end{enumerate}

$K_{\tau}$, $K_I$, and $K_j$, $j$=\{1..k\}, defined in equations
(\ref{eq:extinction}), (\ref{eq:emission}), and (\ref{eq:abundance}),
can be combined into one superkernel $K(x, a)$, given by:
\begin{eqnarray}
K(x, a) &  = & [\{K^{[1]}_{\tau}(\lambda^{\rm{ext}}, a),\ldots,
     K^{[n]}_{\tau}(\lambda^{\rm{ext}}, a)\},
  \{K^{[1]}_{I}(\lambda^{\rm{em}}, a),\ldots,
     K^{[n]}_{I}(\lambda^{\rm{em}}, a)\}, \nonumber \\
              &    & \{K^{[1]}_{1}(a),\ldots,K^{[n]}_{1}(a)\},\ldots,
	             \{K^{[1]}_{k}(a),\ldots,K^{[n]}_{k}(a)\}],
  \label{eq:K}
\end{eqnarray}
where $x$ is a generalized variable: $x=\{\lambda^{\rm{ext}}, \lambda^{\rm{em}},
j\}$, which runs the wavelengths of the extinction data:
$\lambda^{\rm{ext}}=\lambda^{\rm{ext}}_{\rm{min}},\ldots,\lambda^{\rm{ext}}_{\rm{max}}$,
then the wavelengths of the emission spectrum data:
$\lambda^{\rm{em}}=\lambda^{\rm{em}}_{\rm{min}},\ldots,\lambda^{\rm{em}}_{\rm{max}}$,
and, finally, the number of the abundance constraints: $j = 1,\ldots,k$.
Since equations (\ref{eq:extinction}), (\ref{eq:emission}), and (\ref{eq:abundance})
are linearly dependent on the grain size distribution $f_i(a)$, we can combine
them into a single integral equation as: 
\begin{equation}
  D(x) =
    \int \! \> K(x, a) \> F(a) \> {\rm{d}}a,
  \label{eq:int_eq1}
\end{equation}
where $D(x)$ consists of the observational constraints:
\begin{equation}
D(x)=[\tau(\lambda^{\rm{ext}}), I_{\lambda}(\lambda^{\rm{em}}), N_j]/N_{\rm{H}},
  \label{eq:D}
\end{equation}
and $F(a)$ is an array of the size distribution functions given by: 
\begin{equation}
 F(a)=[f_1(a), \ldots, f_n(a)],
  \label{eq:F}
\end{equation}

In the following we describe the observational constraints and the dust model
assumptions needed, respectively, to define $D(x)$  and calculate
$K(x, a)$. In addition to $D(x)$  and $K(x, a)$, the inversion of equation
(\ref{eq:int_eq1}) will require knowledge of the uncertainties,
$\sigma(x)$, in $D(x)$, which are defined in terms of the uncertainties in
the different observational constraints as:
\begin{equation}
\sigma(x) = [\sigma_{\tau}(\lambda^{\rm{ext}}), \sigma_I(\lambda^{\rm{em}}),
    \sigma_{N_j}]
  \label{eq:sigma}
\end{equation}

\subsection{Observational Constraints, $D(x)$ and Uncertainties,
   $\sigma(x)$}
       \label{subsection:observ_constraints}

\begin{figure}
\begin{center}
\plotone{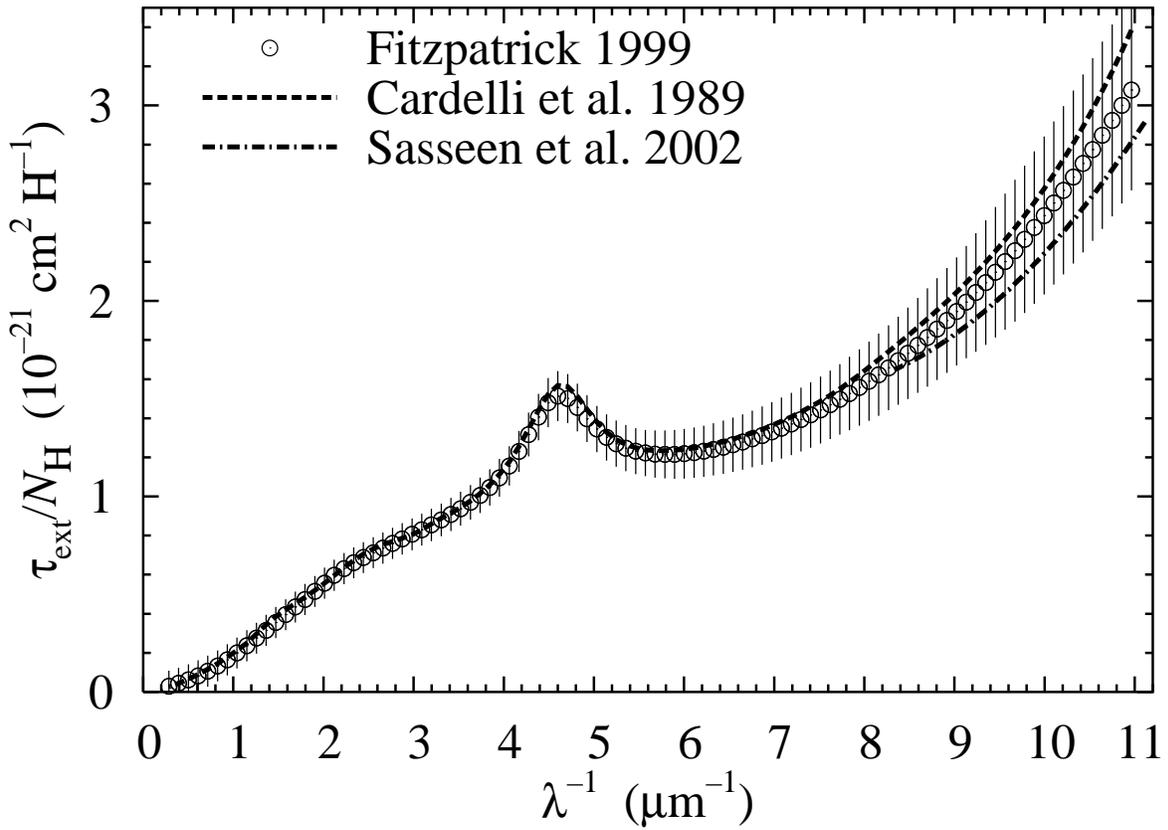}
\caption{ The mean extinction curve of the diffuse interstellar medium
  with $R_V$=3.1 from \citet{fitzpatrick99} together with the curve
  from \citet{ccm89} and the new FUV curve from \citet{sasseen01} }
   \label{fig:extinction1}
\end{center}
\end{figure}

We used the latest available mean extinction curve for $R_{\rm{V}}$=3.1
from \citet{fitzpatrick99} to characterize the average extinction from
the diffuse ISM. In contrast to previous extinction curves
[e.g. by \citet{ccm89} or \citet{sm79}], the new curve has been
constructed to reproduce the detailed wavelength behavior of
the extinction for $R_{\rm{V}}$=3.1 by properly taking into account
the bandpass effects in optical/IR data and the observed broadband,
intermediate-band, and narrow-band extinction measurements. Note
that a new far-UV Galactic extinction curve from 910 to 1200~{\AA}
reported by \citet{sasseen01} shows a good consistency with
Fitzpatrick's curve. Figure \ref{fig:extinction1} displays
the extinction curve in the form of $\tau_{\rm{ext}}/N_{\rm{H}}$
that we used for our modeling. It was derived by us from the original
$E(\lambda-V)/E(B-V)$ curve by using the ratio
$N_{\rm{H}}/E(B-V) = 5.8\times10^{21}$~H cm$^{-2}$
typical for the diffuse ISM \citep{bohlin79}. Also shown in the figure
is the new far-UV extinction curve from 910 to 1200~{\AA} reported by
\citet{sasseen01} and the 912 {\AA} to 3.5~$\mu$m extinction curve
of \citet{ccm89}. We note that \citet{ld01b} used
a different normalization of the curve at $I$(0.9~$\mu$m) band, given by:
$A(I)/N_{\rm{H}}=2.6{\times}10^{-22}$~cm$^{2}$ \citep{draine89}.
This produces a factor of 1.12 difference between their adopted extinction
curve and ours. The uncertainties on Figure~\ref{fig:extinction1} are
the observed dispersion of the curve. The estimate in the UV-through-optical
segment is based on the {\it ANS} satellite work by \citet{savage85} covering
about 1000 sightlines. The uncertainty in the infrared part was estimated
supposing the $R_{\rm{V}}$ dispersion of 0.4 to 0.5, which is
consistent with the observed scatter of the extinction curve at
1500~{\AA} \citep{fitzpatrick99}.

\begin{figure}
\begin{center}
\plotone{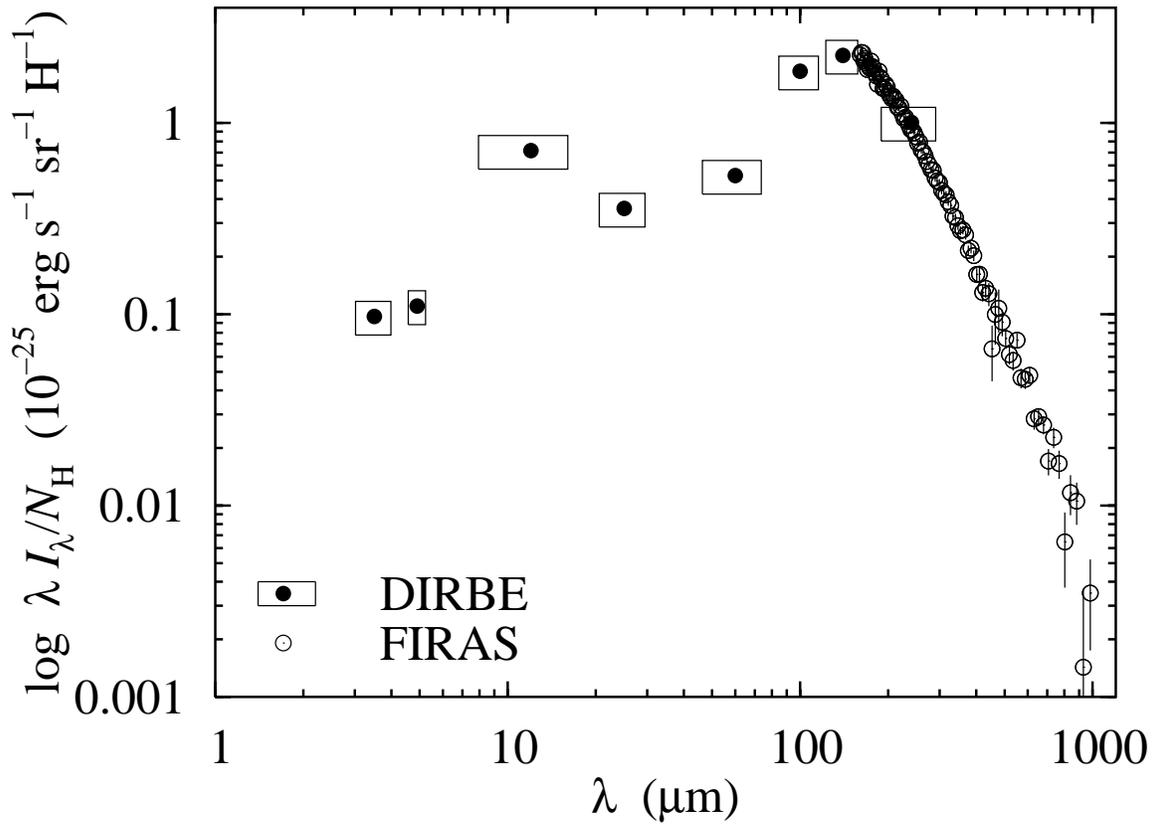}
\caption{ Specific intensity per hydrogen atom of dust emission at
  high Galactic latitudes, $|b| \geq 25^{\circ}$. Observational data are
  from DIRBE \citep{arendt98} and FIRAS \citep{dwek97} }
   \label{fig:emission1}
\end{center}
\end{figure}

The {\it COBE}  all sky survey provided the most comprehensive spectrum of dust
emission from the diffuse ISM  in eight  DIRBE bands at 3.5, 4.9, 12, 25, 60,
100, 140, and 240~$\mu$m, and in the FIRAS channels spanning
the $\sim$ 200 and 1000~$\mu$m wavelength range \citep{dwek97,arendt98}.
Figure~\ref{fig:emission1} shows the average IR emission from the diffuse
ISM used by us to constrain the dust models. The uncertainties of
the DIRBE fluxes are from \citet{dwek97}.

\begin{figure}
\begin{center}
\plotone{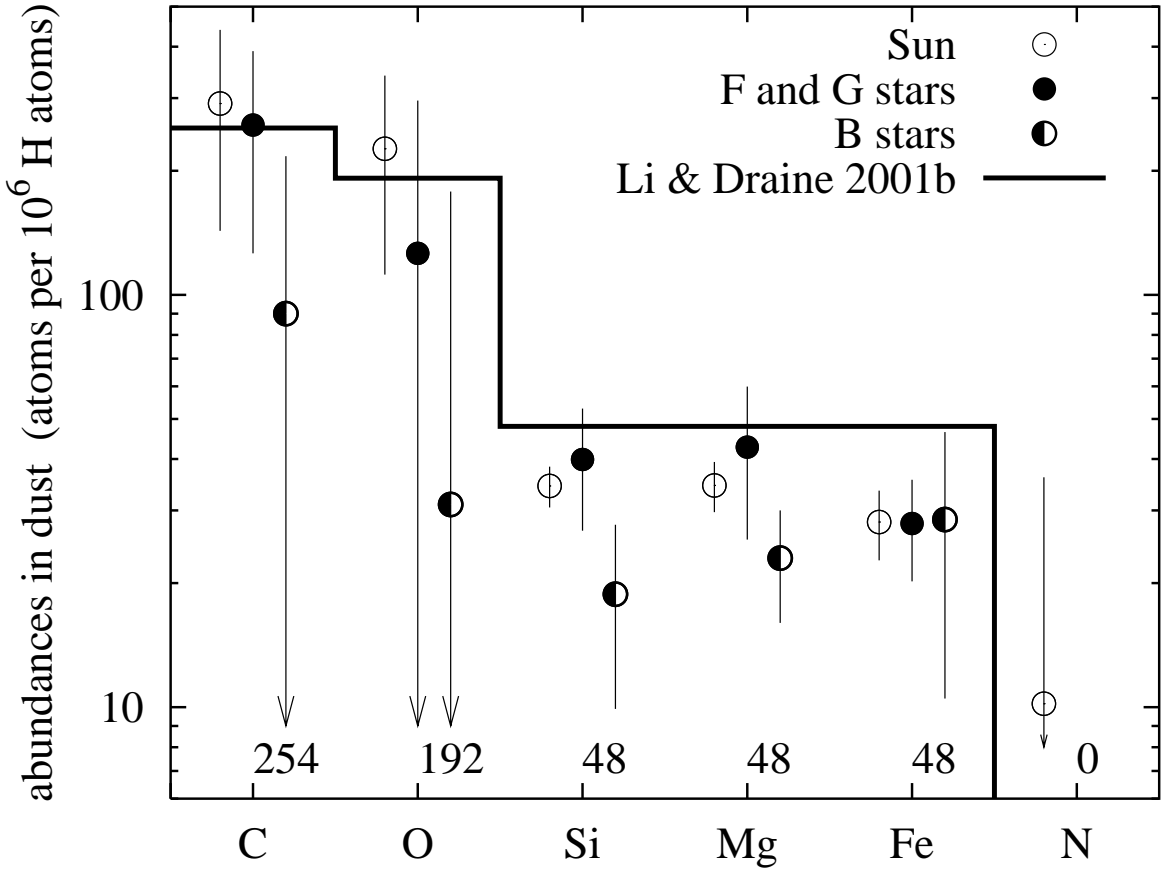}
\caption{ The elemental abundances in dust derived by assuming
  the interstellar abundances to be solar those for F and G stars,
  and B stars. Numbers along the abscissa are the \citet{ld01b}
  abundances. 
}
   \label{fig:abund_ld01}
\end{center}
\end{figure}

Since the chemical composition of the dust cannot be directly measured, it is
common to estimate the abundance of an element locked up in dust by
subtracting  the observed gas phase abundance of that element from an adopted
measure of its total ISM abundance. The latter is  determined from stellar
surface abundances which are believed to represent the total abundance of
that element in the gas and dust phases of the ISM from which those stars
have formed. Due to the uncertainties in the choice of a representative set
of average ISM abundances we have used three different measurements based on solar
\citep{holweger01}, B~star \citep{sw96,sofia01}, and F and G~star
\citep{sofia01} abundances to characterize the standard abundances of
the elements in the ISM. Figure~\ref{fig:abund_ld01} and
Table~\ref{tab_dust_abund} summarizes the abundances of the most abundant
elements expected to be in the dust: C, O, Si, Mg, Fe, and N.
Also shown in the table are the measured gas phase abundances of
these elements. The dust phase abundances used to constrain the dust models
are simply the ISM minus the gas phase abundances for the different
elements. Because of the large discrepancy between the two available
estimates of the gas-phase C of 140$\pm$20 \citep{cardelli96}
and 75$\pm$25 \citep{dwek97}, we adopted a single estimate of 108$\pm$16
that is a straight average of the two values. 

\subsection{Dust Model and Grain Emissivities, $K(x, a)$}
       \label{subsection:dust_model}

Several dust compositions were used to characterize the interstellar dust
population: (1) {\it bare silicate and graphite grains} with optical
constants of \citet{dl84}, \citet{ld93}, and \citet{wd01};
(2) {\it polycyclic aromatic hydrocarbons (PAHs)} with specific
density of 2.24~g cm$^{-3}$, and absorption cross sections from
\citet{ld01b}; and (3) {\it amorphous carbon dust}. We tried three
different types of amorphous carbon: ACAR, BE, and
ACH2\footnote{ACAR sample was produced in arc discharge between amorphous
carbon electrodes in an Ar atmosphere at 10~mbar;
ACH2 sample was produced in arc discharge between amorphous
carbon electrodes in an H$_2$ atmosphere at 10~mbar;
BE sample was produced by burning of benzene in air under
normal consitions \citep{col95}
}
with the optical constants
from \citet{zubko96}. In addition to these dust constituents,
we also considered composite particles composed of silicate, amorphous carbon,
and variations of the following components: (1) {\it organic refractory material}
consisting of [C$_{25}$H$_{25}$O$_5$N] with optical constants from \citet{lg97};
(2) {\it water ice}, with optical constants from \citet{warren84};
and (3) {\it voids}. Table \ref{tab_dust_const} summarizes the properties
of the different dust constituents used in this paper.

Like \citet{ld01b}, we treat PAHs and graphite as different dust
components. However, we allow for the extension of the graphite
grain size distribution to very small particles. Very small graphite
particles are essentially dehydrated PAHs. In contrast, Li \& Draine
extend the graphite size distribution only down to 50~{\AA},
and adopt PAH-like properties for carbonaceous particles with sizes
below $\sim$20~{\AA}. As we will show in \S5.1,  as a result
of this difference we derive a carbonaceous grain size distribution
that is considerably simpler than that of \citet{ld01b}.

To build $K(x, a)$, we need to know the extinction and absorption cross
sections: $Q^{[i]}_{\rm{ext}}(\lambda, a)$, $Q^{[i]}_{\rm{abs}}(\lambda, a)$,
emissivities $E^{[i]}_{\lambda}(\lambda, a)$, densities $\rho_i$,
and mass fractions $\alpha_{ji}$.
The densities used are listed in Table \ref{tab_dust_const},
the mass fractions are easily calculated by using atomic and
molecular masses of dust constituents.

For the bare silicate, graphite, and amorphous carbon particles,
the extinction and absorption cross sections were calculated using Mie theory
for spherical dust particles \citep{bh83}. For the composite material,
at first step, we used an effective medium theory (EMT) to calculate its
dielectric function, and, at second step, we used Mie theory to calculate
the absorption and scattering cross sections with the effective dielectric
function. The effective dielectric function was calculated by making use of
an EMT based on the Bruggeman approach with adjusted effective
shape factors (EMT-O), that gives the following equation for the effective
dielectric function $\varepsilon_{\rm{eff}}$ \citep{sho95}:
\begin{equation}
 {\sum_k}{\sum_{l=1}^4} \>\> f_l^k {{\varepsilon^k - \varepsilon_{\rm{eff}}}
   \over
   {\varepsilon_{\rm{eff}}+(\varepsilon^k - \varepsilon_{\rm{eff}})L_l^k}} = 0
    \label{eq:epsilon_eff}
\end{equation}
with $L_l^k=\{0,1/2,1,1/3\}$, $f_1^k=(5/9)\> f^k \sin^2(\pi f^k)$,
$f_2^k=(2/9)\> f^k \sin^2(\pi f^k)$, $f_3^k=(2/9)\> f^k \sin^2(\pi f^k)$,
$f_4^k=f^k \cos^2(\pi f^k)$, where $f^k$ and $\varepsilon^k$ are
the volume fraction and isotropic dielectric function of component $k$,
and $L_l^k$ are some shape factors for ellipsoidal grains.
Equation (\ref{eq:epsilon_eff}) is an approximation to different experiments
and theories. The EMT-O approach takes into consideration the effects of
shape distribution, proximity, connected particles and percolation,
and thus it is highly suitable for treating the composite particles
which are the aggregates of clustered grains.
For example, the percolated structure of the composites is simulated
by ``particles'' with shape factor 0. It is easy to show that in the limit
when cells of all components of the composites have spherical shape
($L_l^k=\frac{1}{3}$ and $f_l^k=f^k$), the EMT-O rule
(eq. \ref{eq:epsilon_eff}) reduces to the classical Bruggeman rule
\citep{ossenkopf91}:
\begin{equation}
 {\sum_k} \>\> f^k {{\varepsilon^k - \varepsilon_{\rm{eff}}}
   \over
   {\varepsilon^k + 2\varepsilon_{\rm{eff}}}} = 0
    \label{eq:epsilon_eff_b}
\end{equation}

The grain emissivities, $E^{[i]}_{\lambda}(\lambda, a)$,
were calculated by using the thermal-discrete approximation (TDA)
to the general statistical-mechanical approach developed by \citet{dl01}.
The authors have shown that the TDA is quite good for calculating
the overall emission spectrum of very small grains of typical sizes
less than about 200 {\AA} that experience temperature spikes.
For larger grains, the grain emissivity can be easily calculated by
using Planck function with the equilibrium grain temperature.

To calculate the energy balance of interstellar
dust grains, we assume that the dust grains are heated by absorption
of photons from the interstellar radiation field (ISRF) and are
cooled via their own emission. The ISRF is defined by its mean
intensity $J_{\lambda}^{\rm{ISRF}}(\lambda)$. We adopted the empirical
representation of $J_{\lambda}^{\rm{ISRF}}(\lambda)$ for the solar
neighborhood from \citet{mathis83}
\begin{equation}
J_{\lambda}^{\rm{ISRF}}(\lambda) = J_{\lambda}^{\rm{UV}}(\lambda) +
   \sum_{j=1}^3 W_j \> B_{\lambda}(\lambda, T_j) +
      B_{\lambda}(\lambda, T_{\rm{b}}),
      \label{eq:J_ISRF}
\end{equation}
which includes the UV component, $J_{\lambda}^{\rm{UV}}(\lambda)$,
three effective blackbody sources with dilution factors:
$W_1$=10$^{-14}$, $W_2$=10$^{-13}$, $W_3$=4$\times$10$^{-13}$,
and temperatures: $T_1$=7500~K, $T_2$=4000~K, $T_3$=3000~K,
and the cosmic microwave background radiation,
$B_{\lambda}(\lambda, T_{\rm{b}})$ with its temperature
$T_{\rm{b}}$=2.73~K  \citep{mather94}.

\section{THE METHOD OF SOLUTION: REGULARIZATION}
     \label{section:regularization}

Equation (\ref{eq:int_eq1}) is a Fredholm integral equation
of first kind, and its solution is a typical ill-posed inverse problem
\citep{tikhonov95}. The inversion entails solving for the size
distribution function $F(a)$ given a set of observations
$D(x)$ and a kernel $K(x, a)$ which is calculated using an adopted
grain model. As a rule, applying a kernel $K$ to a function $F$ is
a smoothing operation. Consequently, the solution can be extremely
sensitive to small changes in input data. Many specialized tools have
been developed to tackle ill-posed equations. They utilize a priori
information about the solution, such as its continuity, positivity,
monotonic behavior, convexity, points of extremum, to name a few. 
Most popular are the method of regularization
\citep{numerical_recipes,tikhonov95}, the Backus-Gilbert method
\citep{bg70}, and the maximum entropy method (MEM)
\citep[e.g.]{narayan86}.

To solve our specific problem, equation (\ref{eq:int_eq1}),
we prefer to use the method of regularization (MR) which
requires minimum input information: the data $D(x)$,
their uncertainties $\sigma(x)$, and a stabilizing
functional. The stabilizing functional is additional input
information. It does not assume a parametric form for the solution,
but it does assume that the solution will have a certain
smoothness to it. In this sense, the MR has an advantage
over the MEM, which additionally requires a default solution.
Like the stabilizing functional in the MR
(see eq. \ref{eq:phi_s} below),
the default serves to find a stable solution in the MEM.
However, as a rule, it is practically impossible to guess
the default solution in our case, as we will see later
in {\S}~\ref{section:results}. Thus, the solutions derived
with the MEM are more biased than those derived with the MR.

The MEM was used for modeling of interstellar extinction and
polarization by Kim, Martin, \& Hendry (1994), \citet{km94,km95,km96}.
A power law with an exponential cutoff was used as a default solution
in all these papers. The use of such a default was partly
justified because the authors performed their calculations largely
within the framework of then-standard graphite-silicate model by
\citet{dl84} which postulates power law grain-size distributions for
both components. However, there are still problems with
the default even in this case. For example, it is not clear how
to choose the value of the exponential cutoff parameters that
regulate the larger grain-size distribution:
figures 2 and 6 from \citet{km96} can best serve to demonstrate
the problem. At the same time, \citet{z97} and Zubko, Kre{\l}owski,
\& Wegner (1996, 1998) have shown that the MR can be successfully
applied for modeling of both circumstellar and interstellar extinction.
There is no need for any default solution. Since the essential
details of our numerical implementation of MR have already been
described elsewhere \citep{z97}, here we present only the key
features of the method.

In the regularization approach, it is strictly proved that
solving the integral equation (\ref{eq:int_eq1}) is equivalent
to minimization of the following smoothing functional \citep{tikhonov95}:
\begin{equation}
\Psi[F^{(\alpha)}(a)] = \Phi_{\rm{d}}[F^{(\alpha)}(a)] +
  \alpha \Phi_{\rm{s}}[F^{(\alpha)}(a)]
    \label{eq:phi}
\end{equation}
where the two parts are: (1) the discrepancy functional:
\begin{equation}
\Phi_{\rm{d}}[F^{(\alpha)}(a)] = {1 \over n_D} \sum_{j=1}^{n_D}
   \Biggl[ {
   { \int \! K(x_j,a) F^{(\alpha)}(a)\> {\rm{d}}a
      - D(x_j)
   }
   \over
   {w(x_j)} }
   \Biggr]^2 ,
    \label{eq:phi_d}
\end{equation}
and (2) the Tikhonov's stabilizing functional or stabilizer:
\begin{equation}
\Phi_{\rm{s}}[F^{(\alpha)}(a)] = \int \! \{[F^{(\alpha)}(a)]^2 + \beta \>
   [{\rm{d}}F^{(\alpha)}(a)/{\rm{d}}a]^2 \} \> {\rm{d}}a,
    \label{eq:phi_s}
\end{equation}
$\alpha$ is the parameter of regularization ($\alpha \ge 0$),
$\beta$ is a dimensional adjusting parameter ($\beta \ge 0$),
and $w(x)$ is the weight function. $n_D$ is the total
number of data points to fit. In our case, it includes
the numbers of points of the extinction curve and emission spectrum,
plus 6 points of the abundance constraints (C, Si, O, Mg, Fe, N).
Generally, functional $\Phi_{\rm{s}}[F^{(\alpha)}(a)]$ is a measure
of smoothness of the solution and can contain derivatives of higher
orders. However, the practice shows that keeping specific form
(\ref{eq:phi_s}) with the first derivative is sufficient for
solving most ill-posed problems reducing to integral equation
(\ref{eq:int_eq1}) \citep{tikhonov95}. This is also the case
for our present study. If we measure $F$ in $\mu$m$^{-1}$~H$^{-1}$
and $a$ in $\mu$m, then $\alpha$ is expessed in $\mu$m~H$^2$,
and $\beta$ in $\mu$m$^2$. Normally, the weights of both contributors
to (\ref{eq:phi_s}) are chosen to be equal: $\beta$=1
\citep{tikhonov95}. We also adopted this approach.

The solution is given by $F^{(\alpha)}$ that iteratively minimizes
$\Psi$ for an $\alpha$ that satisfies the discrepancy rule:
\begin{equation}
\Phi_{\rm{d}}[F^{(\alpha)}(a)] = \Phi_{\rm{d}}[F^{(\alpha=0)}(a)] +
  {1 \over n_D}
  \sum_{j=1}^{n_D}
    \Bigl[
        { \sigma(x_j) \over w(x_j) }
    \Bigr]^2
      \label{eq:dis_rule}
\end{equation}
Note that equation (\ref{eq:dis_rule}) for $\alpha$ is solved
simultaneously with the minimization of $\Psi$. It has been proved
that $\Phi_{\rm{d}}$ is a monotonic function of $\alpha$:
the larger $\alpha$ the larger $\Phi_{\rm{d}}$ \citep{tikhonov95}.
Thus, there exists a single unique solution for both $\alpha$ and
$F^{(\alpha)}(a)$. 

In the original formulation of the regularization approach
\citep{tikhonov95}, the weight function was set to
$w(x)$$\equiv$1. This formulation was used in the previous
papers on modeling of interstellar extinction by one of us
\citep{zkw96,zkw98}. However, we experimented with various
expressions of $w(x)$ and found out that the choice
of $w(x)$=$\sigma(x)$ is much more suitable for
simultaneous fitting of various data sets
(in our case: extinction + emission + abundances), and
especially those having the large dynamic range of values.
With this choice of $w(x)$, the discrepancy functional
$\Phi_{\rm{d}}$ resembles (but does not coincide with)
the reduced $\chi^2$ function, and the discrepancy rule
(\ref{eq:dis_rule}) reads:
\begin{equation}
 \Phi_{\rm{d}}[F^{(\alpha)}(a)] =
 \Phi_{\rm{d}}[F^{(\alpha=0)}(a)] + 1
      \label{eq:dis_rule_2}
\end{equation}

The introduction of Tikhonov's stabilizer is decisive in finding
a mathematically stable and unique solution, that is
a well-behaved function that does not have wild oscillations
or discontinuities. If we would try to minimize the discrepancy
functional $\Phi_{\rm{d}}$ itself, when
the number of unknown parameters (the values of function $F$
at some discrete grid of grain sizes) exceeds
the number of measurements or observations $n_D$, we would not
obtain a unique solution for $F$: the functional $\Phi_{\rm{d}}$
is degenerate in this case. However, the smoothing functional
$\Phi_{\rm{s}}$ is always nondegenerate. This guarantees
that the combined functional $\Psi$ is nondegenerate in any
case [see chapter 18.4 in Numerical Recipes \citep{numerical_recipes}
for more details].
In addition, because the functional $\Psi$ is a strongly convex
one \citep{tikhonov95}, minimization of $\Psi$ will lead to
a stable and unique solution for $F$. It results in some
smoothing of the solution, and the extent of the smoothing,
which is expressed by the parameter of regularization $\alpha$,
is defined by using the uncertainties $\sigma(x)$
through the discrepancy rule (\ref{eq:dis_rule_2}). 

To minimize the smoothing functional $\Psi$, we use the iterative
technique of conjugate gradient projections on nonnegative
vector set, described in detail by \citet{tikhonov95}.
This provides even more stability to the solution, because
the oscillations with changing sign are a priori excluded.

To estimate the uncertainty of the solution, we used the Monte
Carlo-like computer simulation. Input observational data: extinction,
emission, and elemental abundances have been perturbed by
Gaussian noise with a standard deviation equal to
the errors of the observations. By using the 'new'
observational data, we calculated the grain-size distributions
with the procedure described above. The variance of our results
have been evaluated using an ensemble of such distributions.
For the models reported here, we used the ensemble of 100
size distributions. We also experimented with larger ensembles of
500 and 1000 size distributions, but found that the variance in
these cases is mostly similar to that for the case of
100 distributions. 

In the discrepancy rule (eq. \ref{eq:dis_rule}), the functional
$\Phi_{\rm{d}}[F^{(\alpha)}]$ is calculated for function
$F^{(\alpha)}$ corresponding to the parameter of regularization
$\alpha$, whereas $\Phi_{\rm{d}}[F^{(\alpha=0)}]$ is the minimum
value of $\Phi_{\rm{d}}$ corresponding to the case of
no regularization: $\alpha=0$. Generally, rule (\ref{eq:dis_rule})
is valid for an incompatible problem (eq. \ref{eq:int_eq1}) which has
no exact solution: $\Phi_{\rm{d}}[F^{(\alpha=0)}] \neq 0$.
This is the case for the models reported here.

\section{MODELING DETAILS}
      \label{section:mod_details}

\subsection{Classes of Dust Models}

We have fit the observational constraints with five different dust
compositions, from which we created five different classes of dust models.
The five different dust constituents are: (1) PAHs; (2) graphite;
(3) amorphous carbon of types  ACAR, BE, and ACH2; (4) silicates,
MgFeSiO$_4$; and (5) composite particles containing different proportions
of silicates, organic refractory material (C$_{25}$H$_{25}$O$_5$N),
water ice (H$_2$O), and voids.

These five dust compositions were combined to create five different
classes of dust models. The first class consists of PAHs, and bare
graphite and silicate grains, identical to the carbonaceous/silicate
model recently proposed by \citet{ld01b}. The second class of models
contains in addition to PAHs, bare graphite and silicate grains,
composite particles. The third and fourth classes of models comprise
of the first and second classes but with the graphite grains completely
replaced by amorphous carbon grains. In the fifth class of models
the only carbon is in PAHs, that is, it comprises of only PAHs,
bare silicate, and composite particles. 

We designated models including only PAHs and bare grains as BARE,
and those containing additional composite particles as COMP. 
The BARE and COMP models were further subdivided according to
the composition of the bare carbon particles used in the model: graphite,
amorphous carbon, and no carbon, designated respectively by
-GR, -AC, and -NC. For all model classes, we performed the fit using
the three different sets of ISM abundances listed in
Table~\ref{tab_dust_abund}: solar, B~stars, and F and G~stars
(correspondingly designated as -S, -B, and -FG submodels).
So, for example, a BARE model with graphite grains derived by assuming
the solar ISM abundances is designated as BARE-GR-S, or a COMP model
with no carbon obtained by assuming the B star abundances is designated
as COMP-NC-B.

\subsection{Computational Details}

For computational purposes, we used a discrete set of wavelengths to
calculate $D(x)$: a set of 100 wavelengths equally spaced in
$\lambda^{-1}$ between 1/3.0 and 1/0.912 {\micron}$^{-1}$ to characterize
the extinction; eight wavelengths at 3.5, 4.9, 12, 25, 60,
100, 140, and 240~{\micron}, corresponding to the nominal wavelengths
of the DIRBE filters, and 93 wavelengths between 160 and 1000 {\micron},
corresponding to the FIRAS channels, to characterize the IR emission. 
For each dust component, we constructed a grid of 30--100 grain radii,
logarithmically distributed between $a_{\rm{min}}$  and $a_{\rm{max}}$. 
We adopted $a_{\rm{min}}$=3.5~{\AA} for PAHs, graphite, and silicate grains,
because smaller grains are photolytically unstable \citep{gd89}.
For composite grains, we performed calculations for larger
values of $a_{\rm{min}}$=100--500~{\AA}, since composite grains, being
aggregates of smaller particles, are a priori expected to be larger.
For the composite grains the results do not depend drastically on
$a_{\rm{min}}$, so we adopted $a_{\rm{min}}$=200~{\AA} for these particles.
Note that for the grains of sizes larger than 200~{\AA}, the contribution
of temperature fluctuations to the grain energy balance becomes negligible,
and we can use the equilibrium temperature approach for calculating
the emissivity of the grains. For all components, we chose the upper
limit $a_{\rm{max}}$=5~{\micron}, but found that the size distribution
is essentially zero for  $a>1$ {\micron} for all the models even
with $a_{\rm{max}}$ up to 20--50~{\micron}.

\section{RESULTS}
      \label{section:results}

\subsection{Comparison with Constraints}
      \label{subsection:comp}

\begin{figure}
\begin{center}
\epsscale{1.0}
\plottwo{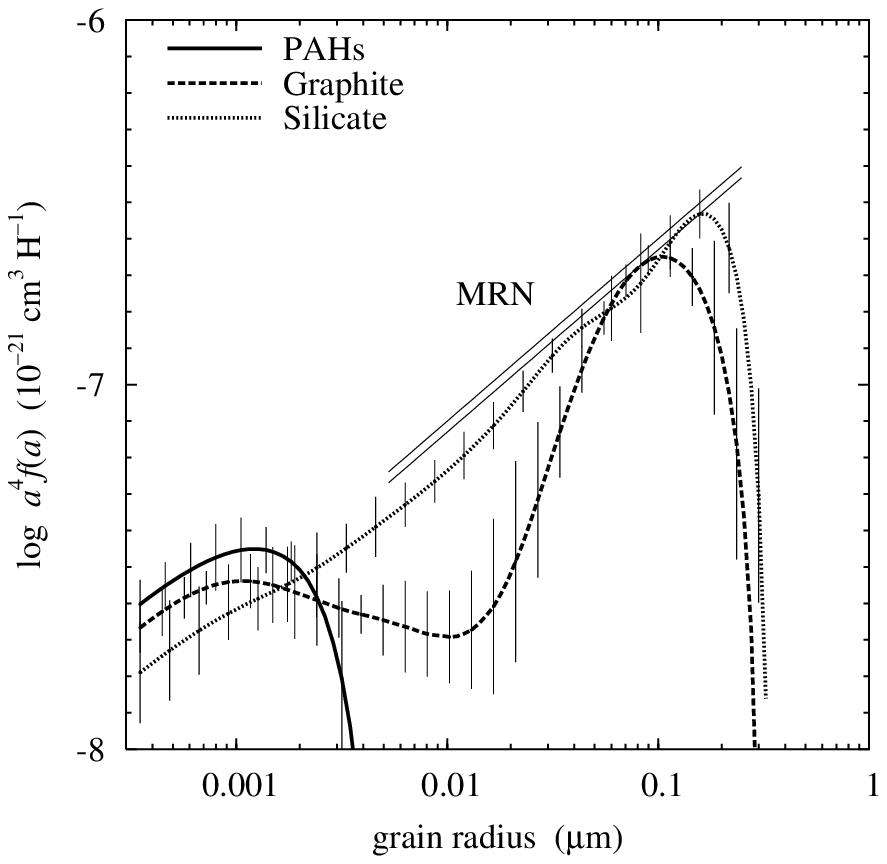}{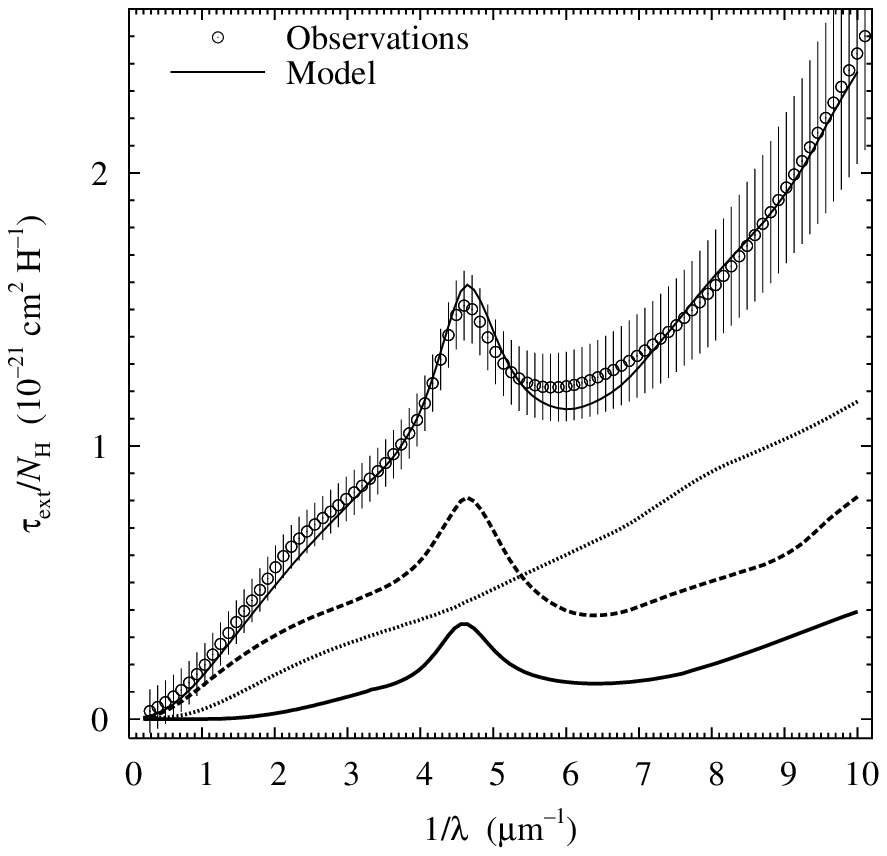}
\epsscale{2.2}
\plottwo{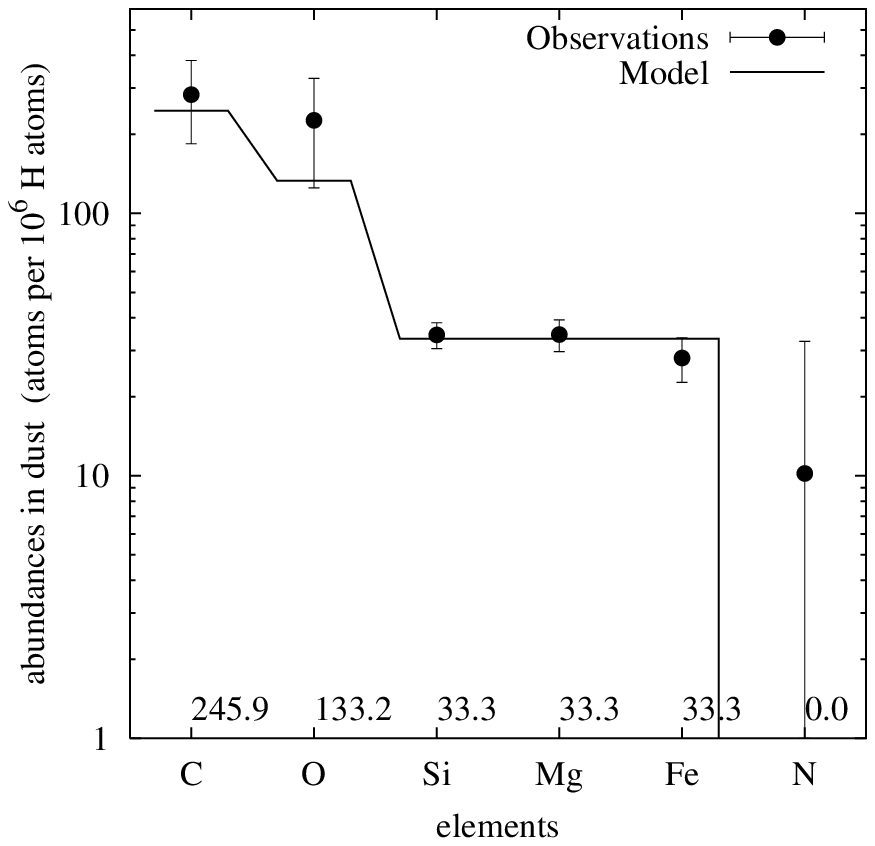}{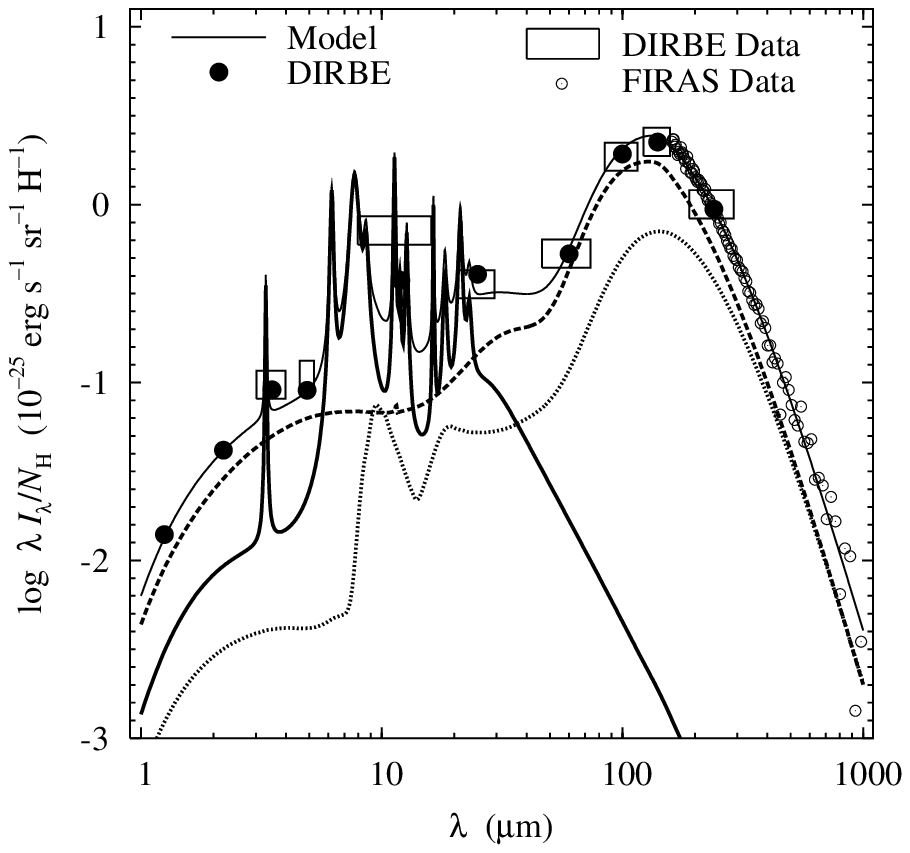}
\caption{ BARE-GR-S dust model: the size distributions (top left),
  extinction curve (top right), elemental requirements (bottom left),
  and emission spectrum (bottom right). Two straight lines are
  the MRN size distributions for silicate (upper line) and
  graphite (lower line).
}
   \label{fig:bare-s-gr}
\end{center}
\end{figure}

\begin{figure}
\begin{center}
\epsscale{1.0}
\plottwo{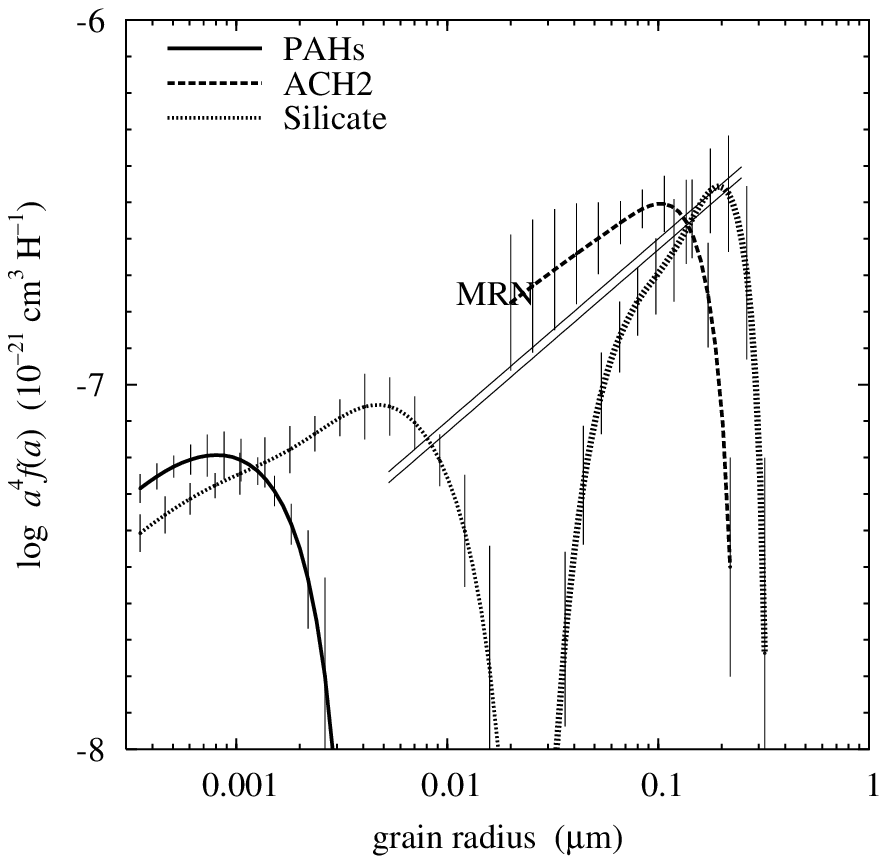}{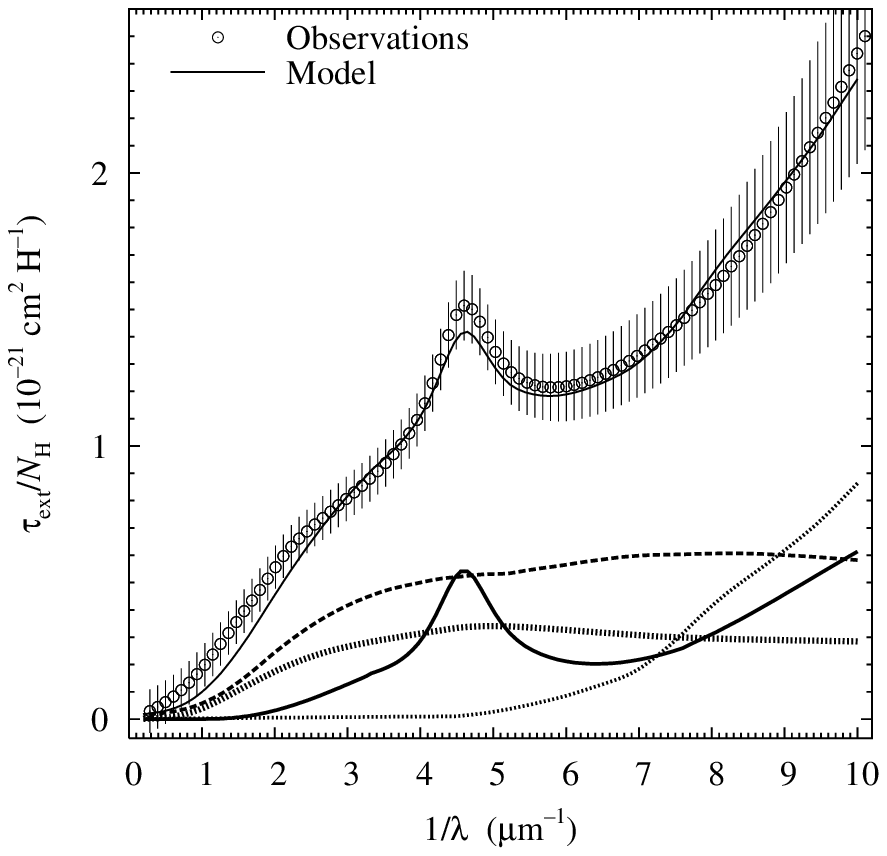}
\epsscale{2.2}
\plottwo{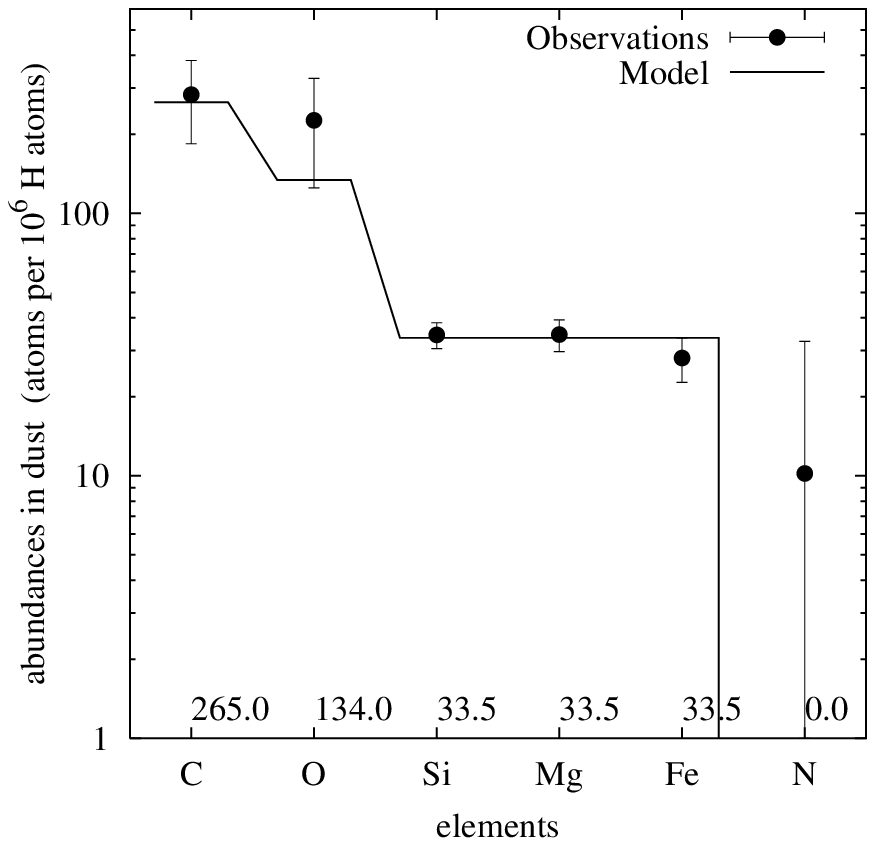}{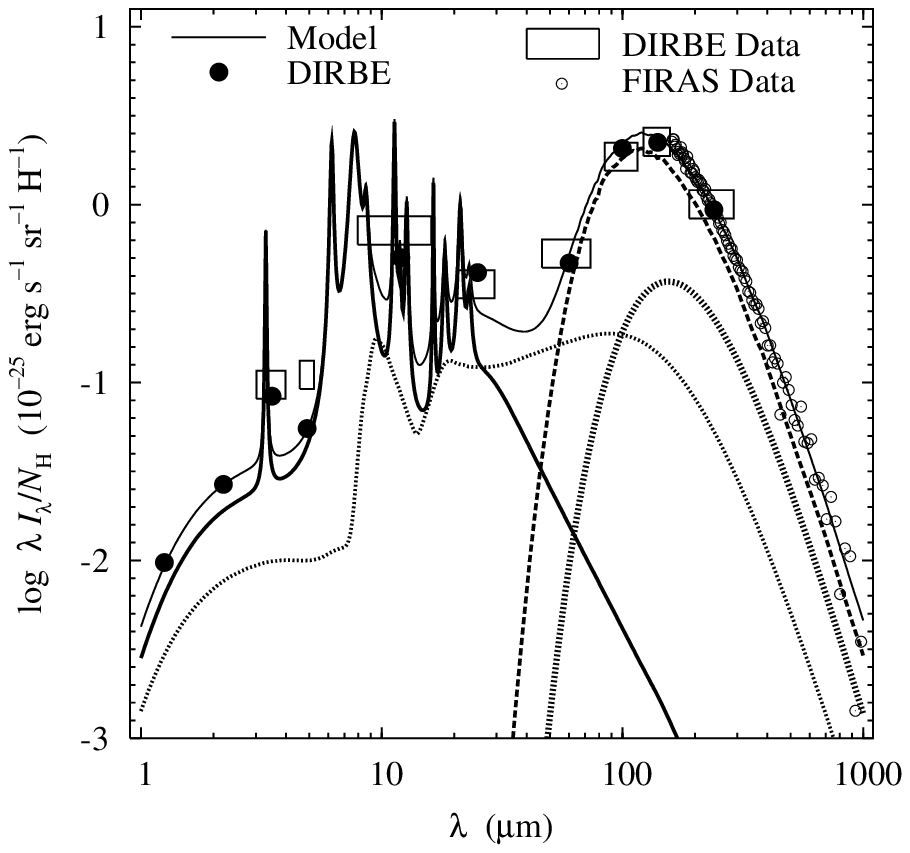}
\caption{ BARE-AC-S dust model: the size distributions (top left),
  extinction curve (top right), elemental requirements (bottom left),
  and emission spectrum (bottom right). Various populations of
  the same dust component are depicted by the lines of various
  width. Two straight lines are
  the MRN size distributions for silicate (upper line) and
  graphite (lower line).
}
   \label{fig:bare-s-ac}
\end{center}
\end{figure}

\begin{figure}
\begin{center}
\epsscale{1.0}
\plottwo{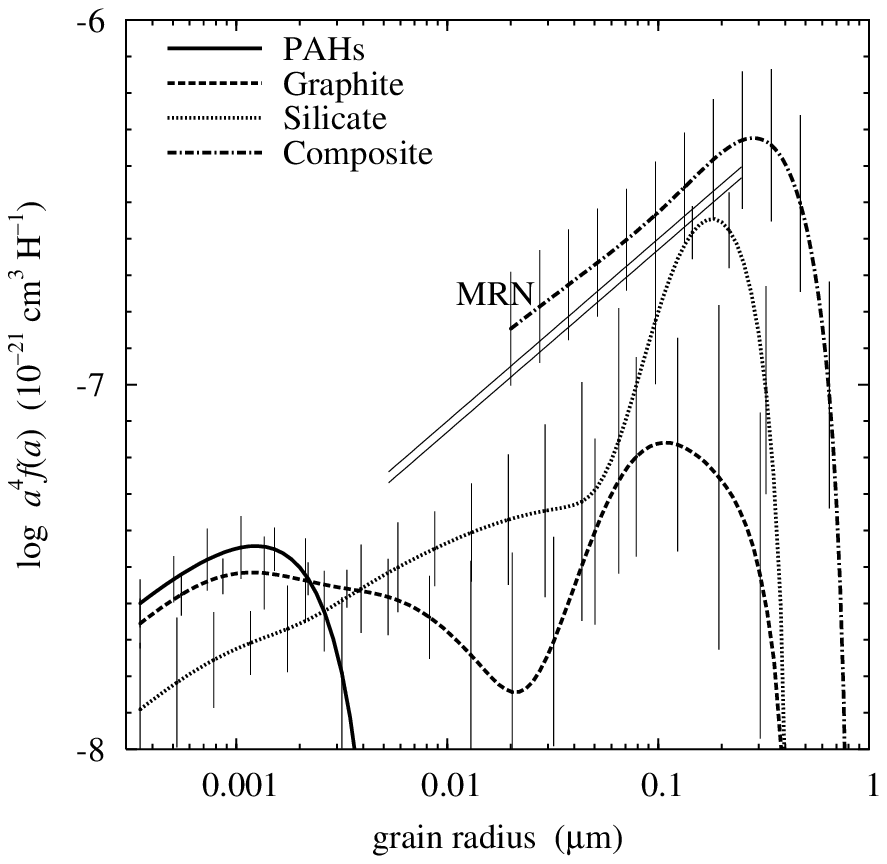}{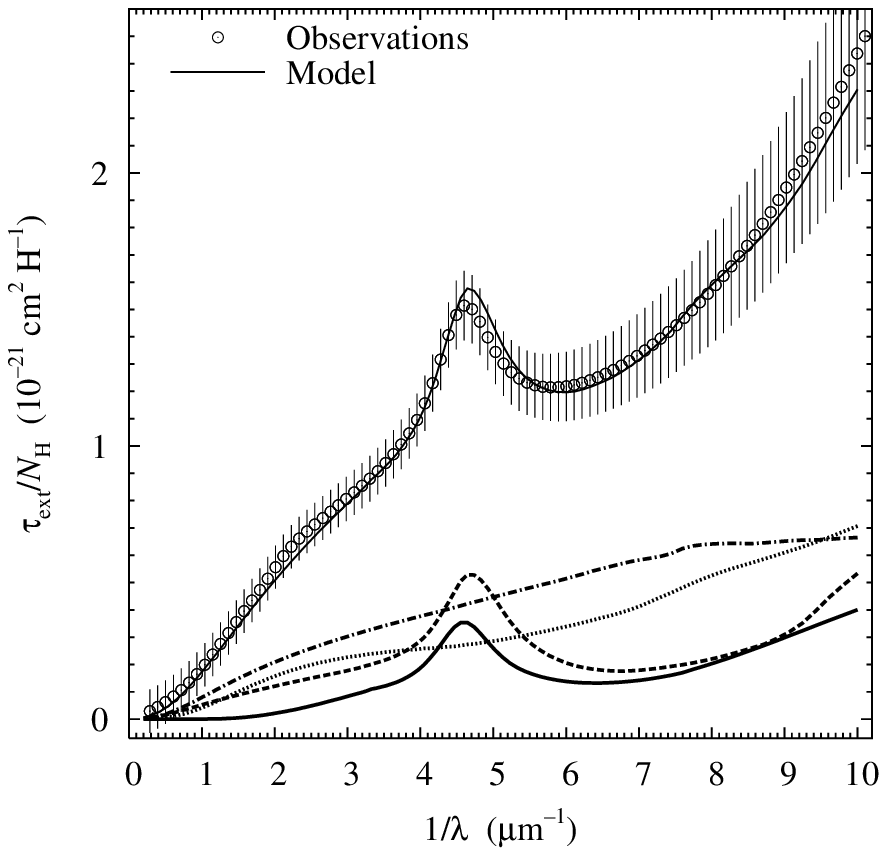}
\epsscale{2.2}
\plottwo{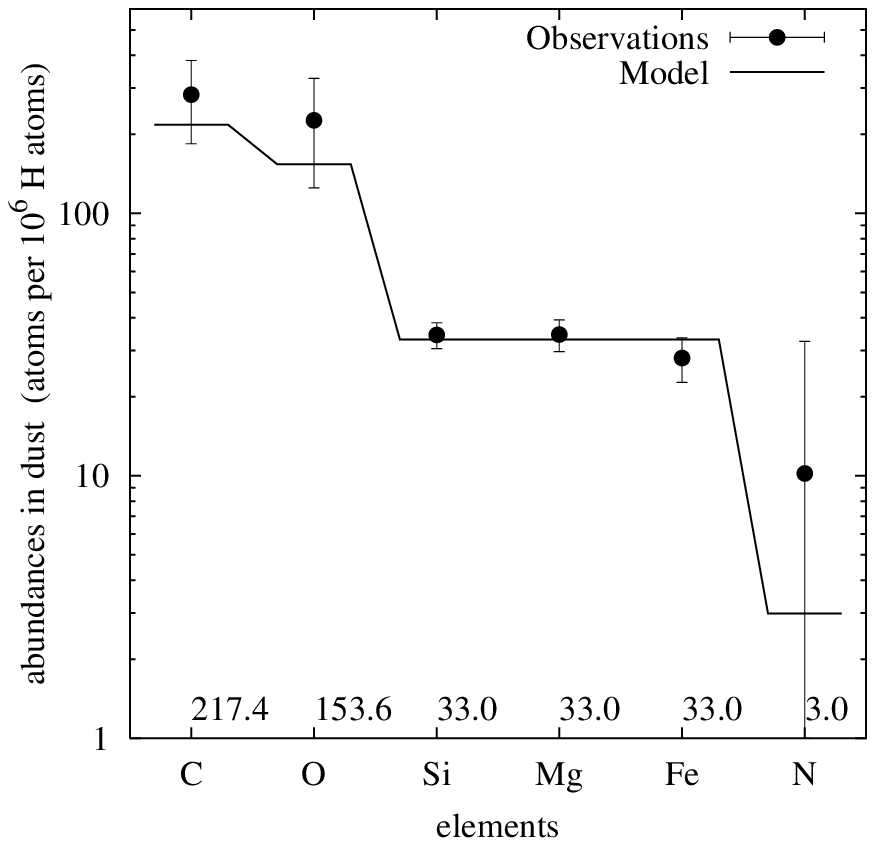}{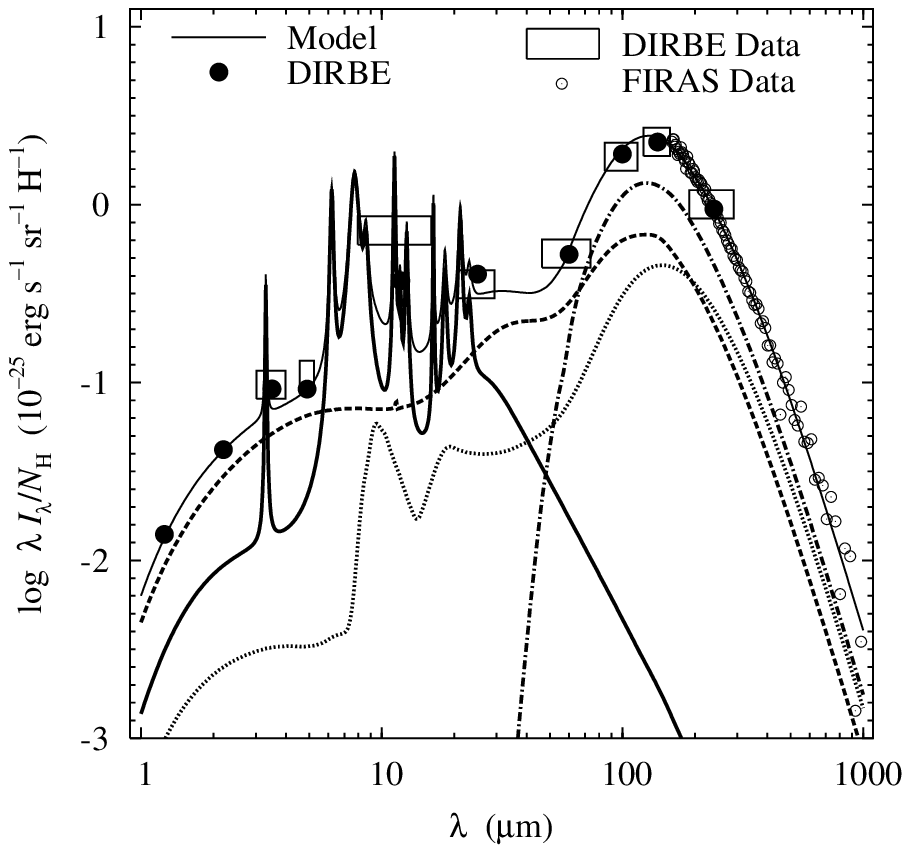}
\caption{ COMP-GR-S dust model: the size distributions (top left),
  extinction curve (top right), elemental requirements (bottom left),
  and emission spectrum (bottom right). Two straight lines are
  the MRN size distributions for silicate (upper line) and
  graphite (lower line).
}
   \label{fig:comp-s-gr}
\end{center}
\end{figure}

\begin{figure}
\begin{center}
\epsscale{1.0}
\plottwo{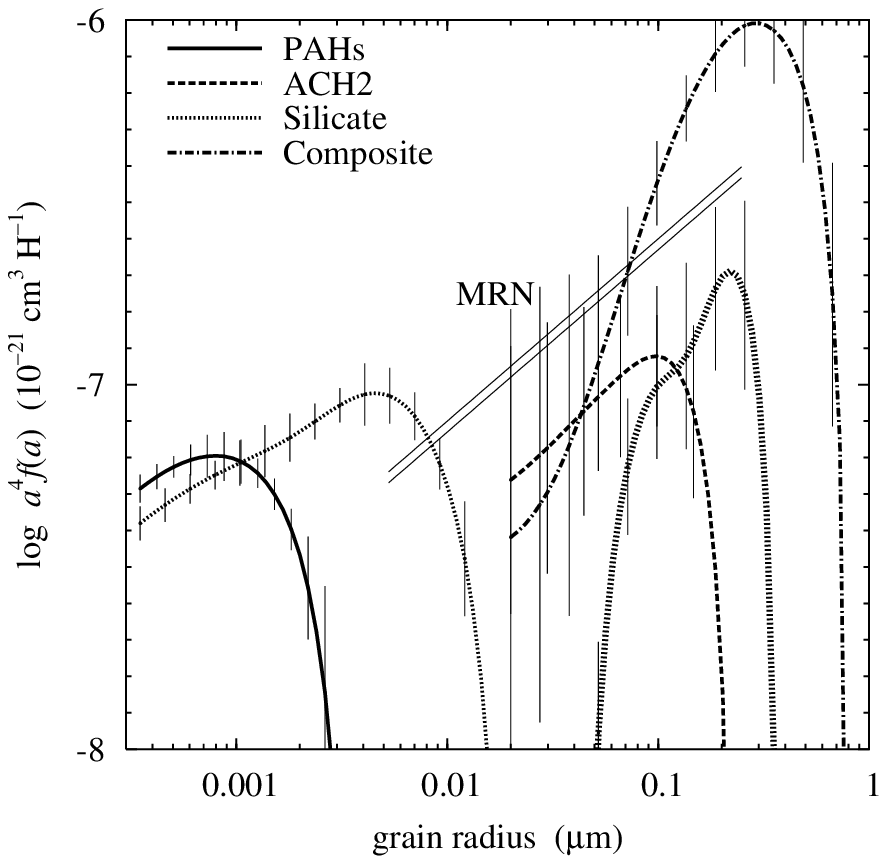}{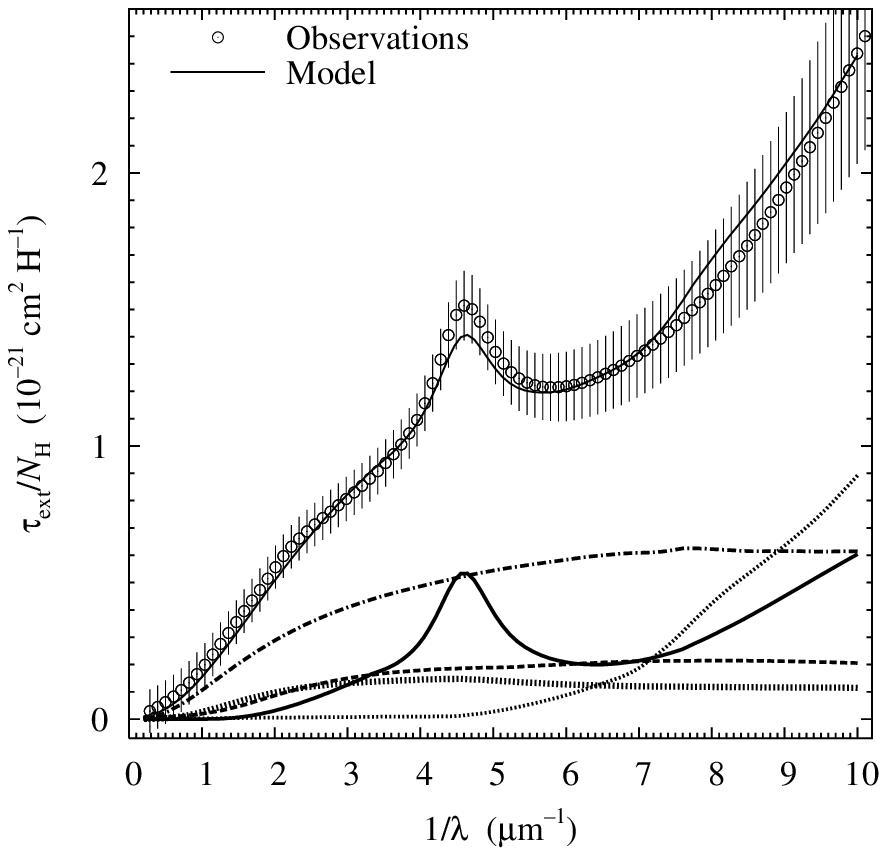}
\epsscale{2.2}
\plottwo{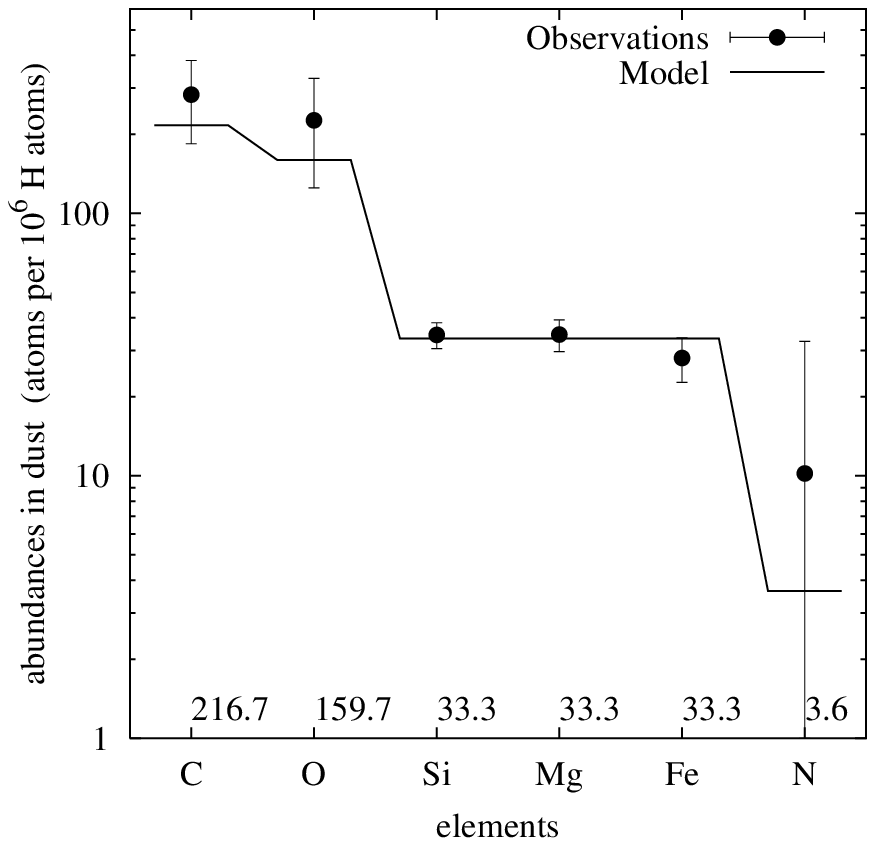}{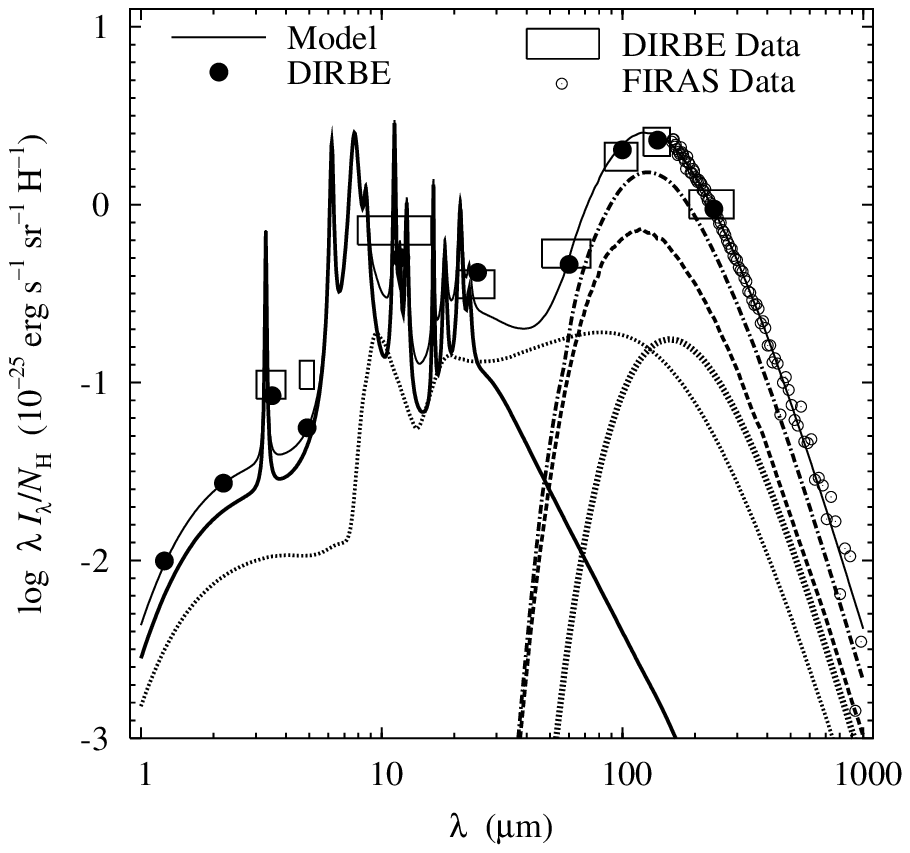}
\caption{ COMP-AC-S dust model: the size distributions (top left),
  extinction curve (top right), elemental requirements (bottom left),
  and emission spectrum (bottom right). Various populations of
  the same dust component are depicted by the lines of various
  width. Two straight lines are
  the MRN size distributions for silicate (upper line) and
  graphite (lower line).
}
   \label{fig:comp-s-ac}
\end{center}
\end{figure}

\begin{figure}
\begin{center}
\epsscale{1.0}
\plottwo{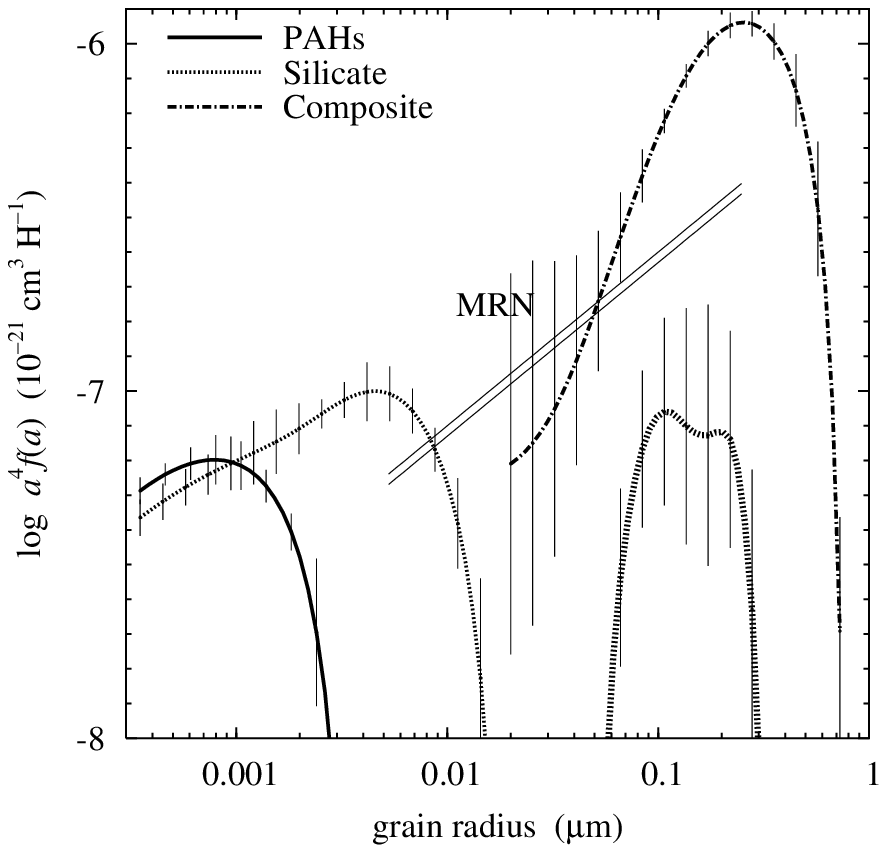}{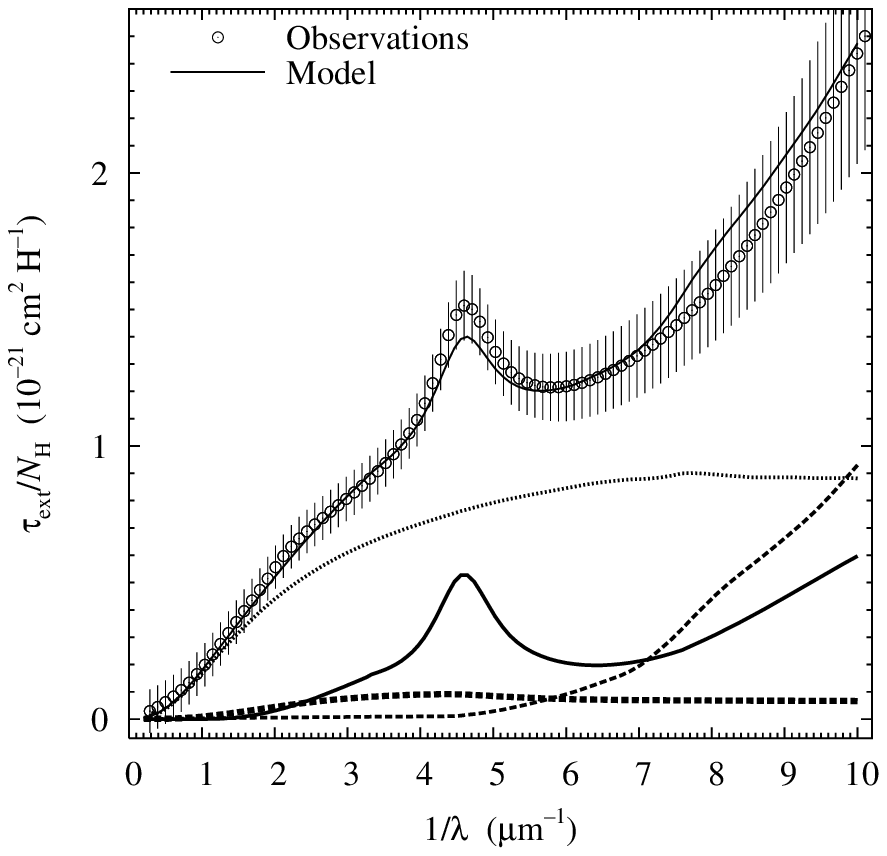}
\epsscale{2.2}
\plottwo{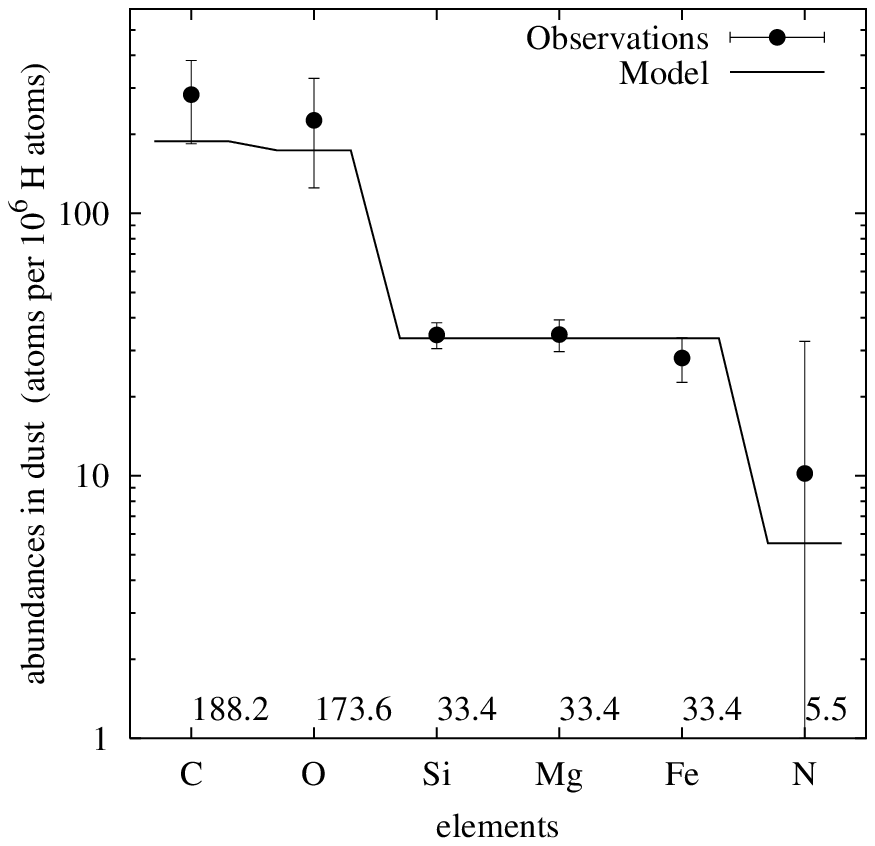}{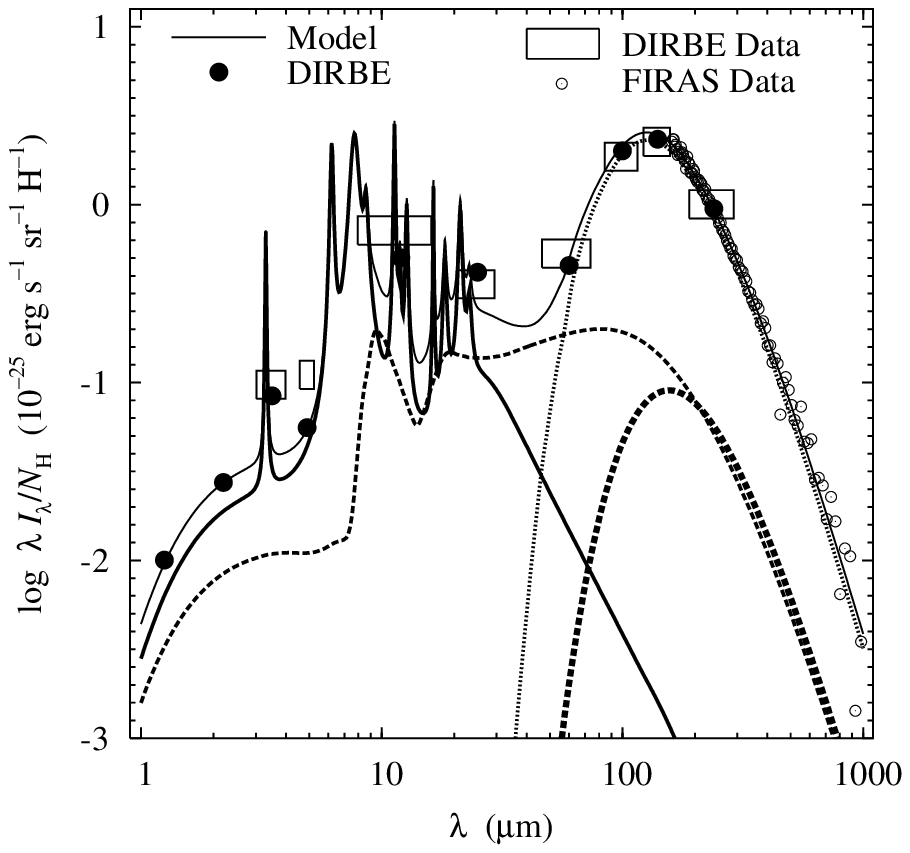}
\caption{ COMP-NC-S dust model: the size distributions (top left),
  extinction curve (top right), elemental requirements (bottom left),
  and emission spectrum (bottom right). Various populations of
  the same dust component are depicted by the lines of various
  width. Two straight lines are
  the MRN size distributions for silicate (upper line) and
  graphite (lower line).
}
   \label{fig:comp-s-nc}
\end{center}
\end{figure}

\begin{figure}
\begin{center}
\epsscale{1.0}
\plottwo{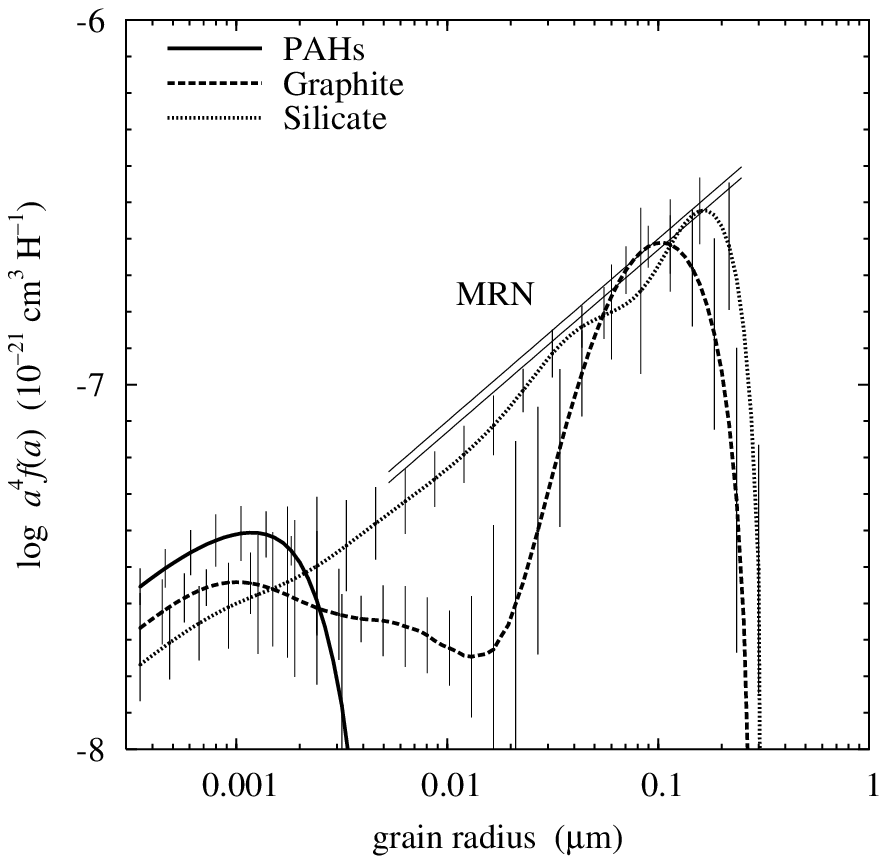}{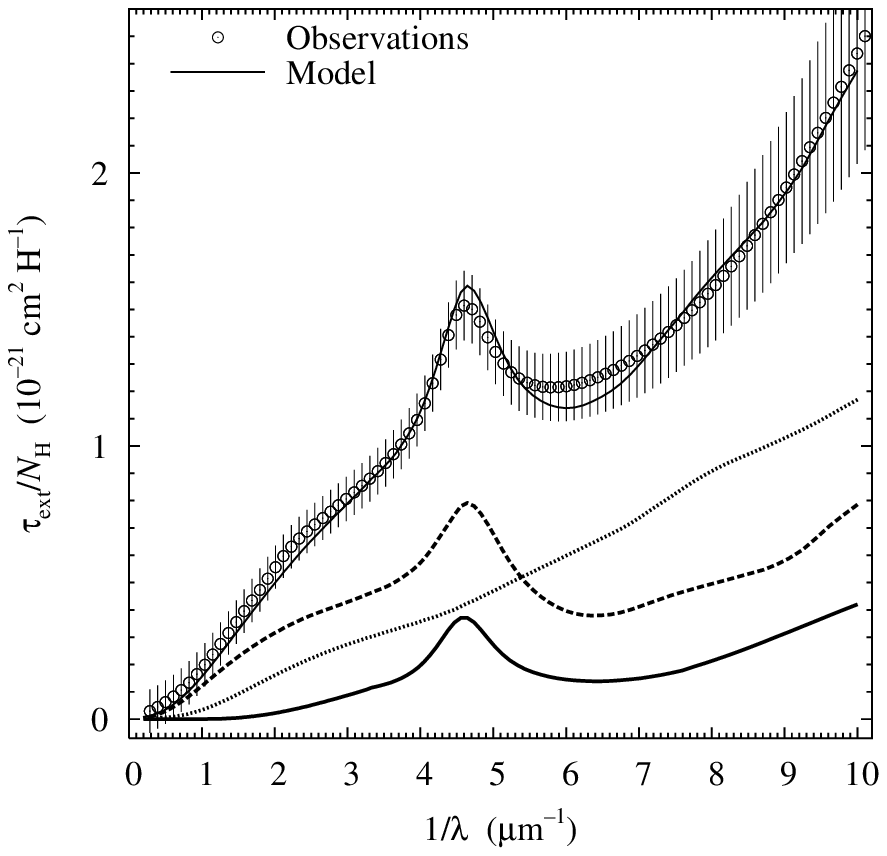}
\epsscale{2.2}
\plottwo{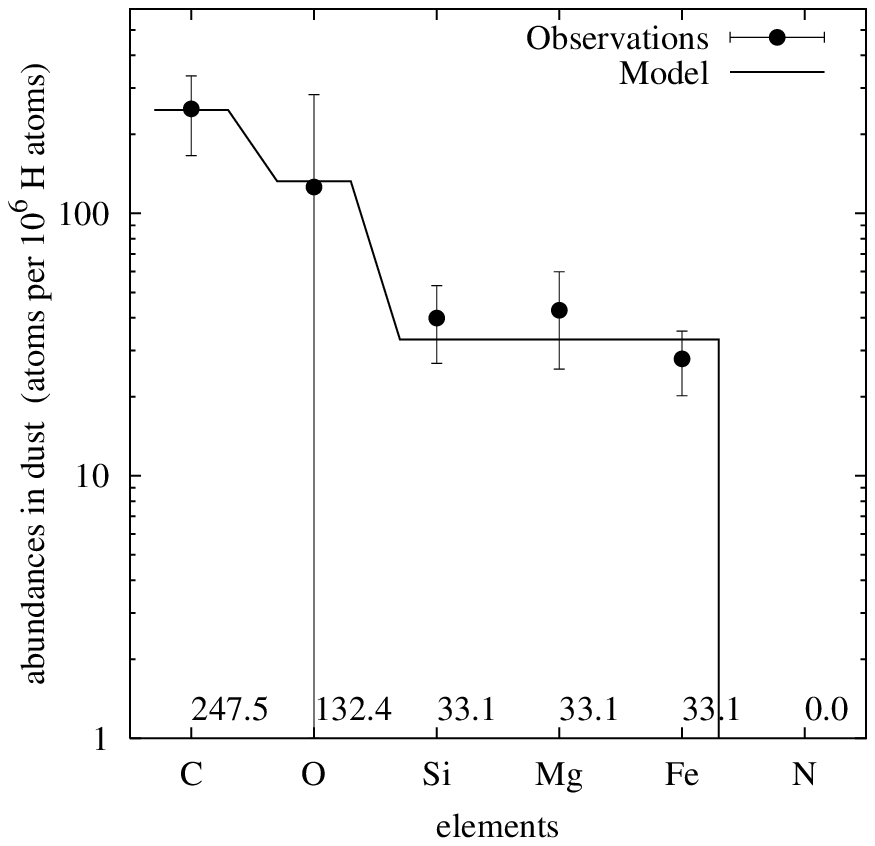}{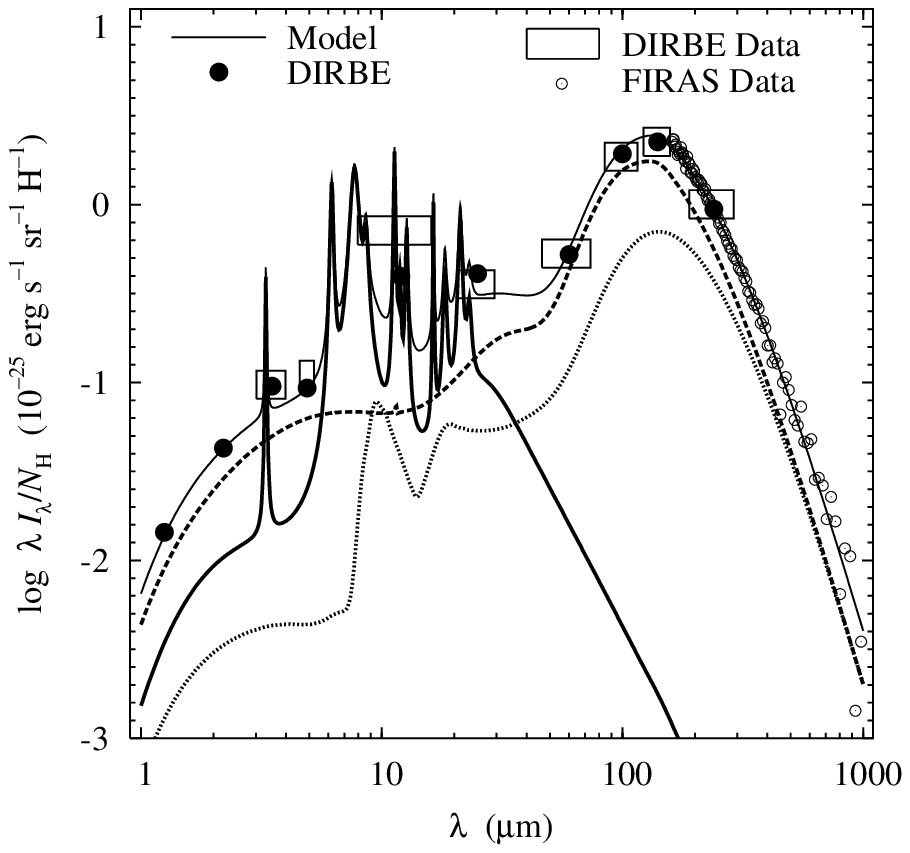}
\caption{ BARE-GR-FG dust model: the size distributions (top left),
  extinction curve (top right), elemental requirements (bottom left),
  and emission spectrum (bottom right). Two straight lines are
  the MRN size distributions for silicate (upper line) and
  graphite (lower line).
}
   \label{fig:bare-fg-gr}
\end{center}
\end{figure}

\begin{figure}
\begin{center}
\epsscale{1.0}
\plottwo{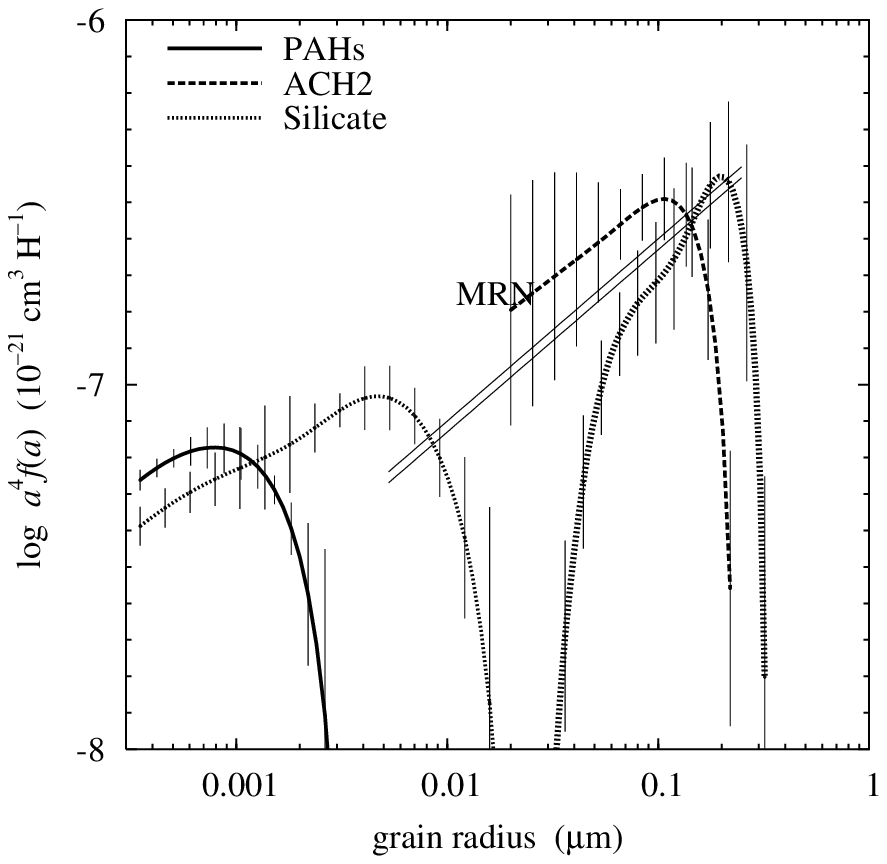}{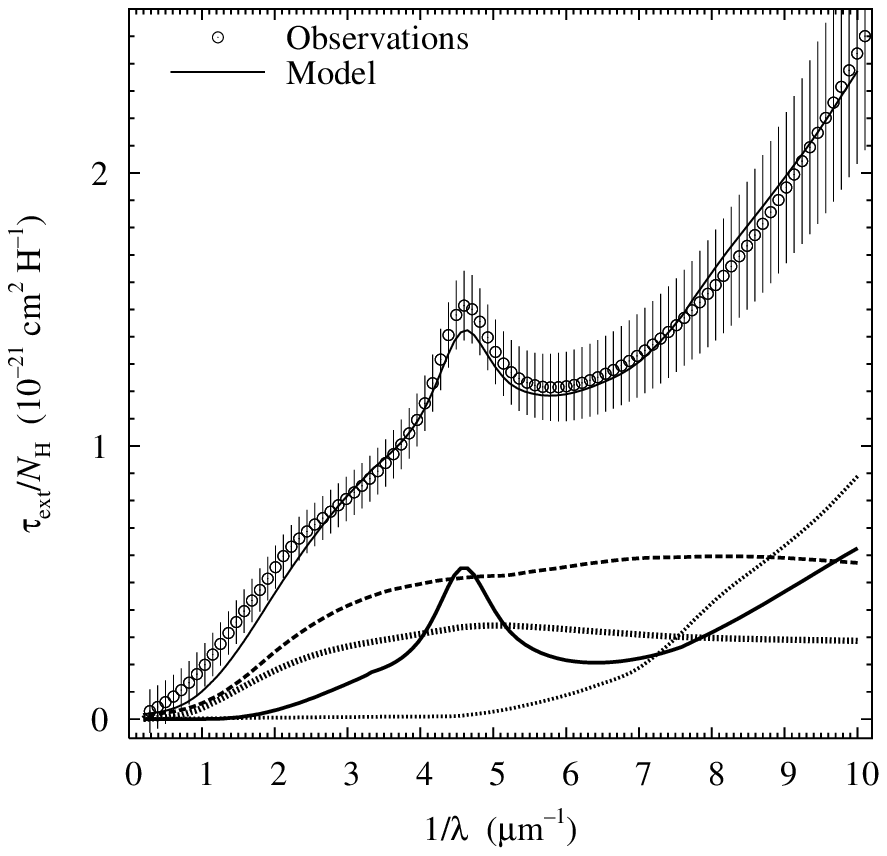}
\epsscale{2.2}
\plottwo{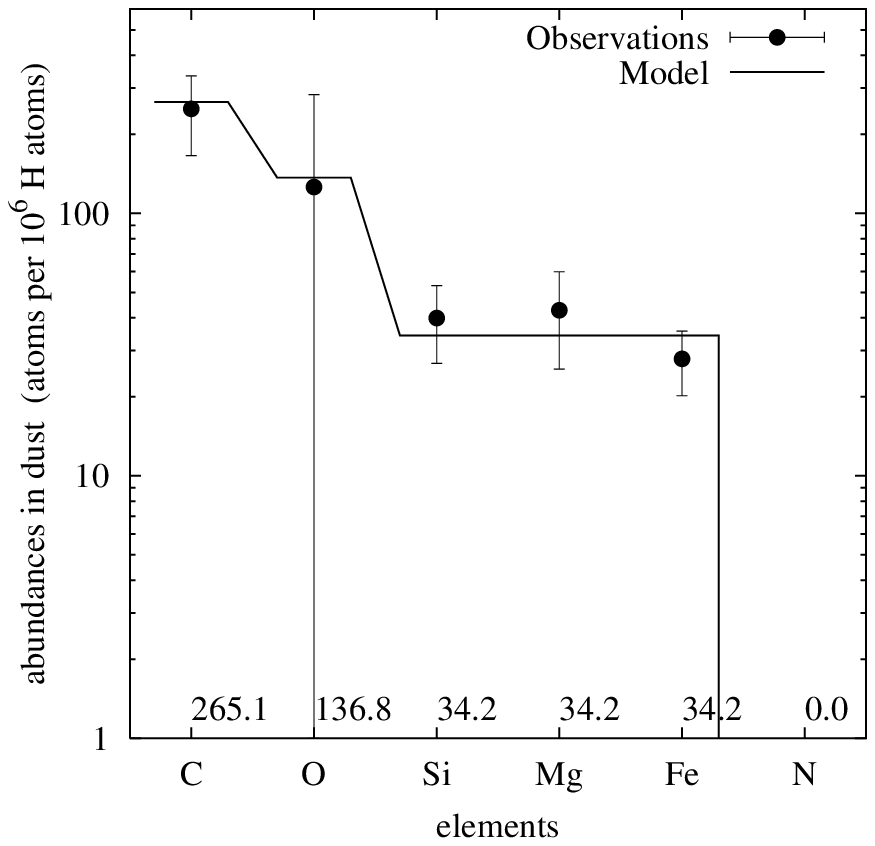}{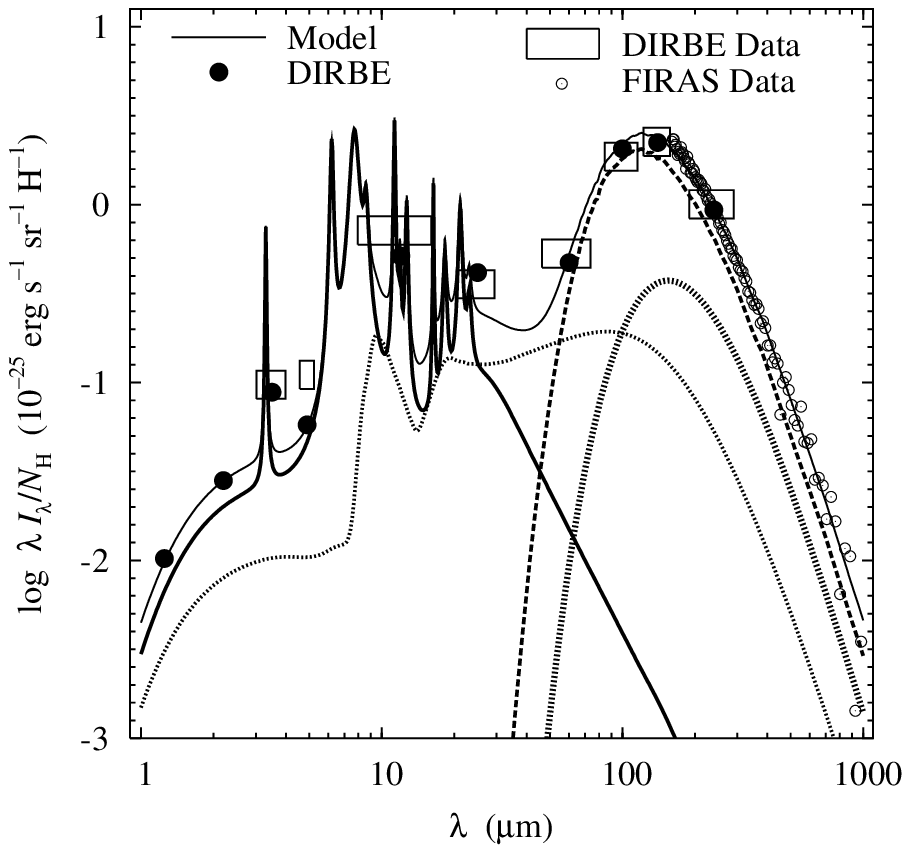}
\caption{ BARE-AC-FG dust model: the size distributions (top left),
  extinction curve (top right), elemental requirements (bottom left),
  and emission spectrum (bottom right). Various populations of
  the same dust component are depicted by the lines of various
  width. Two straight lines are
  the MRN size distributions for silicate (upper line) and
  graphite (lower line).
}
   \label{fig:bare-fg-ac}
\end{center}
\end{figure}

\begin{figure}
\begin{center}
\epsscale{1.0}
\plottwo{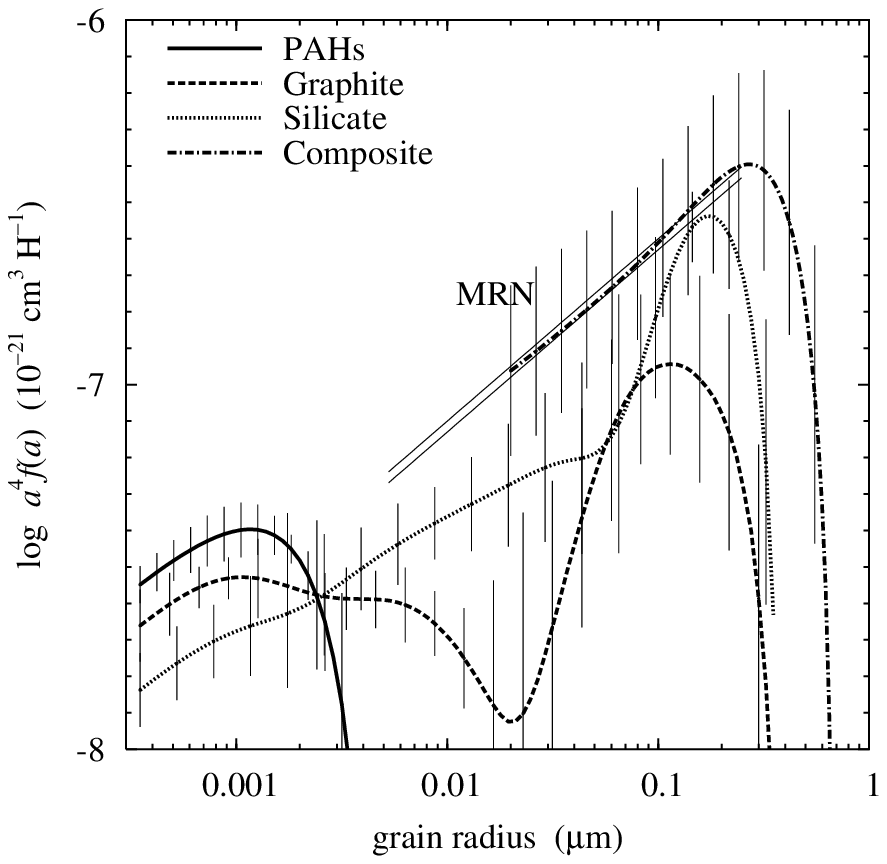}{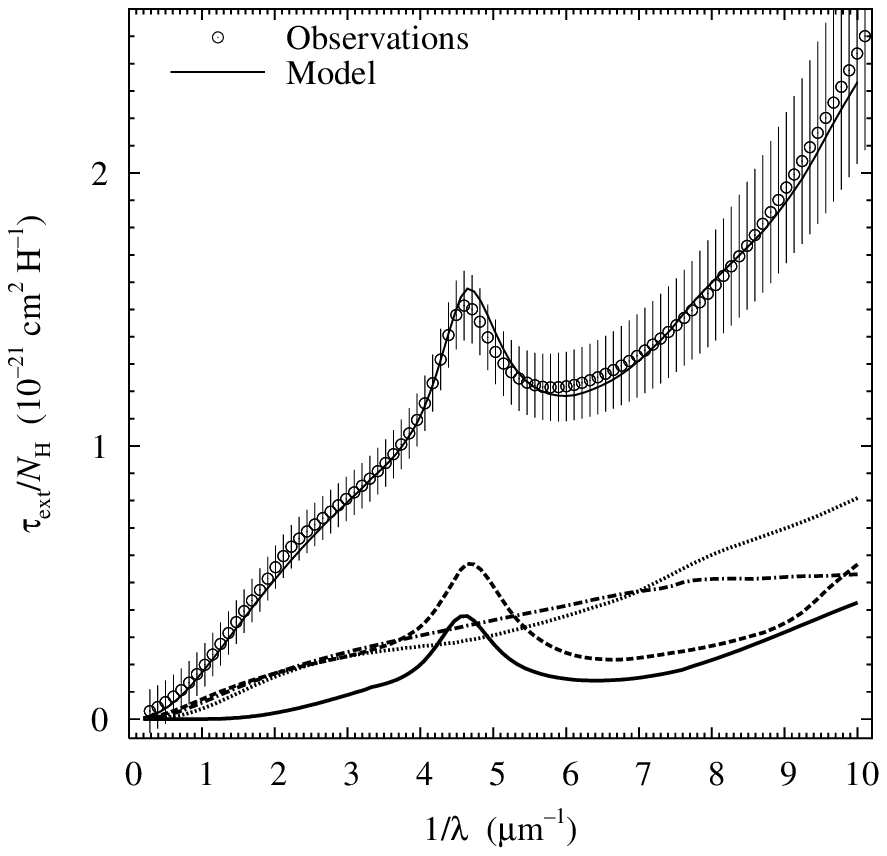}
\epsscale{2.2}
\plottwo{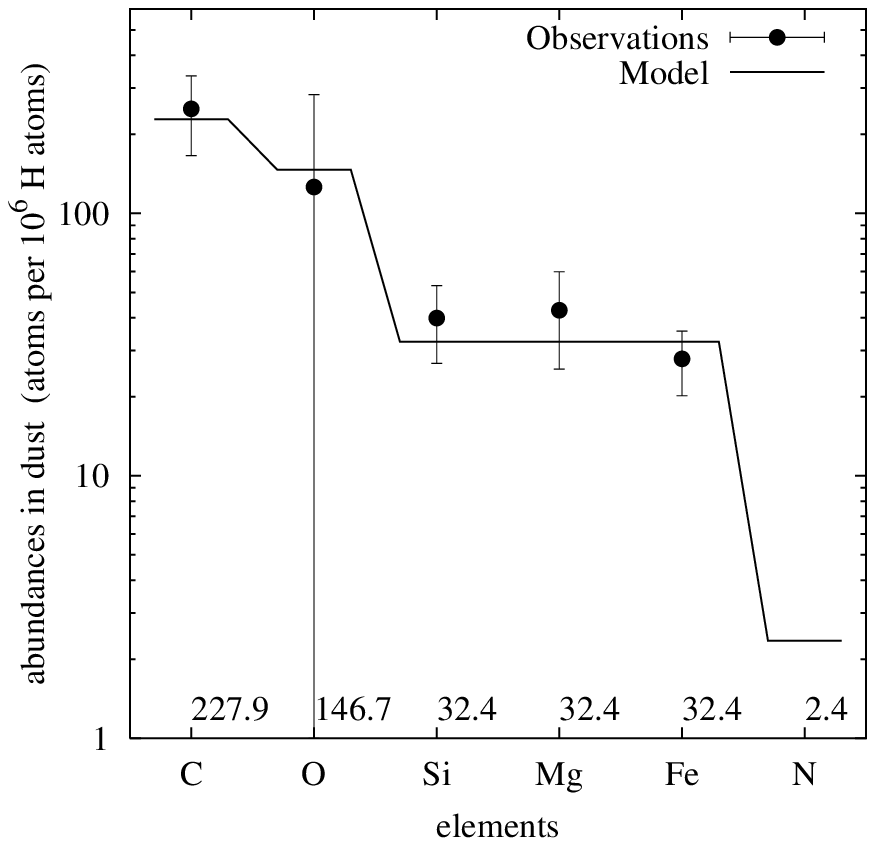}{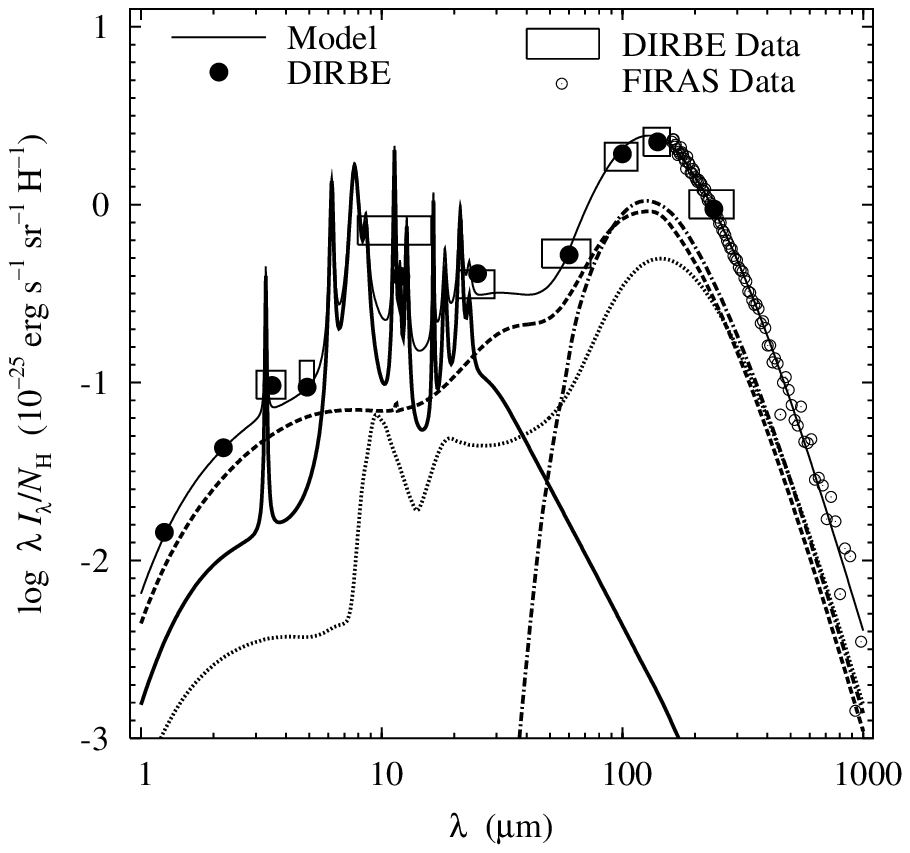}
\caption{ COMP-GR-FG dust model: the size distributions (top left),
  extinction curve (top right), elemental requirements (bottom left),
  and emission spectrum (bottom right). Two straight lines are
  the MRN size distributions for silicate (upper line) and
  graphite (lower line).
}
   \label{fig:comp-fg-gr}
\end{center}
\end{figure}

\begin{figure}
\begin{center}
\epsscale{1.0}
\plottwo{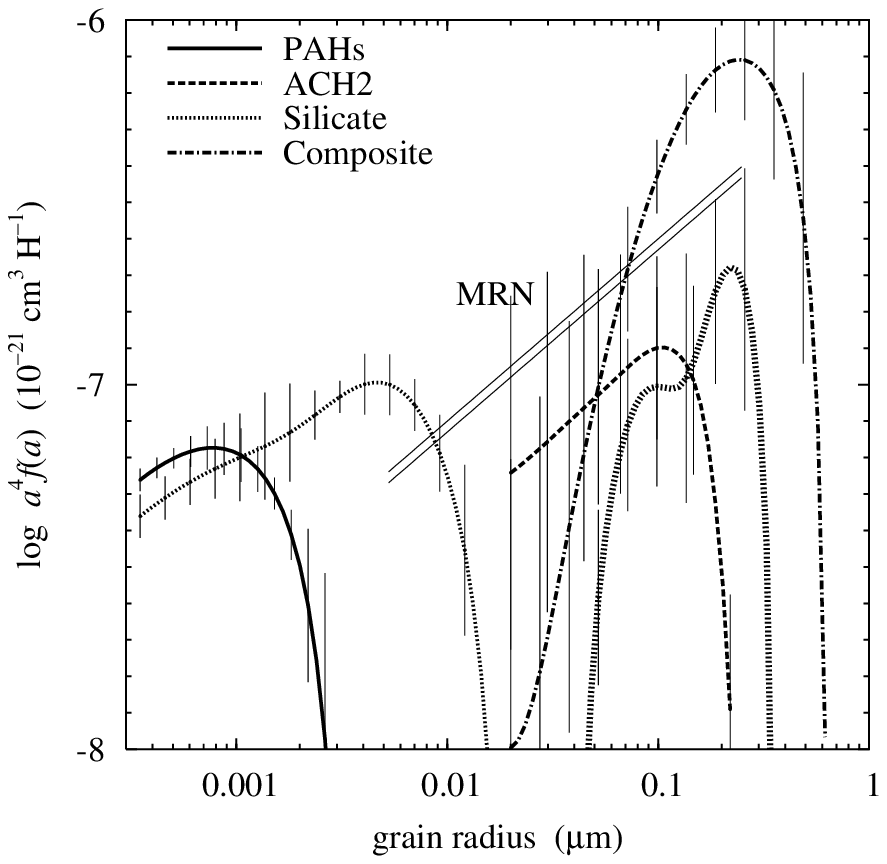}{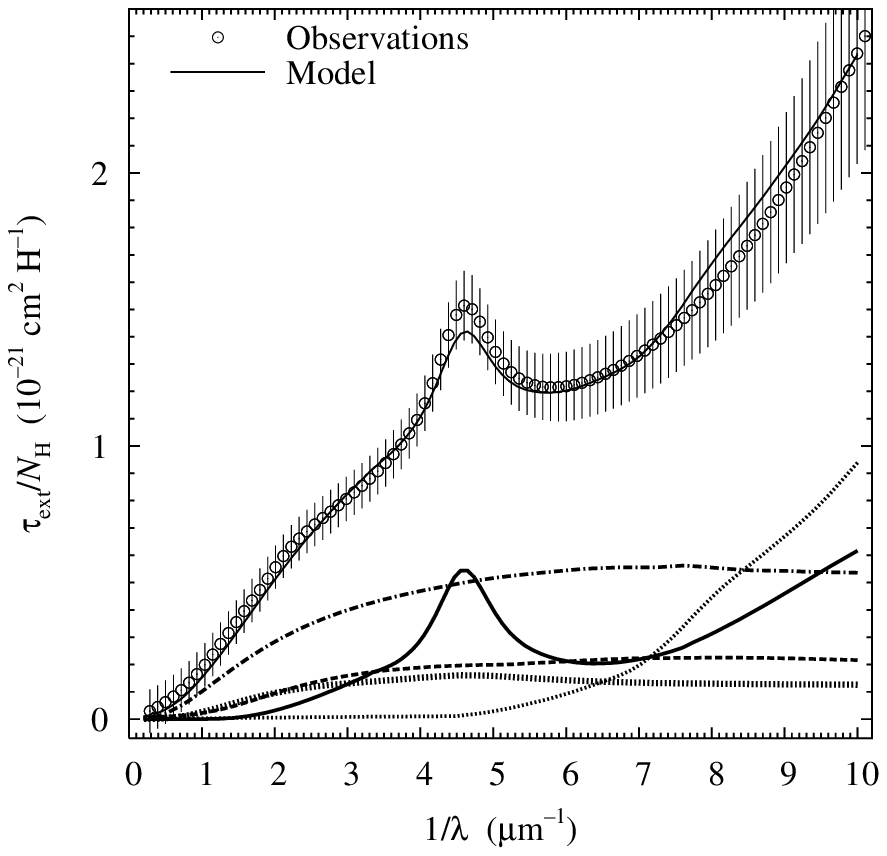}
\epsscale{2.2}
\plottwo{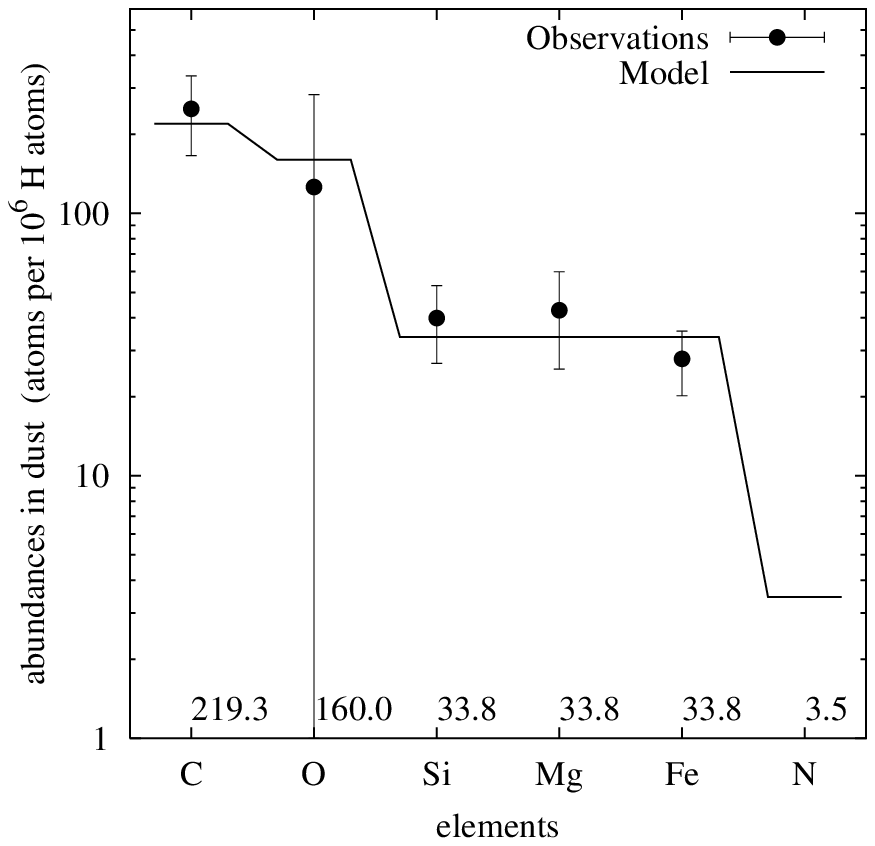}{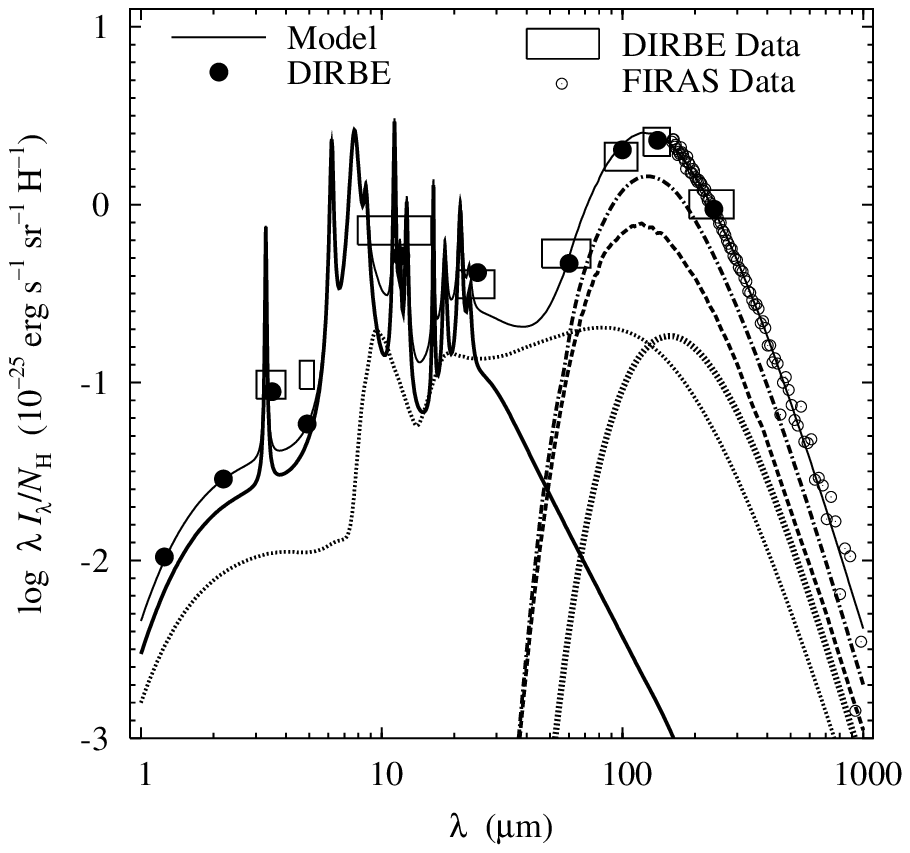}
\caption{ COMP-AC-FG dust model: the size distributions (top left),
  extinction curve (top right), elemental requirements (bottom left),
  and emission spectrum (bottom right). Various populations of
  the same dust component are depicted by the lines of various
  width. Two straight lines are
  the MRN size distributions for silicate (upper line) and
  graphite (lower line).
}
   \label{fig:comp-fg-ac}
\end{center}
\end{figure}

\begin{figure}
\begin{center}
\epsscale{1.0}
\plottwo{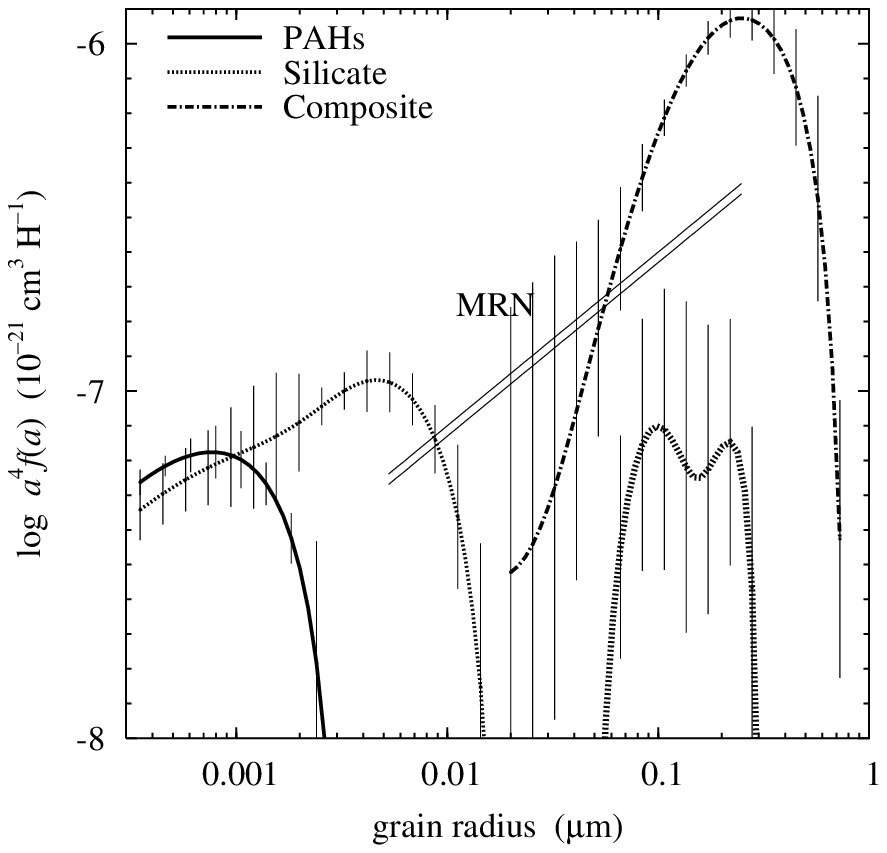}{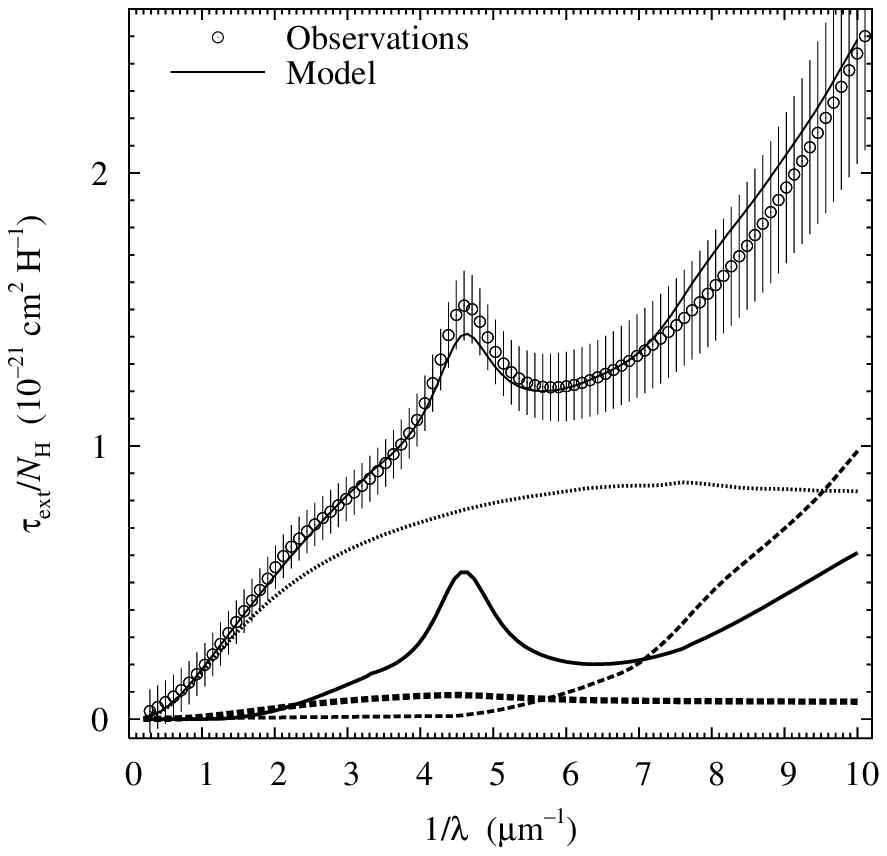}
\epsscale{2.2}
\plottwo{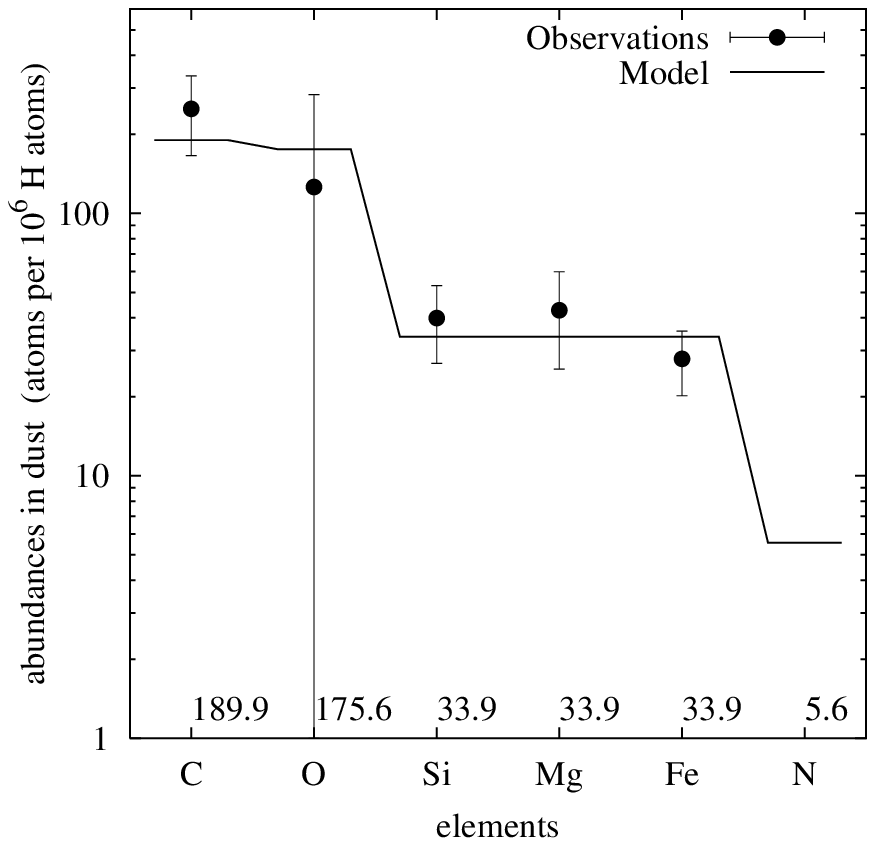}{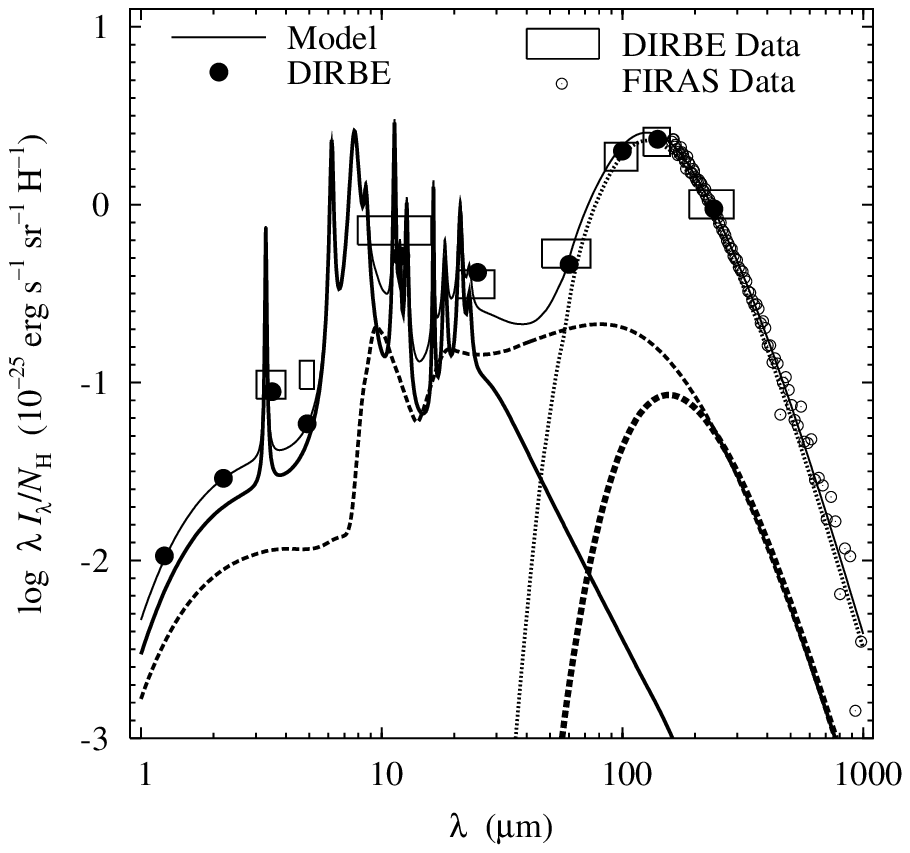}
\caption{ COMP-NC-FG dust model: the size distributions (top left),
  extinction curve (top right), elemental requirements (bottom left),
  and emission spectrum (bottom right). Various populations of
  the same dust component are depicted by the lines of various
  width. Two straight lines are
  the MRN size distributions for silicate (upper line) and
  graphite (lower line).
}
   \label{fig:comp-fg-nc}
\end{center}
\end{figure}

\begin{figure}
\begin{center}
\epsscale{1.0}
\plottwo{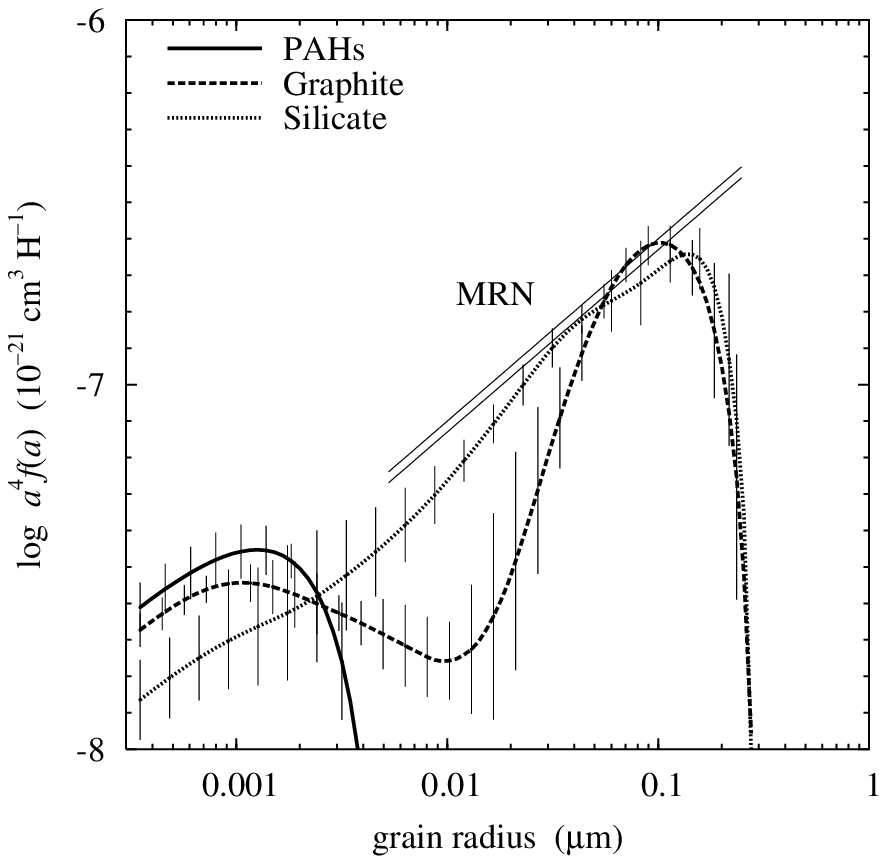}{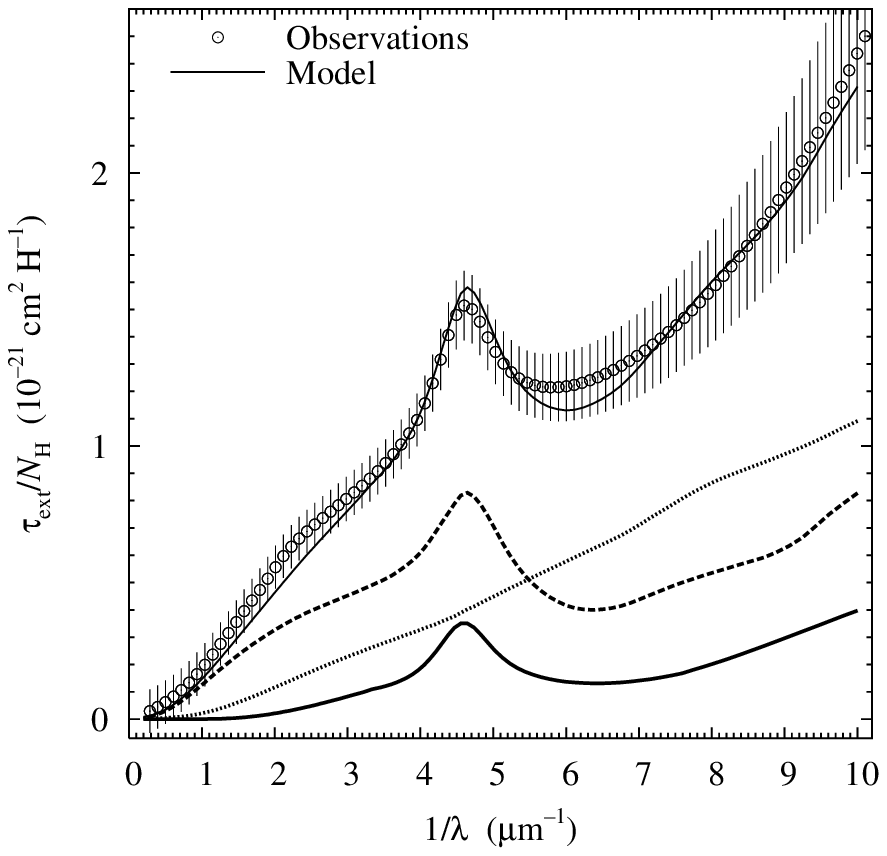}
\epsscale{2.2}
\plottwo{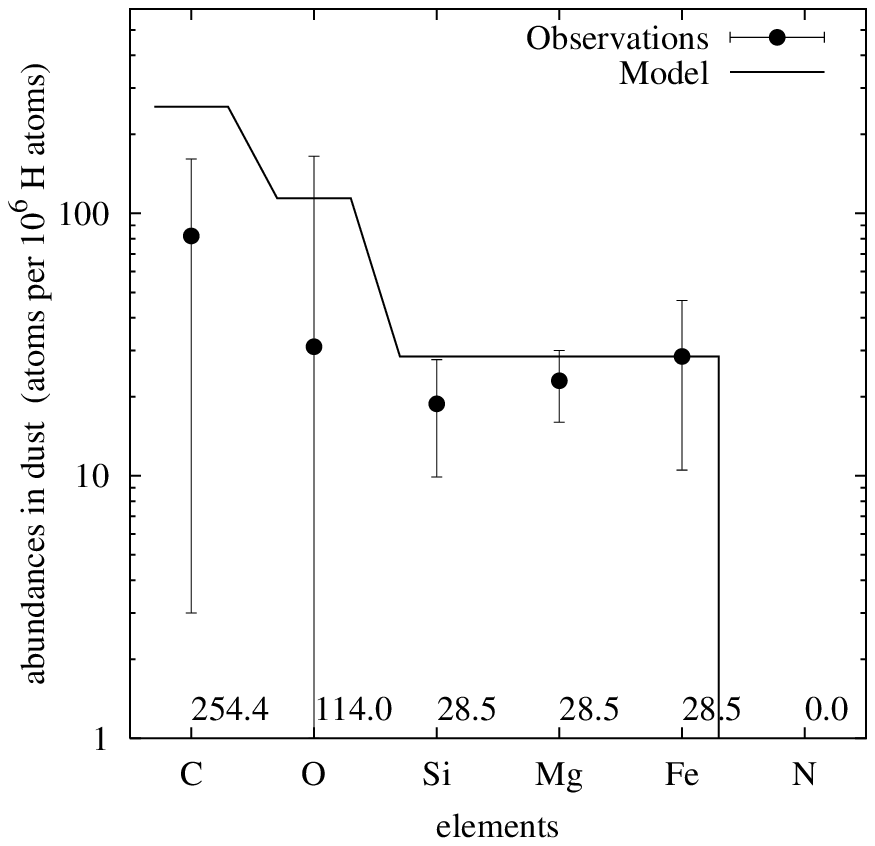}{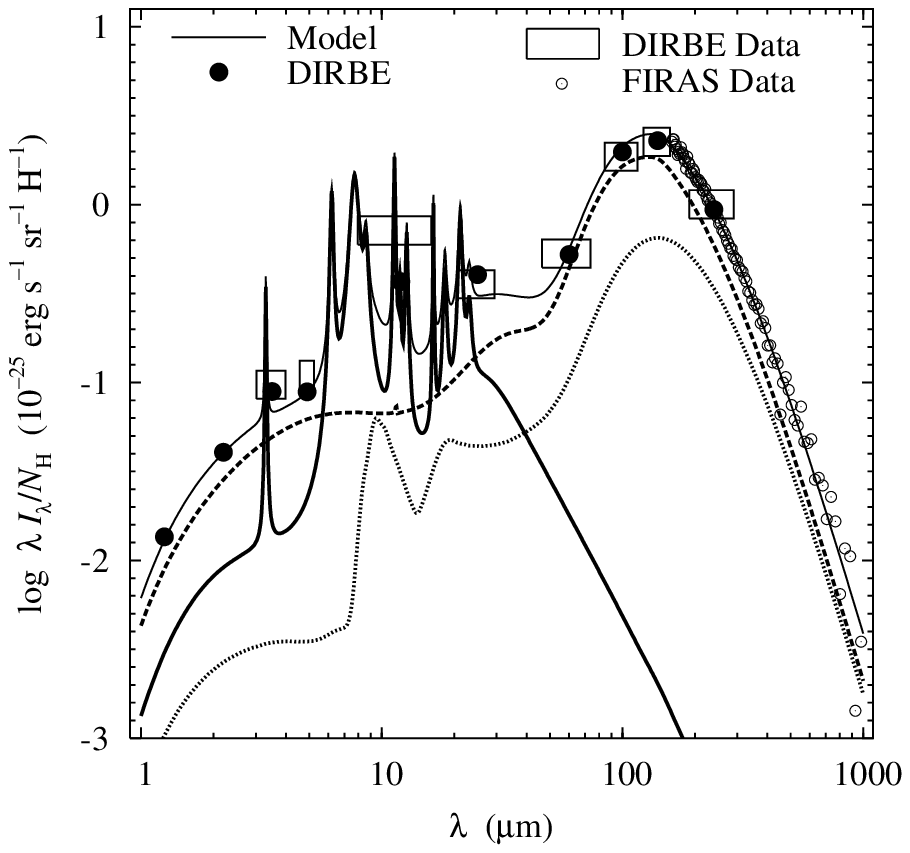}
\caption{ BARE-GR-B dust model: the size distributions (top left),
  extinction curve (top right), elemental requirements (bottom left),
  and emission spectrum (bottom right). Two straight lines are
  the MRN size distributions for silicate (upper line) and
  graphite (lower line).
}
   \label{fig:bare-b-gr}
\end{center}
\end{figure}

\begin{figure}
\begin{center}
\epsscale{1.0}
\plottwo{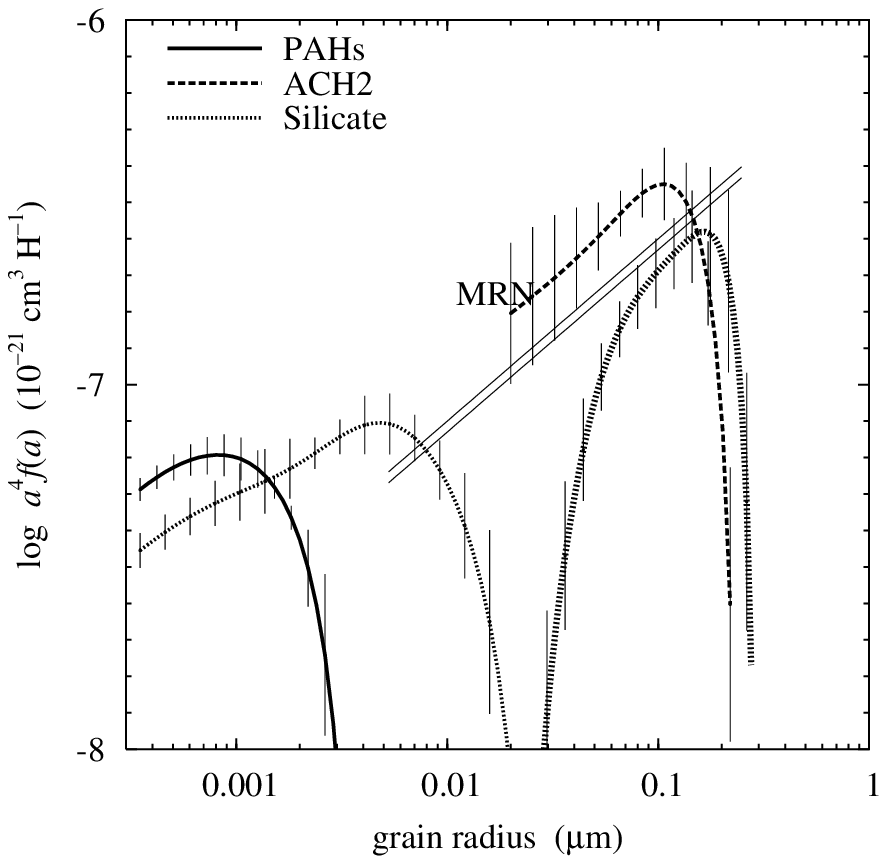}{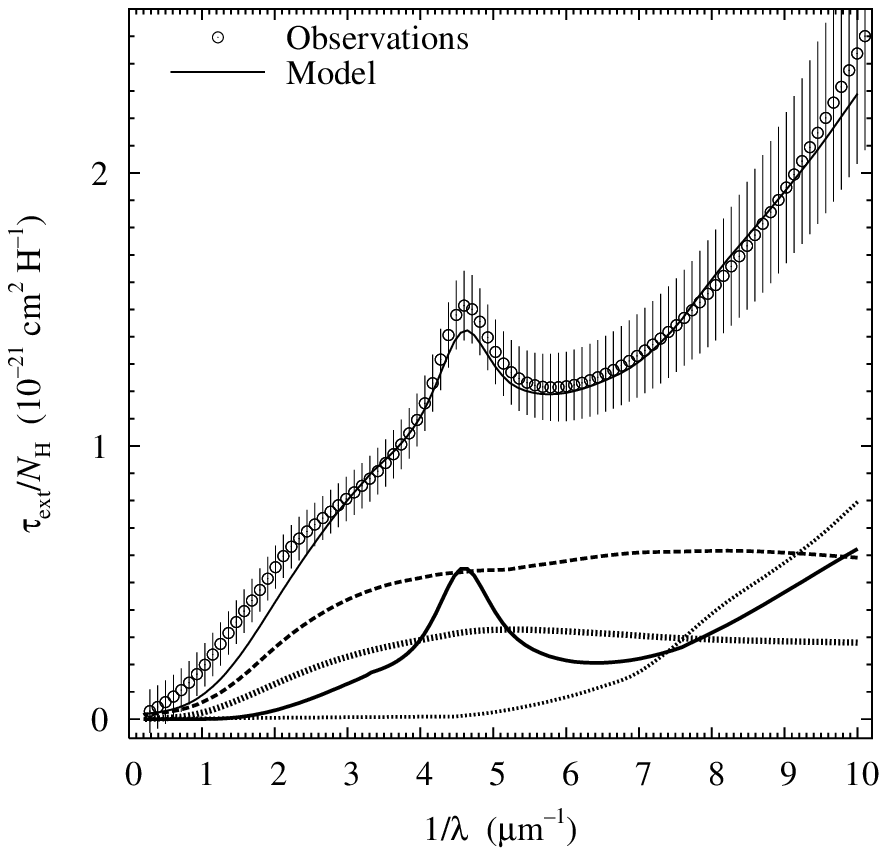}
\epsscale{2.2}
\plottwo{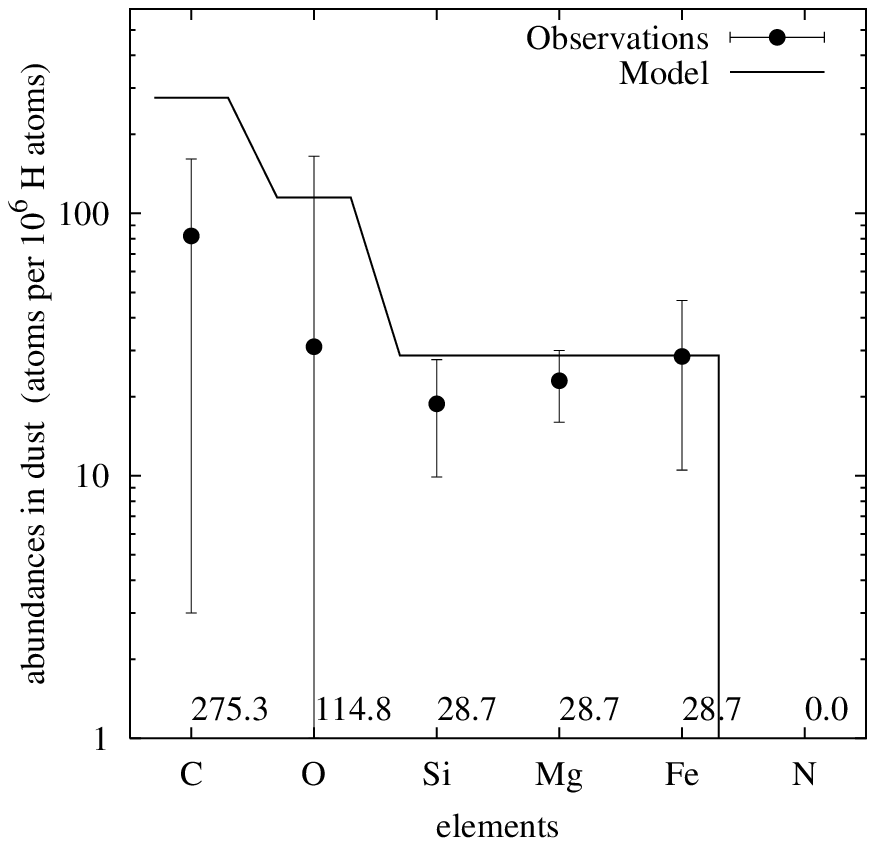}{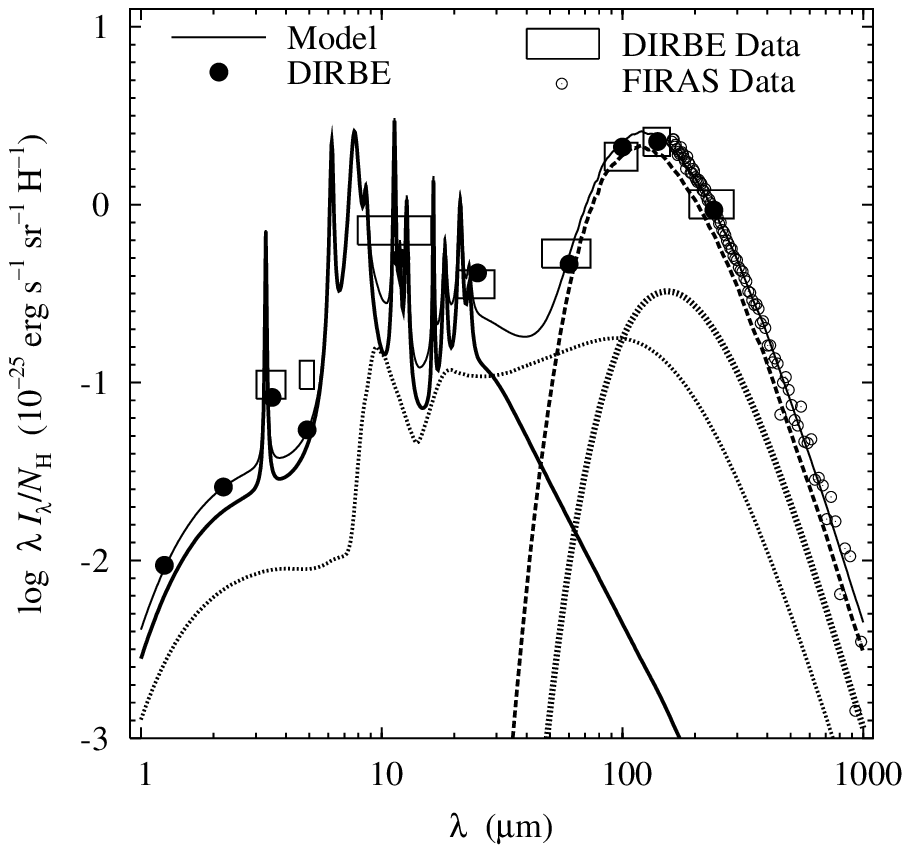}
\caption{ BARE-AC-B dust model: the size distributions (top left),
  extinction curve (top right), elemental requirements (bottom left),
  and emission spectrum (bottom right). Various populations of
  the same dust component are depicted by the lines of various
  width. Two straight lines are
  the MRN size distributions for silicate (upper line) and
  graphite (lower line).
}
   \label{fig:bare-b-ac}
\end{center}
\end{figure}

\begin{figure}
\begin{center}
\epsscale{1.0}
\plottwo{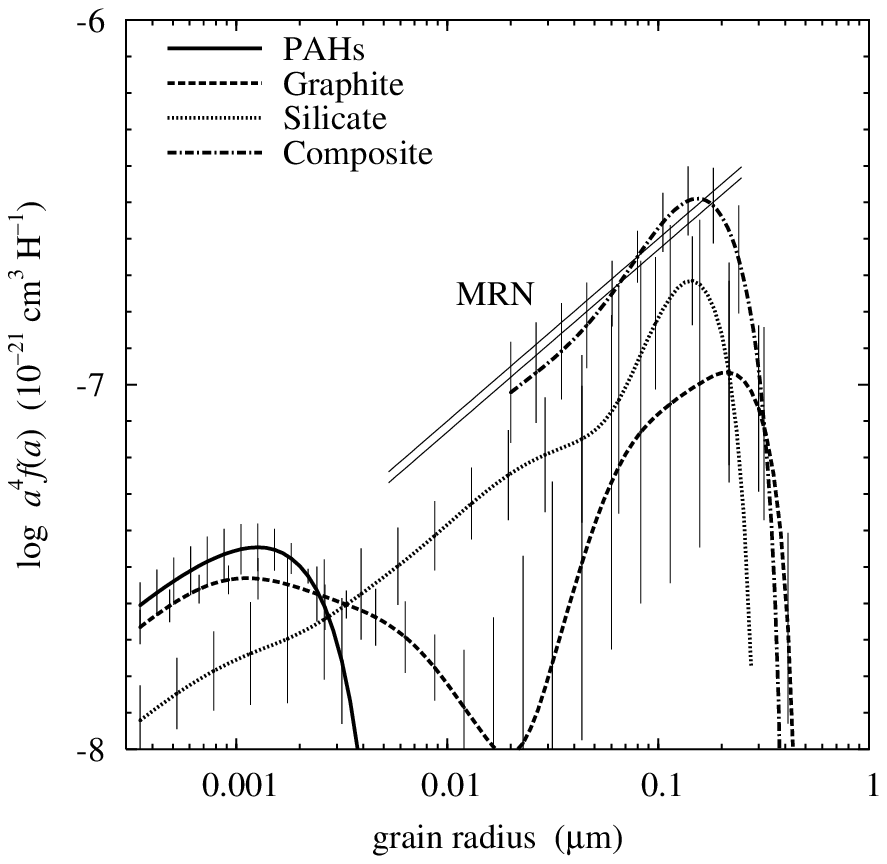}{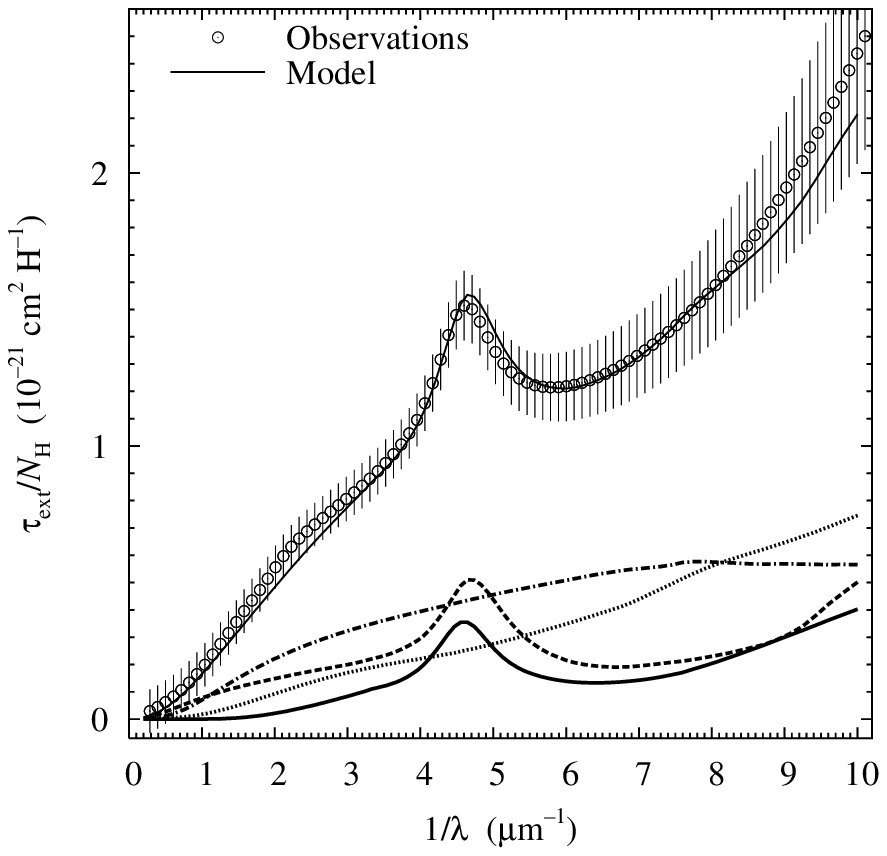}
\epsscale{2.2}
\plottwo{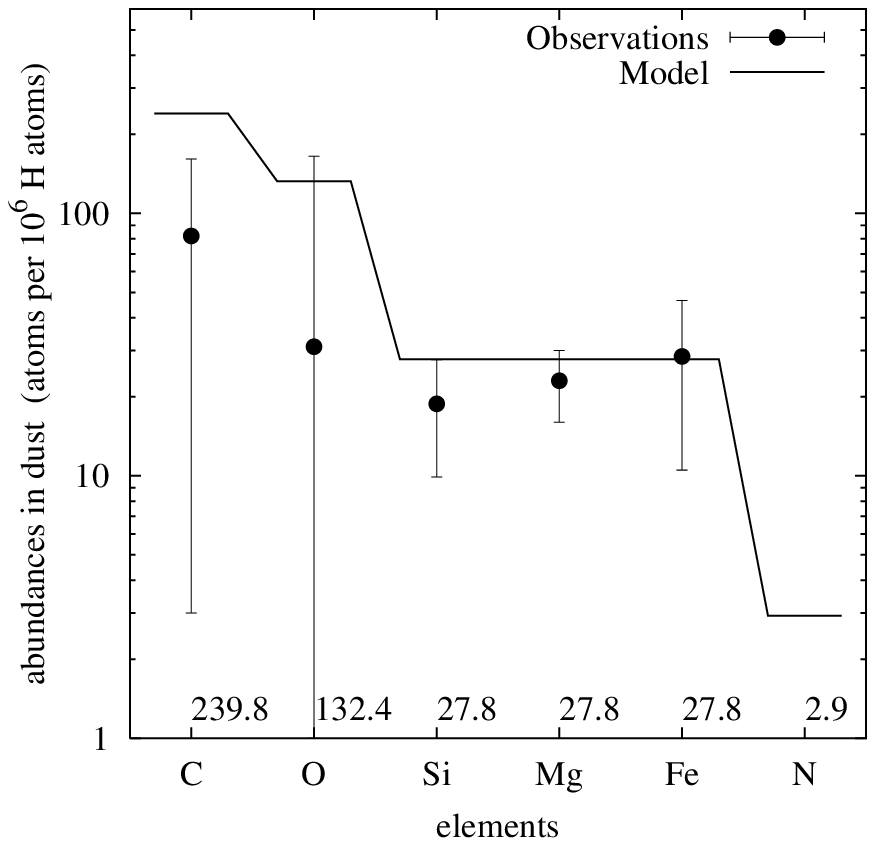}{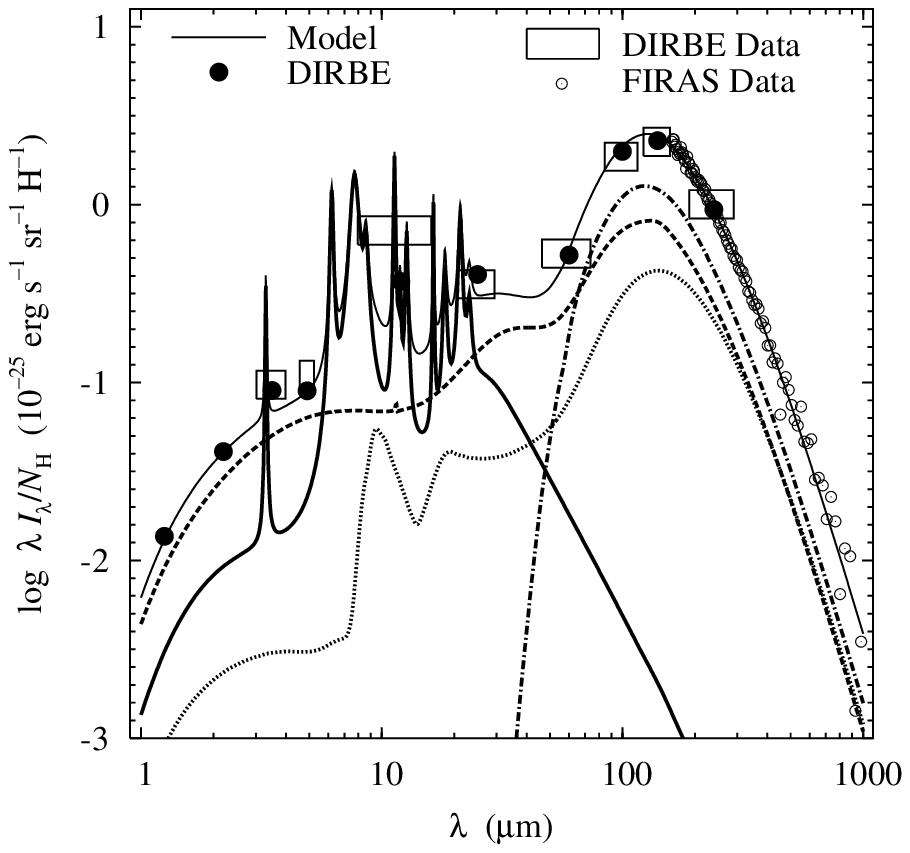}
\caption{ COMP-GR-B dust model: the size distributions (top left),
  extinction curve (top right), elemental requirements (bottom left),
  and emission spectrum (bottom right). Two straight lines are
  the MRN size distributions for silicate (upper line) and
  graphite (lower line).
}
   \label{fig:comp-b-gr}
\end{center}
\end{figure}

\begin{figure}
\begin{center}
\epsscale{1.0}
\plottwo{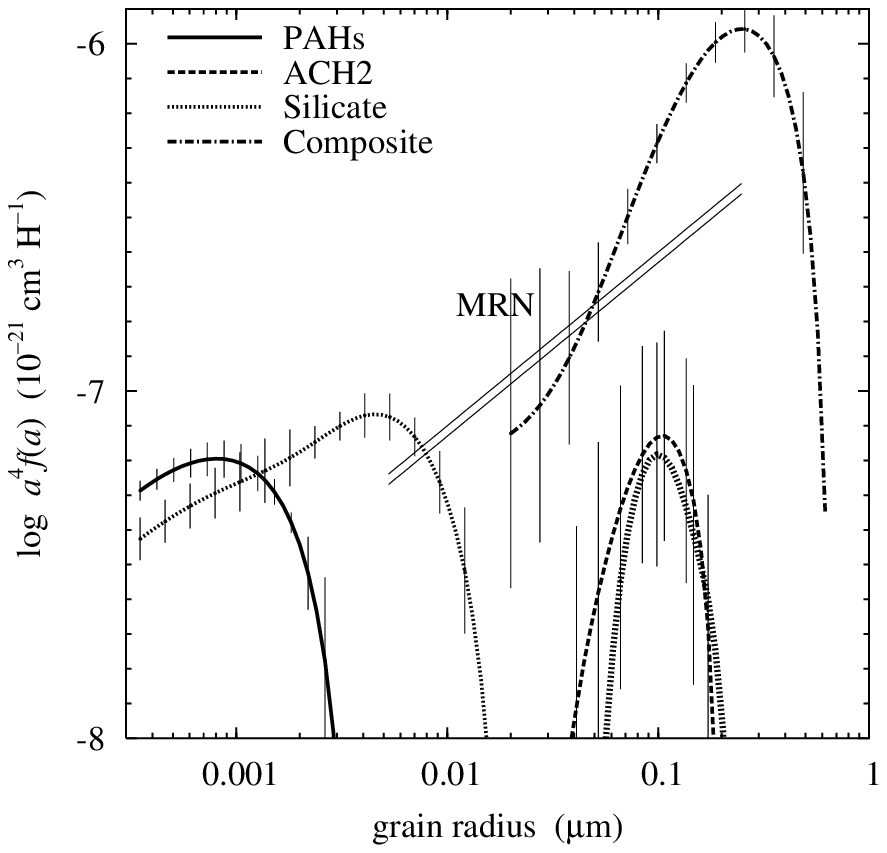}{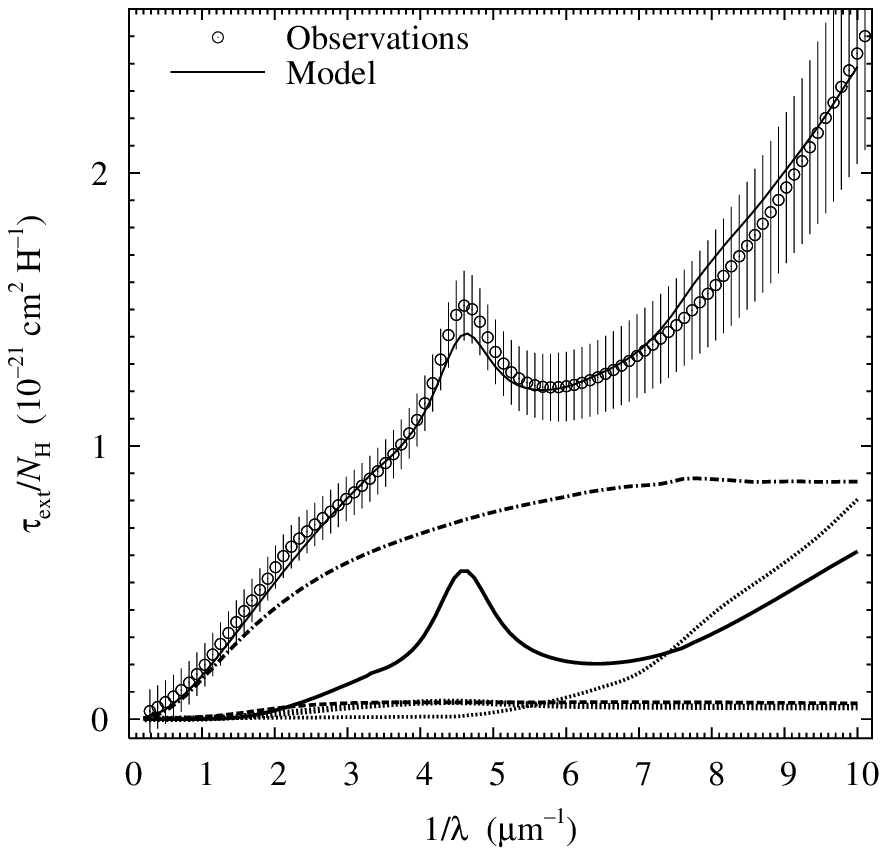}
\epsscale{2.2}
\plottwo{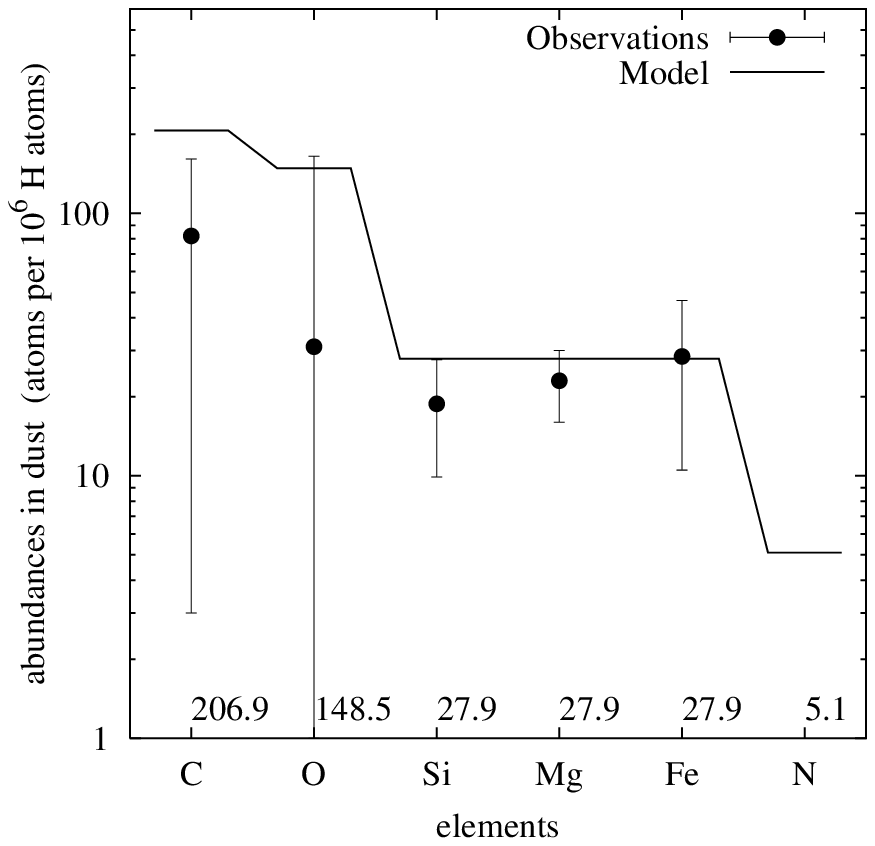}{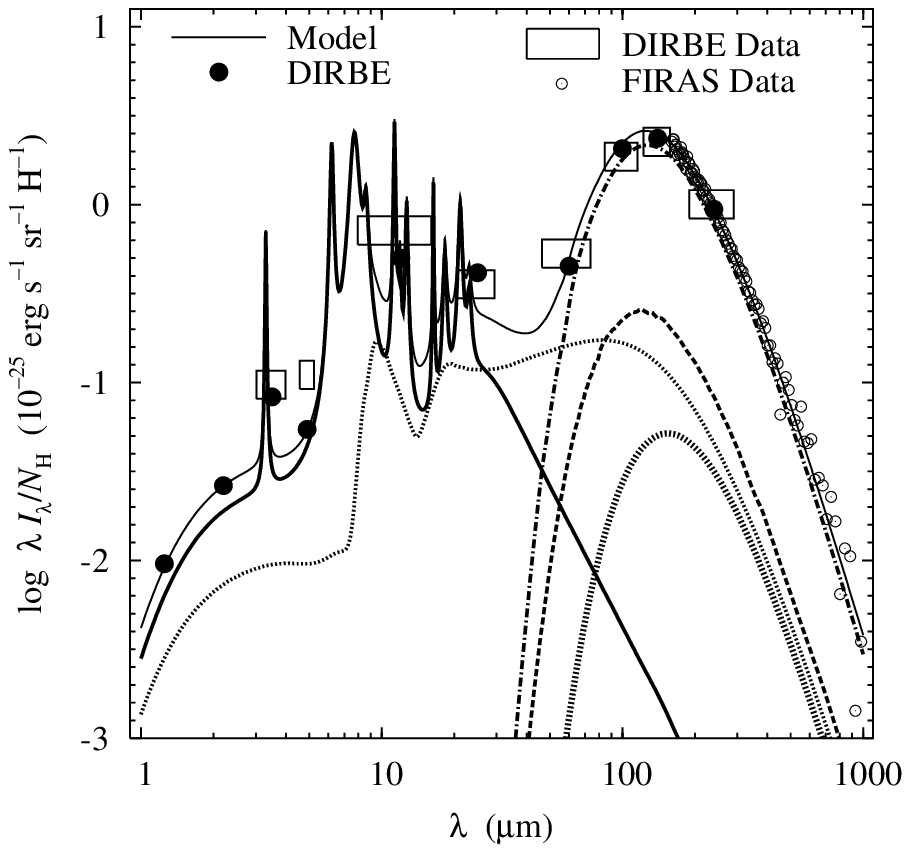}
\caption{ COMP-AC-B dust model: the size distributions (top left),
  extinction curve (top right), elemental requirements (bottom left),
  and emission spectrum (bottom right). Various populations of
  the same dust component are depicted by the lines of various
  width. Two straight lines are
  the MRN size distributions for silicate (upper line) and
  graphite (lower line).
}
   \label{fig:comp-b-ac}
\end{center}
\end{figure}

\begin{figure}
\begin{center}
\epsscale{1.0}
\plottwo{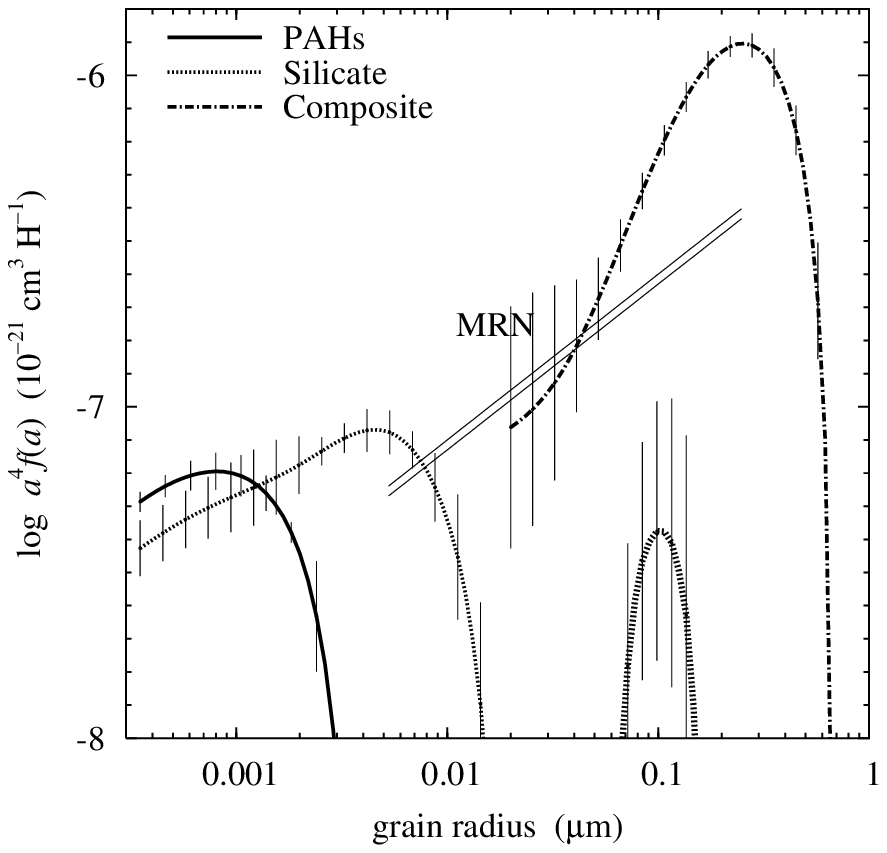}{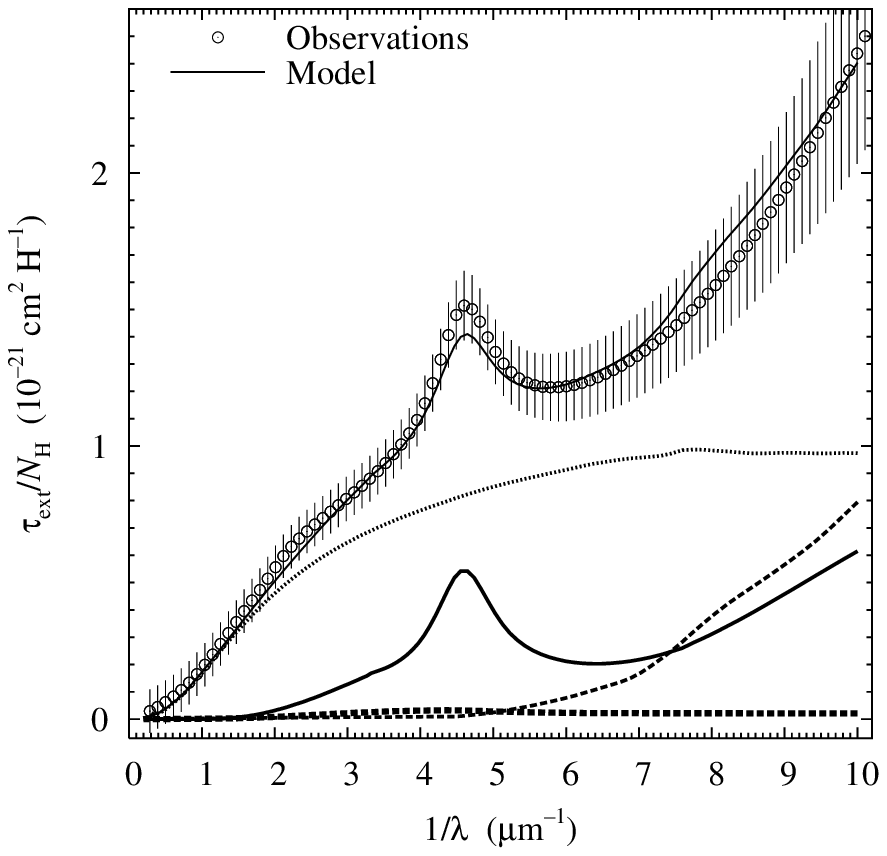}
\epsscale{2.2}
\plottwo{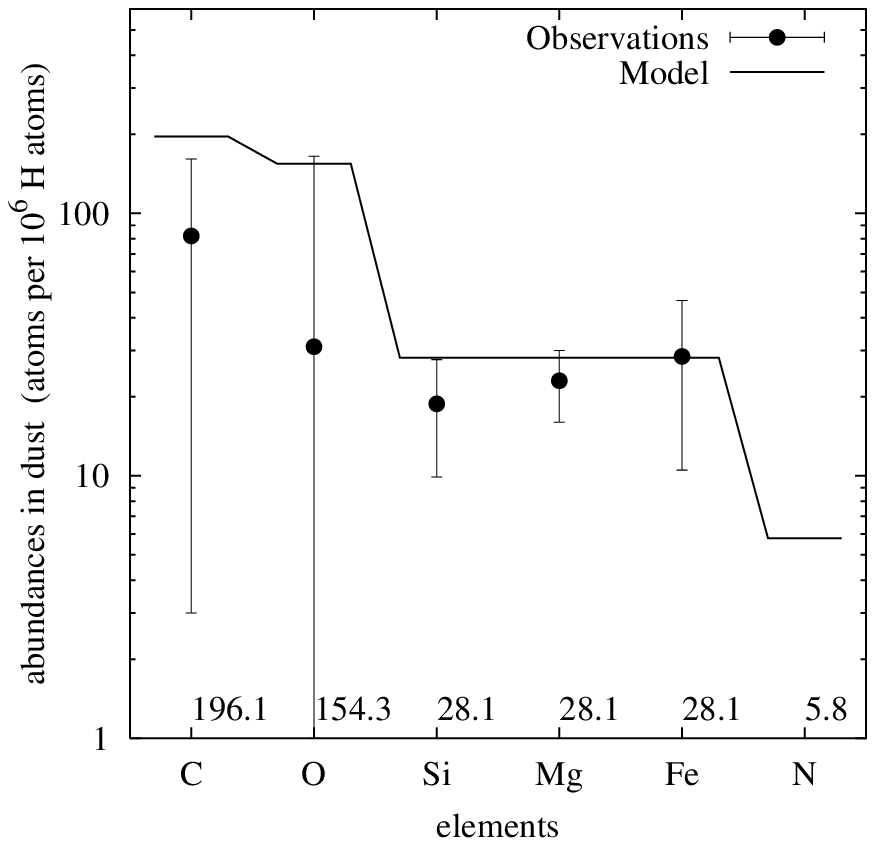}{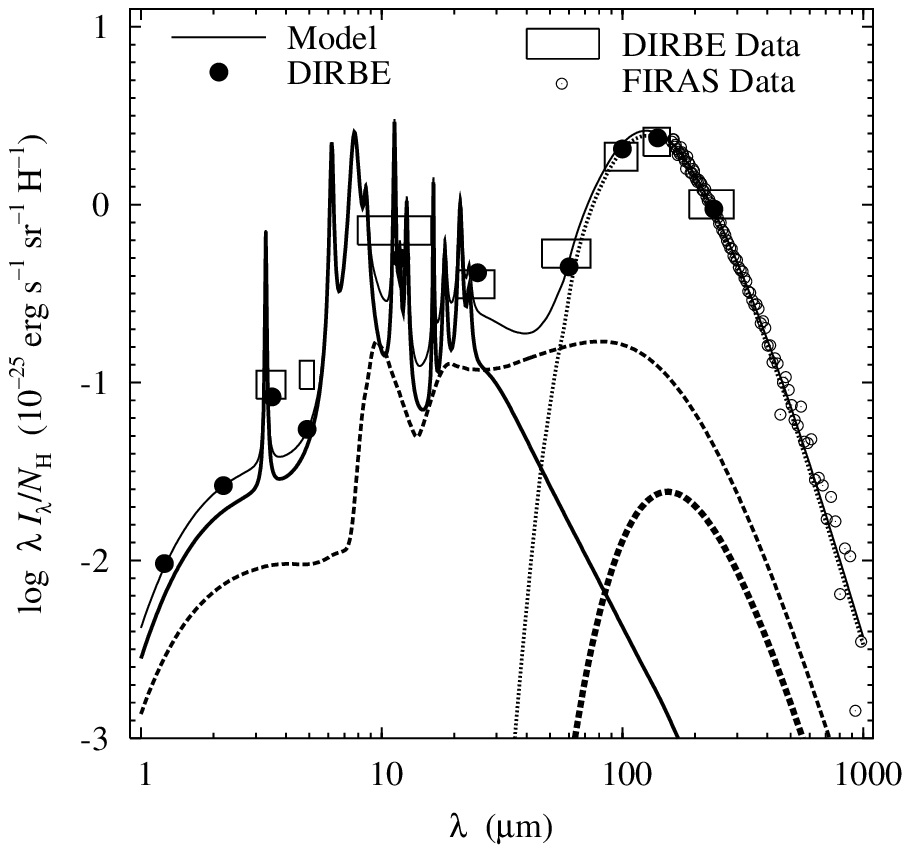}
\caption{ COMP-NC-B dust model: the size distributions (top left),
  extinction curve (top right), elemental requirements (bottom left),
  and emission spectrum (bottom right). Various populations of
  the same dust component are depicted by the lines of various
  width. Two straight lines are
  the MRN size distributions for silicate (upper line) and
  graphite (lower line).
}
   \label{fig:comp-b-nc}
\end{center}
\end{figure}

The main results of our modeling are shown in
Figures~\ref{fig:bare-s-gr}--\ref{fig:comp-b-nc}
which compare the calculations for the BARE and COMP models
to the observations. Each figure consists of four
panels depicting the following results: (1) the grain-size
distributions and their uncertainties for each dust component,
which is the main result of the fitting (inversion);
(2) the calculated extinction curve, with the partial contributions
of each dust component; (3) the calculated abundances of
the different elements in the dust; and (4) the calculated emission
spectrum along with the contributions of each dust component.

For BARE models with graphite grains, the main contribution to
the UV/optical extinction comes from graphite grains and PAHs,
whereas silicate grains dominate in the far UV extinction.
The UV bump is explained by both PAHs and graphite grains.
In emission, PAHs and graphite grains mostly contribute in
the near infrared (3--25 {\micron}), graphite grains prevail
for $\lambda$=25--250 {\micron}, and the contributions of graphite
and silicate grains are comparable for $\lambda > 250$~{\micron}.

For BARE models with amorphous carbon, most optical and UV extinction
is explained by amorphous carbon grains: this contribution
is quite flat. The far UV slope is provided by small silicate
grains and PAHs. The UV bump is solely explained by PAHs.
The near infrared emission is due to PAHs and silicate grains,
the middle infrared ($\lambda$=25--60~{\micron}) is explained by
silicate grains, and the amorphous carbon grains prevail
in the far infrared emission for $\lambda > 60$~{\micron}.

COMP models with both graphite and amorphous carbon grains have some
similarities with the respective BARE models. Note however that
the composites are one of the main contributors to the optical
to far-UV extinction. Emission in the near and mid infrared
(3--60 {\micron}) is similar to BARE models. However, composites
contribute to emission in the far infrared for
$\lambda >$~200~{\micron}.

For COMP models without graphite or amorphous carbon grains,
composites are dominating contributor to the infrared/optical/UV
extinction and far infrared emission. The UV bump is explained
by PAHs. These models generally require less carbon than any other
models.

\begin{figure}
\begin{center}
\epsscale{1.1}
\plottwo{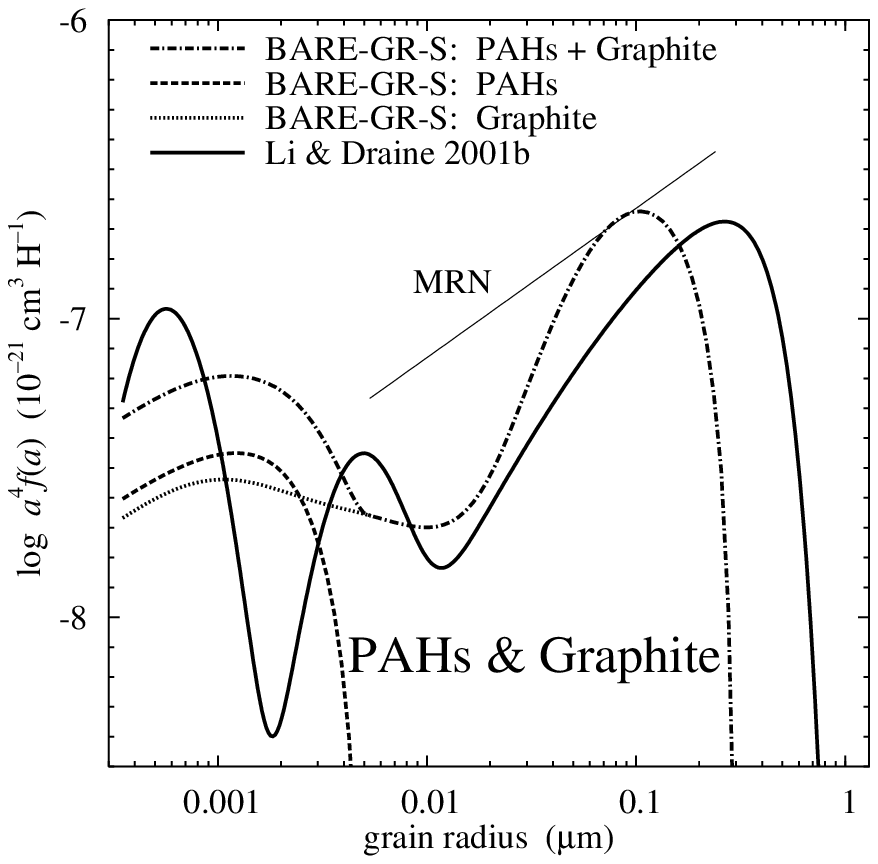}{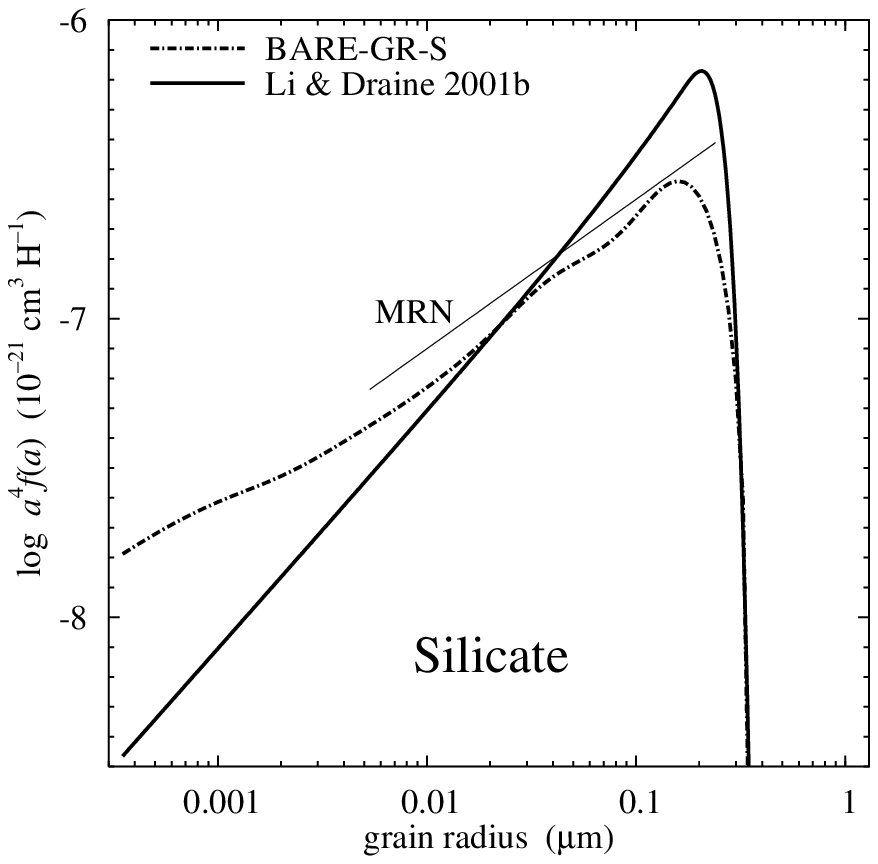}
\caption{ Grain-size distributions for the BARE-GR-S model
  compared with the \citet{ld01b} size distributions.
   \label{fig:size_ld01}
}
\end{center}
\end{figure}

Figure \ref{fig:size_ld01} presents the size distribution
for the BARE-GR-S dust model that comprises of dust particles with
identical compositions and optical properties as the \citet{ld01b} model.
For sake of comparison we also added in the figure the Li \& Draine
grain size distributions. For carbonaceous grains, the Li \& Draine
size distribution is tri-modal with a primary peak at
around 0.3~{\micron}, and secondary peaks at about 50 and 5~{\AA}.

In contrast, our size distribution for the carbonaceous particles is
bi-modal, consisting of a graphitic component peaked at $\sim$0.1~$\mu$m,
and a PAH component, peaked at $\sim$15~{\AA}. The graphitic component
extends to very small grain sizes, representing a population of
dehydrated PAHs. The dehydrated PAHs are needed to produce
the continuum mid-IR emission without the associated PAH features.
This mid-IR emission also masks the presence of the 10~$\mu$m
silicate bump that is generated by the very small silicate particles
in the BARE-GR-S dust model. The additional $\sim$50~{\AA} bump
at the lower limit of the Li \& Draine graphite grain size
distribution may reflect the need to produce a featureless mid-IR
continuum emission in their model.

The silicate grain size distribution in both,  the Li-Draine and
the BARE-GR-S dust models, is essentially a power law with a similar
cutoff at large grain sizes. However, the silicate size distribution
in the BARE-GR-S model has a shallower slope, similar to that of
the MRN model, and consequently contains a larger number of very
small silicate particles compared to the Li-Draine model.

Another contribution to the differences in the grain size distribution
between the BARE-GR-S and the Li \& Draine dust models may be
attributed to the fact that \citet{ld01b} constrained the functional
form of their size distributions, and thus did not derive the optimal
distributions that fits the observational constraints.

\begin{figure}
\begin{center}
\epsscale{1.0}
\plottwo{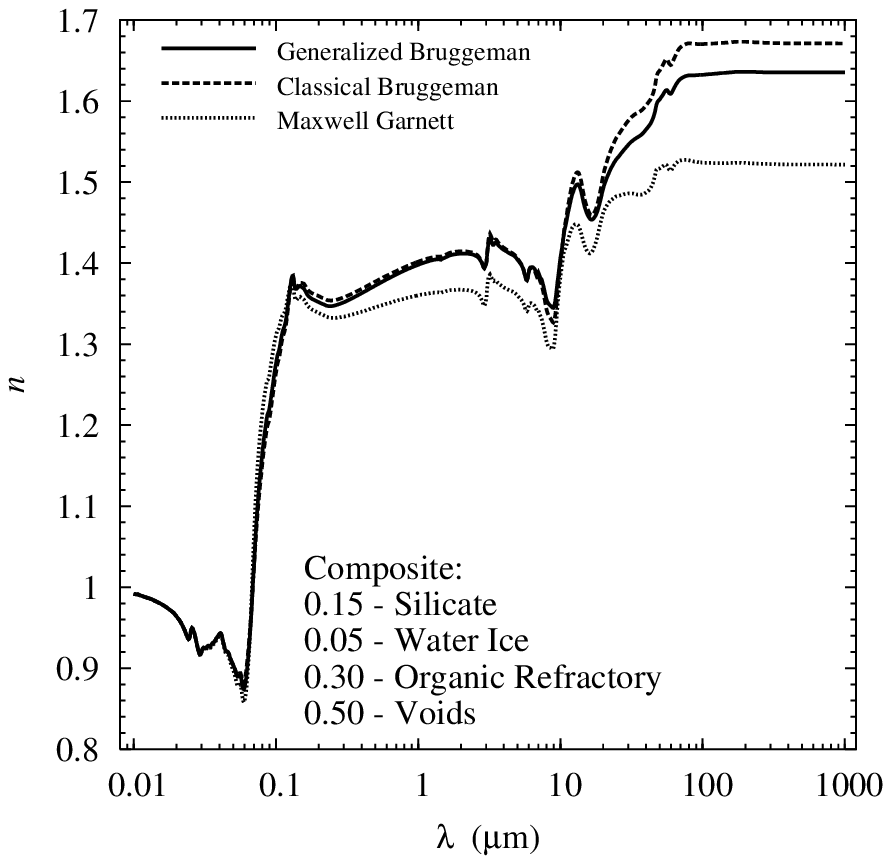}{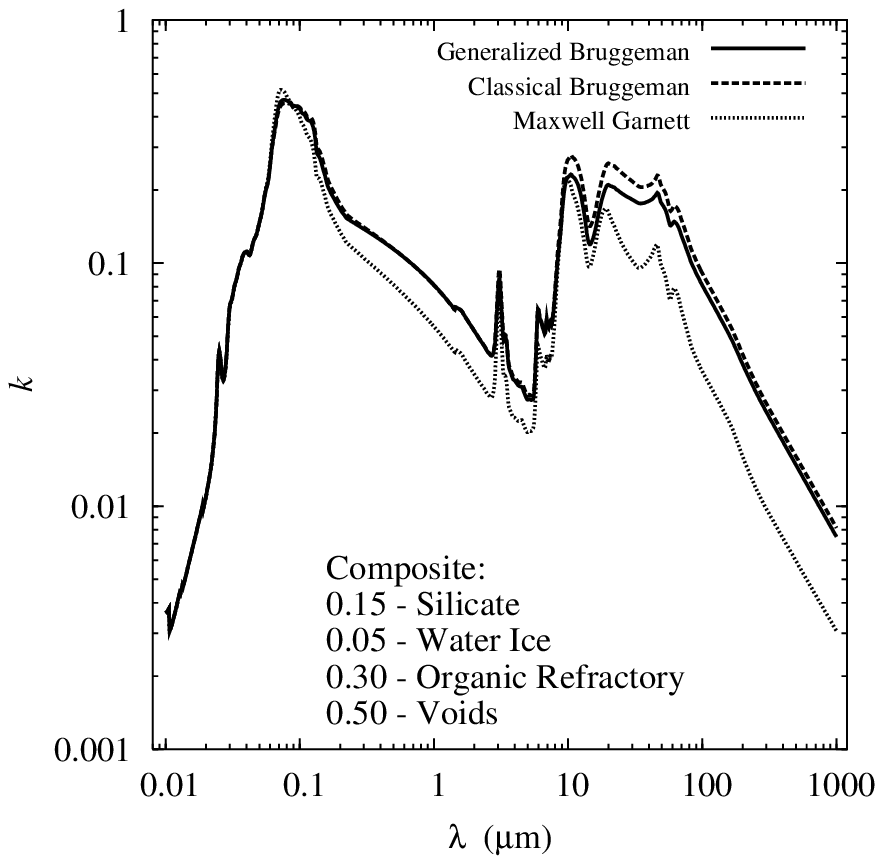}
\epsscale{2.25}
\plottwo{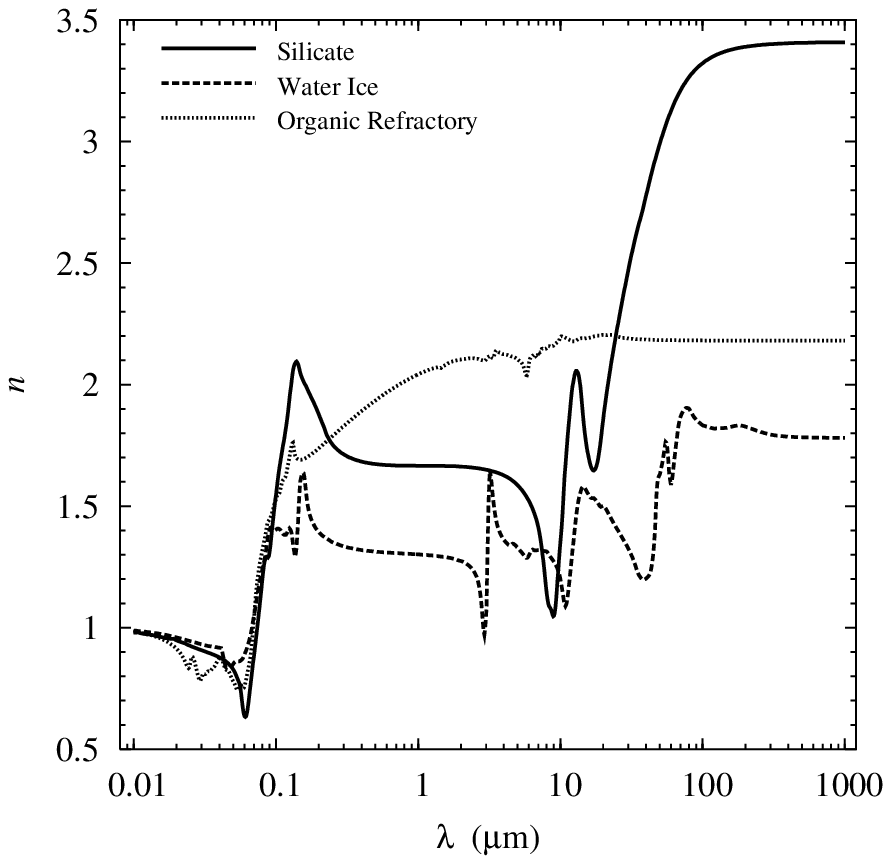}{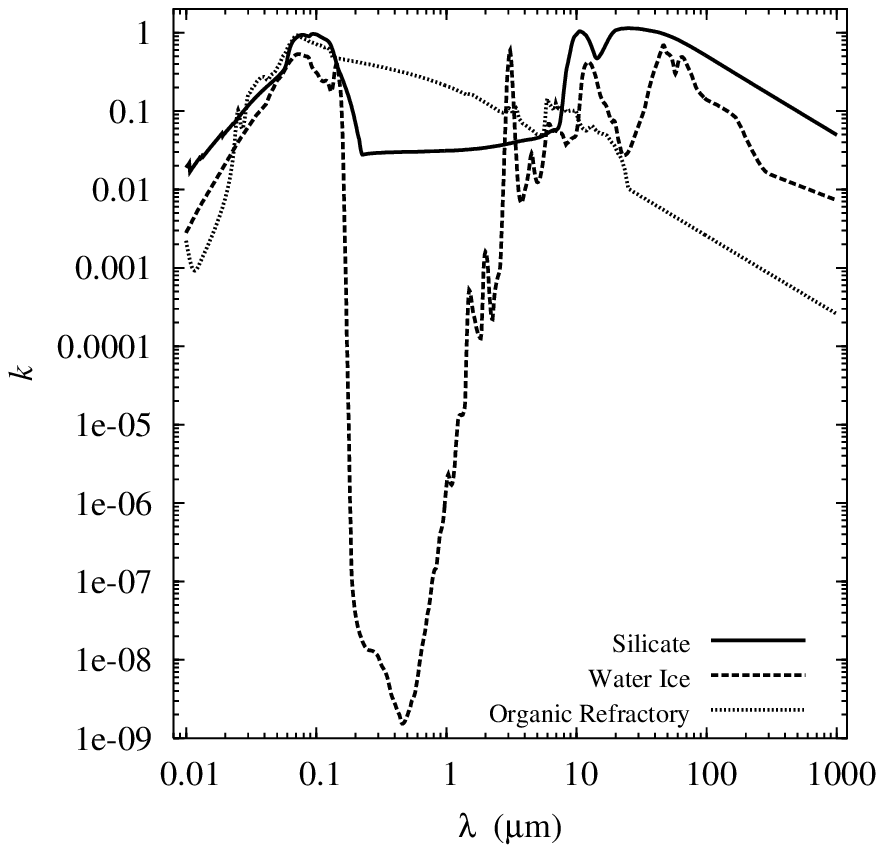}
\caption{ Optical constants for a composite grain (derived with
  various mixing rules) and its constituents.
}
   \label{fig:nk_comp}
\end{center}
\end{figure}

When trying to find the best-fit COMP models, we varied the volume
fractions of composite grains, including their porosities,
to get the mimimum discrepancy. This resulted in a composite model
which has {\em mass} fractions: 0.5 of silicate,
0.457 of organic refractory, and 0.043 of water ice, but with
different porosities listed for each model in Table \ref{tab_vol_fr}.
In most cases, the models for a range of porosities 0.20--0.80
can provide quite good fits. So, for these cases, we present
models for a porosity equal to 0.50, with two exceptions:
(1) for COMP-AC-S, the models fits are good for porosities 0.50--0.70,
so we chose the model with 0.60 as a representative;
(2) for COMP-GR-B, the models with porosities 0--0.20 looks better,
so we present a model with 0.10. We included various types of amorphous
carbon in composites while modeling, but we found that in order
to get good fits the mass fraction of amorphous carbon should
not exceed 1\%. Figure \ref{fig:nk_comp} demonstrates the optical
constants for the composites with porosity 0.50, derived with
the three various mixing rules. Also shown are the optical constants
of the composite constituents: silicate, organic refractory,
and water ice.

\subsection{Quality of Fit}
      \label{subsection:quality}

Table \ref{tab_discr} lists the discrepancies of the models.
It is immediately clear from the Table that the overall quality
of models is highly dependent on the abundance constraint.
The minimum discrepancy is achieved by the group of models that
assume F and G stars ISM abundances with the models assuming
solar and B star abundances following next. Note that
the differences are quite small between the minimum and maximum
discrepancies of neighboring groups of models with
various abundances, that is between  BARE-AC-FG and COMP-GR-S, and
between BARE-AC-S and COMP-GR-B. The discrepancies for our models
are within the range of 10$\pm$2.
The amorphous carbon in the models listed in the table consist
of ACH2 type carbon. We found that it is impossible to get
acceptable fits with ACAR or BE type carbon.

Generally, each of the models provides acceptable fits to
the extinction and emission. Quantitatively, this can be seen
from Table~\ref{tab_discr} by comparing the respective discrepancies.
There are however differences in fitting the abundances in dust.
First of all, the COMP models have smaller discrepancies
($\Phi_{\rm{d}}$) than their BARE counterparts. Probably, this is
mostly because COMP models include one more dust component -- composites,
and thus more fitting parameters. Even the models with the B stars
abundances can provide acceptable fits to the abundances in dust,
while still requiring extra carbon beyond one sigma.
By comparing the discrepancies from Table~\ref{tab_discr},
we can conclude that it is harder to fit B star abundance
constraints than the two other sets.
Note however that larger abundance uncertainties for F and G stars 
make it easier to get good fits for this case.
The COMP models provide somewhat better fits than their
respective BARE counterparts.

\citet{ld01b} adopted an extinction-per-H-atom value that is 1.12
larger than that of our paper. However, this difference does not
account for the large amount of metals consumed in their dust model,
because the model was derived by fitting {\em both} the extinction
and the emission spectrum. As a result, the abundance of the dust
does not scale simply with the extinction curve. To examine
the effect of the extinction curve on the elemental depletions,
we calculated the dust abundances for a BARE-GR-S model optimized
to fit the extinction curve used by \citet{ld01b} instead of the one
used in this paper. The changes in the dust abundances were
quite small, 2.6 and 0.9\% for carbon and silicate dust,
respectively. From this experiment we expect that a Li \& Draine
model fit to the extinction curve used here, will not substantially
alter their derived dust abundances.

\subsection{Characterization of the Models}

\subsubsection{Dust Abundances and Size Distributions}

In Table~\ref{tab_fin_abund}, we list the quantities of chemical
elements, consumed by the models: both total and partial for
each dust component. Table~\ref{tab_mass} contains the dust-to-gas
ratios and the mass per H atom locked up in dust along with
the contributions of each dust constituent in per cent.

From the masses per H atom and the constituent mass percentages
from Table~\ref{tab_mass}, one can calculate the abundances
of water ice required by our models. We found that our COMP
models consume 5--13 ppm of ice (5--7 for COMP-GR models,
7--11 for COMP-AC models, and 12--13 for COMP-NC models).
For comparison, \citet{whittet97} estimated the upper limit
for water ice at the diffuse sightline toward Cygnus OB2 No. 12
to be 2~ppm. However, it is not clear whether this estimate
is applicable to other diffuse lines of sight.
On the other hand, Figure~\ref{fig:ir_ext}
(see \S\ref{section:nir_ext}) demonstrates
that the near-infrared extinction predicted by our
COMP models, especially at the wavelengths around the 3.1 and 6
{\micron} water ice absorption features, is consistent
with the observations for a line of sight toward
the Galactic Center \citep{lutz96}.

We performed analytical approximations for the size distributions
of all our models.  At first step, the size distribution function $f(a)$
(in $\mu$m$^{-1}$~H$^{-1}$) found in a numerical form
was expressed as $f(a)=A \> g(a)$, where $A$ is a normalization
coefficient in H$^{-1}$, $a$ is in $\mu$m, and function $g(a)$
(in $\mu$m$^{-1}$) has a property $\int \! g(a) \> {\rm{d}}a = 1$.
To analytically approximate numerical values of $g(a)$,
we used function $\overline g$:
{\small
\begin{equation}
\log \> {\overline g}(a) =
  c_0 + b_0 \log(a)
      - b_1 \left|\log\left(\frac{a}{a_1}\right)\right|^{m_1}
      - b_2 \left|\log\left(\frac{a}{a_2}\right)\right|^{m_2}
      - b_3 |a - a_3|^{m_3} - b_4 |a - a_4|^{m_4}
    \label{eq:logga}
\end{equation}
}
which contains 14 parameters. Note this is a generic expression:
we normally used 8--11 parameters for fitting.
The parameters of ${\overline g}(a)$ were derived by fitting
the numerical values of $g(a)$ using the nonlinear
least-squares Marquardt-Levenberg algorithm \citep{numerical_recipes}.
Function (\ref{eq:logga}) contains a power law part:
$c_0 + b_0 \log(a)$, cutoff terms:
$- b_i \left|\log(a/a_i\right)|^{m_i}$,
and peaked terms: $- b_i |a - a_i|^{m_i}$. The cutoff and peaked
terms are needed to reproduce the deviations of the model
size distributions from a power law. 

The resulting values of the parameters
of the fitting function (\ref{eq:logga}) are listed in
Tables~\ref{f_bare-s-gr}--\ref{f_comp-b-nc}. The Tables also contain
parameters $a_{\rm{min}}$ and $a_{\rm{max}}$ which define
the range of grain radii for which the approximation is valid,
the normalization constant $A$, and the reduced $\chi^2$
characterizing the quality of the analytical approximation.
Note that $\chi^2_{\rm{red}}$ has no relation to the quality
of the grain-size distribution model. Most approximations have
$\chi^2_{\rm{red}}$ better than of 0.01.

\subsubsection{Near--Infrared Extinction, Albedo, and Asymmetry Parameter}
      \label{section:nir_ext}

\begin{figure}
\begin{center}
\epsscale{1.1}
\plottwo{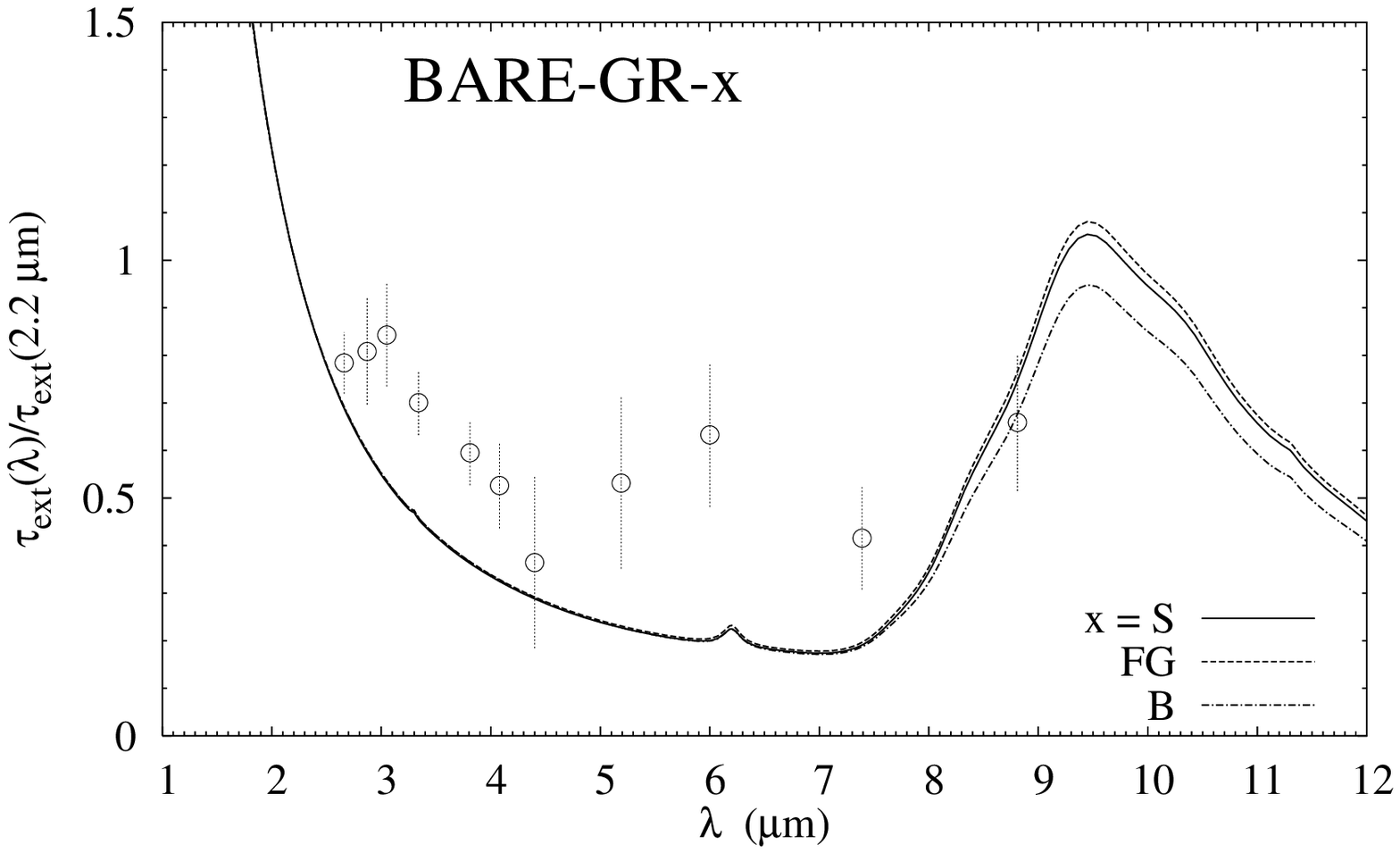}{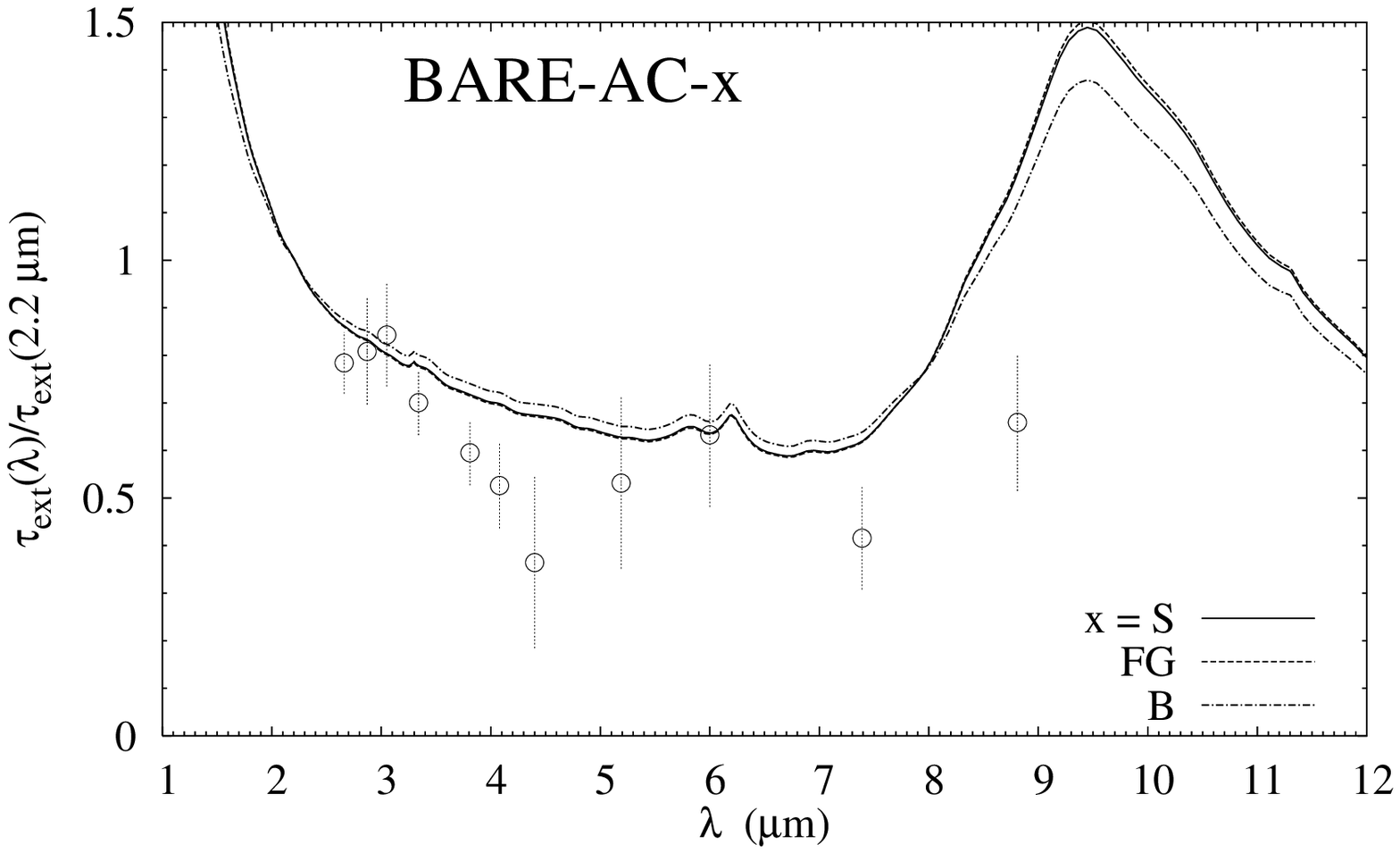}
\epsscale{2.45}
\plottwo{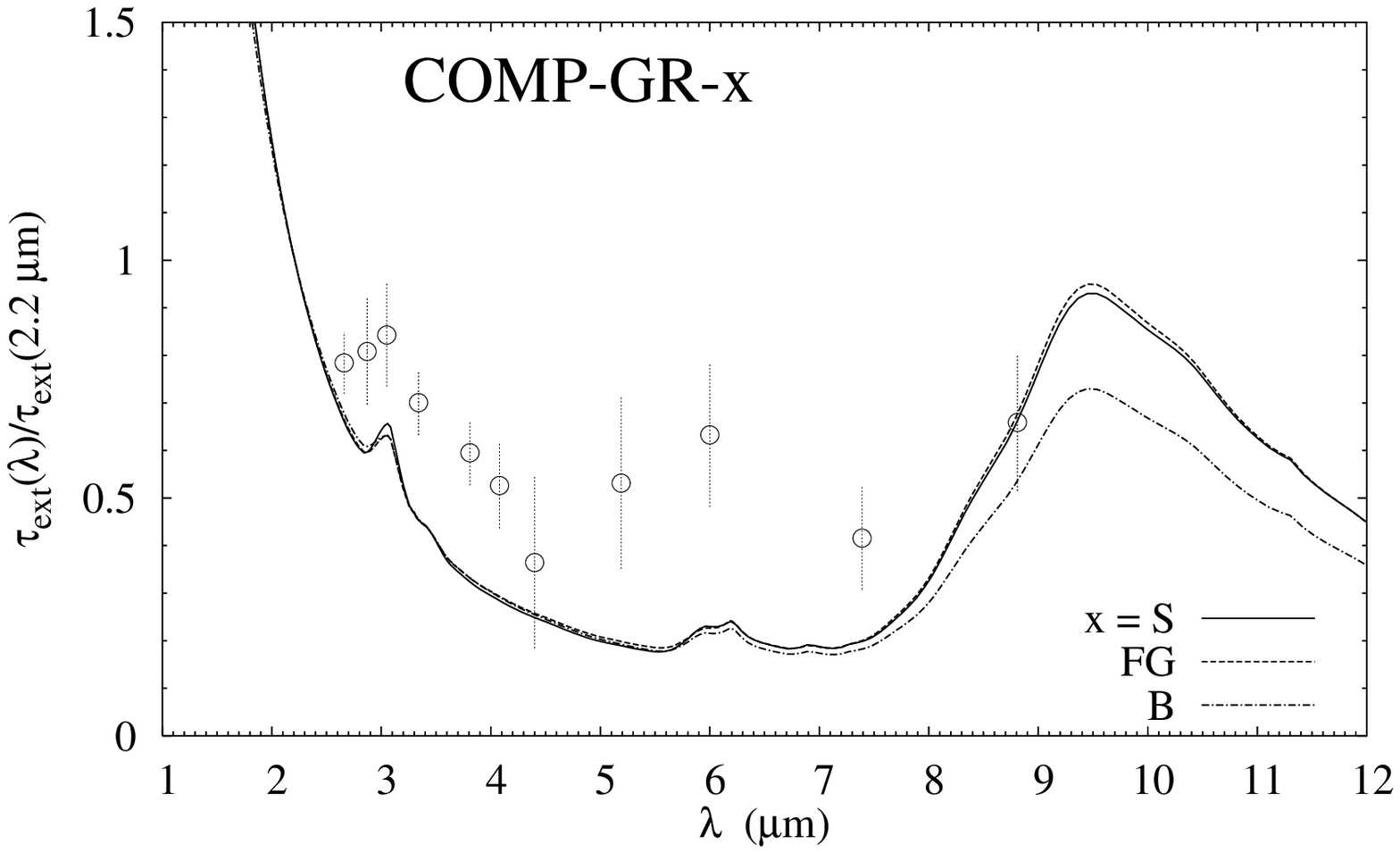}{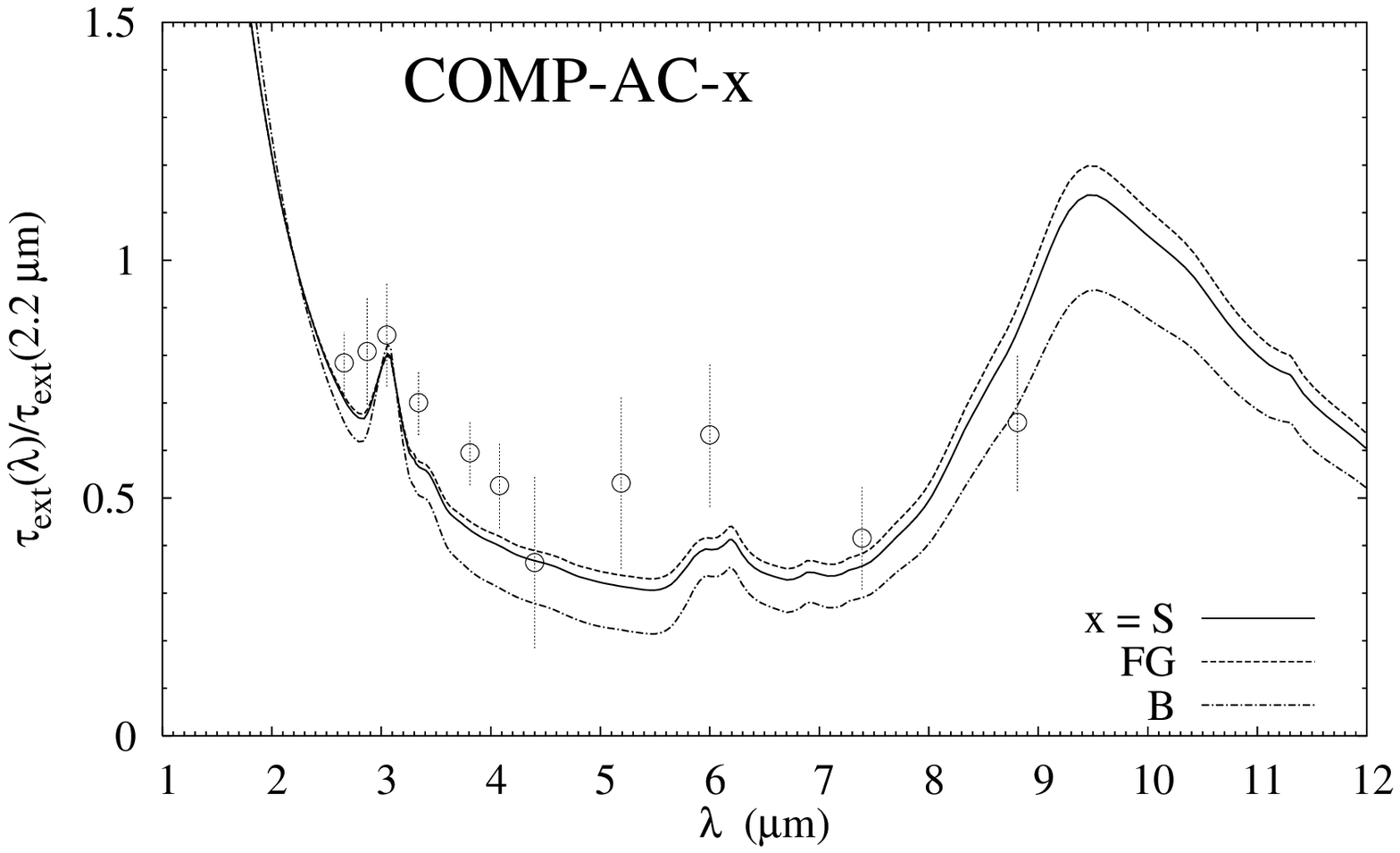}
\epsscale{2.45}
\plotone{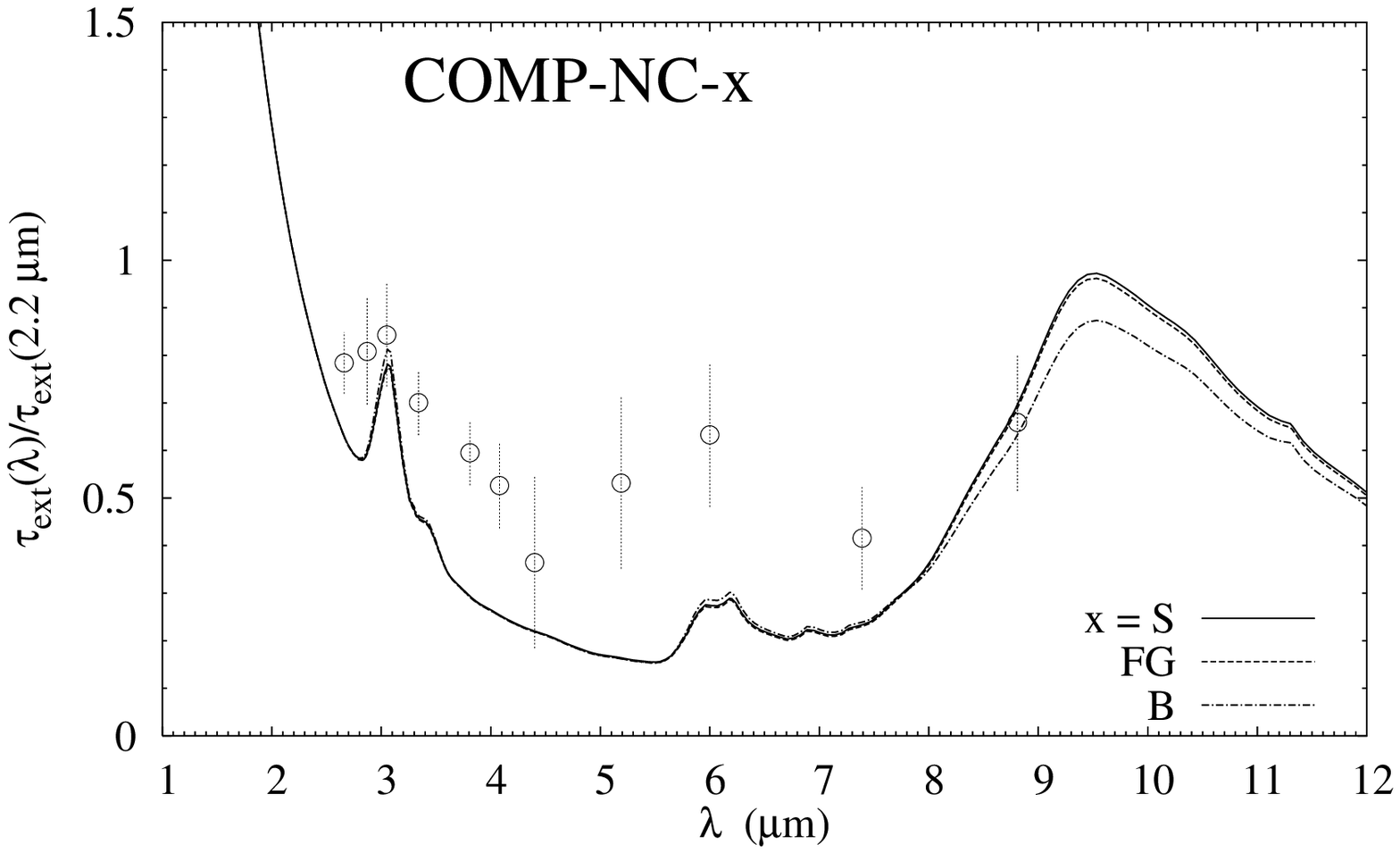}
\caption{ Near infrared extinction curves for the dust models.
    Observational data are for a line of sight toward
    the Galactic Center \citep{lutz96}.
}
   \label{fig:ir_ext}
\end{center}
\end{figure}

Figure~\ref{fig:ir_ext}, shows the near-infrared extinction curves
for the different BARE and COMP models. The observed extinction was taken from
\citet{lutz96} who estimated it by using the hydrogen recombination lines
detected between 2.5 and 9 {\micron} towards the Galactic Center.
Generally, the COMP models are more consistent with the observed
extinction here: the features at around 3--3.5 and 5.5--7.0~{\micron}
are mostly due to water ice and organic refractory.
It is immediately clear from Figure~\ref{fig:ir_ext} that additional
dust components should be present, e.g., other types of silicates,
metal oxides or sulphides, CO$_2$ and methane ices; or else
the optical constants for the existing components are bad.
However, currently, it is impossible to involve these species
into analysis because of lack of the respective
optical constants. Note also that the Galactic Center line of sight
may be very different from the high latitude emission and extinction
that were fit, especially with regard to molecular clouds
and thus probably composite or multilayer grains.

\begin{figure}
\begin{center}
\epsscale{1.1}
\plottwo{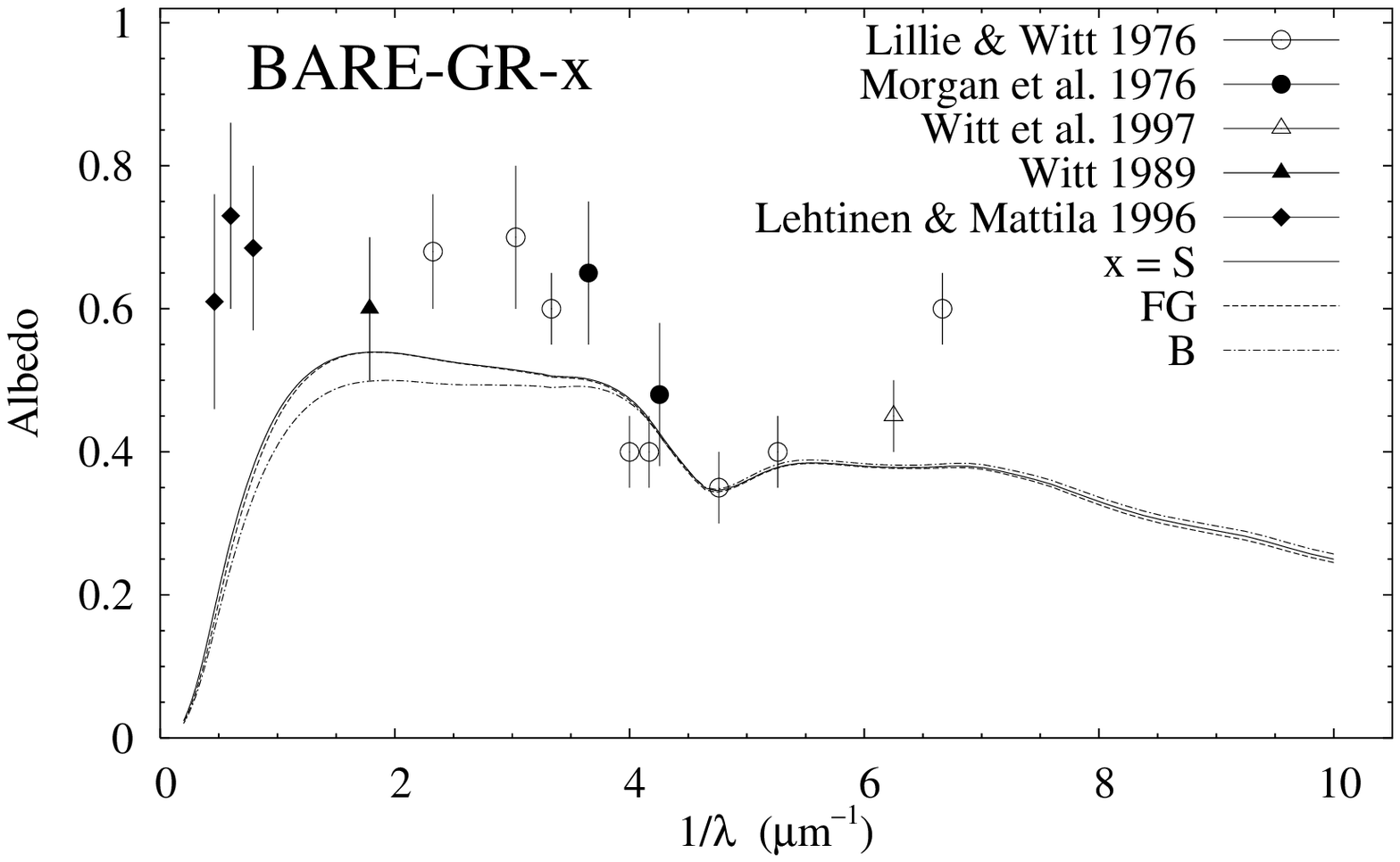}{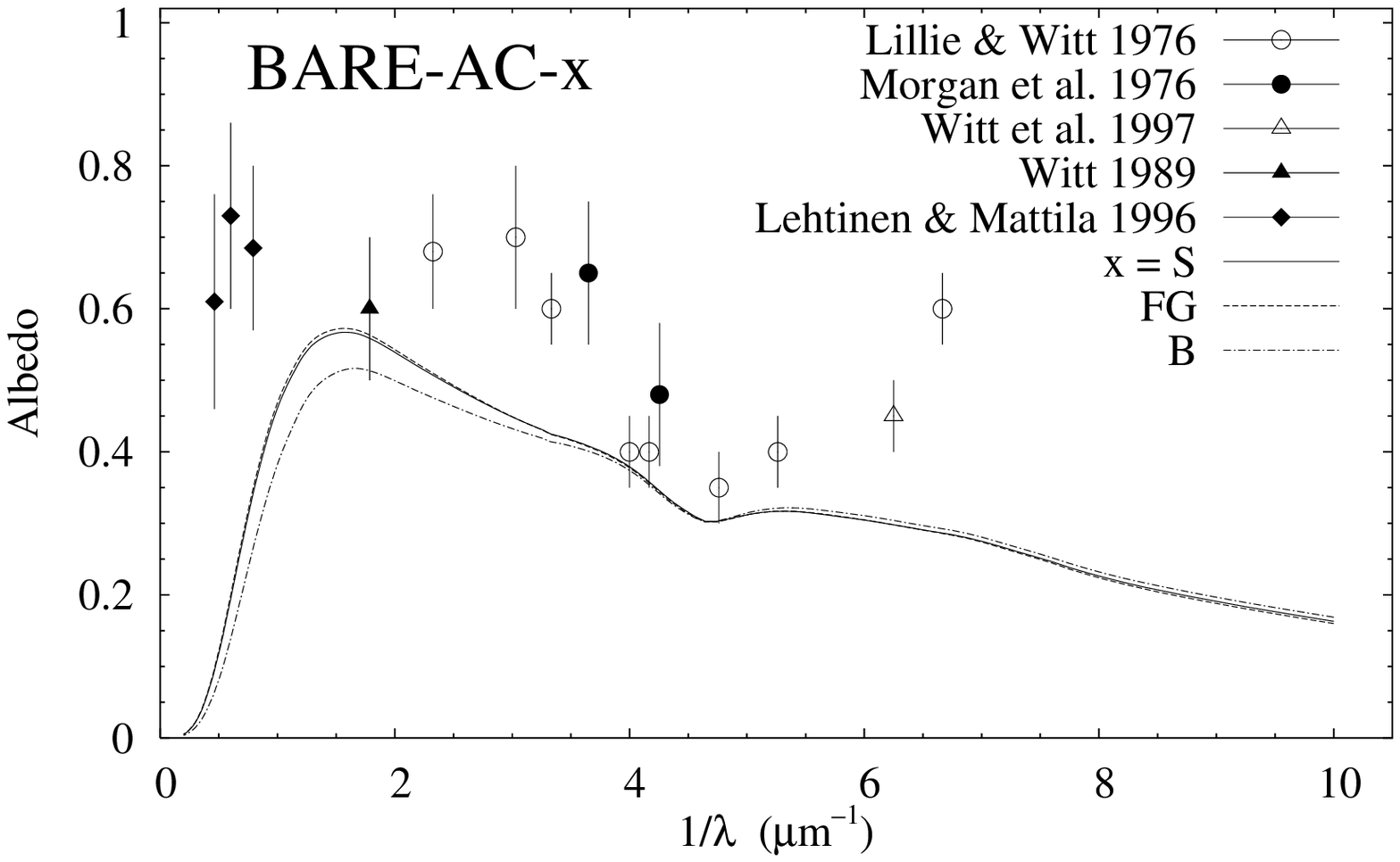}
\epsscale{2.45}
\plottwo{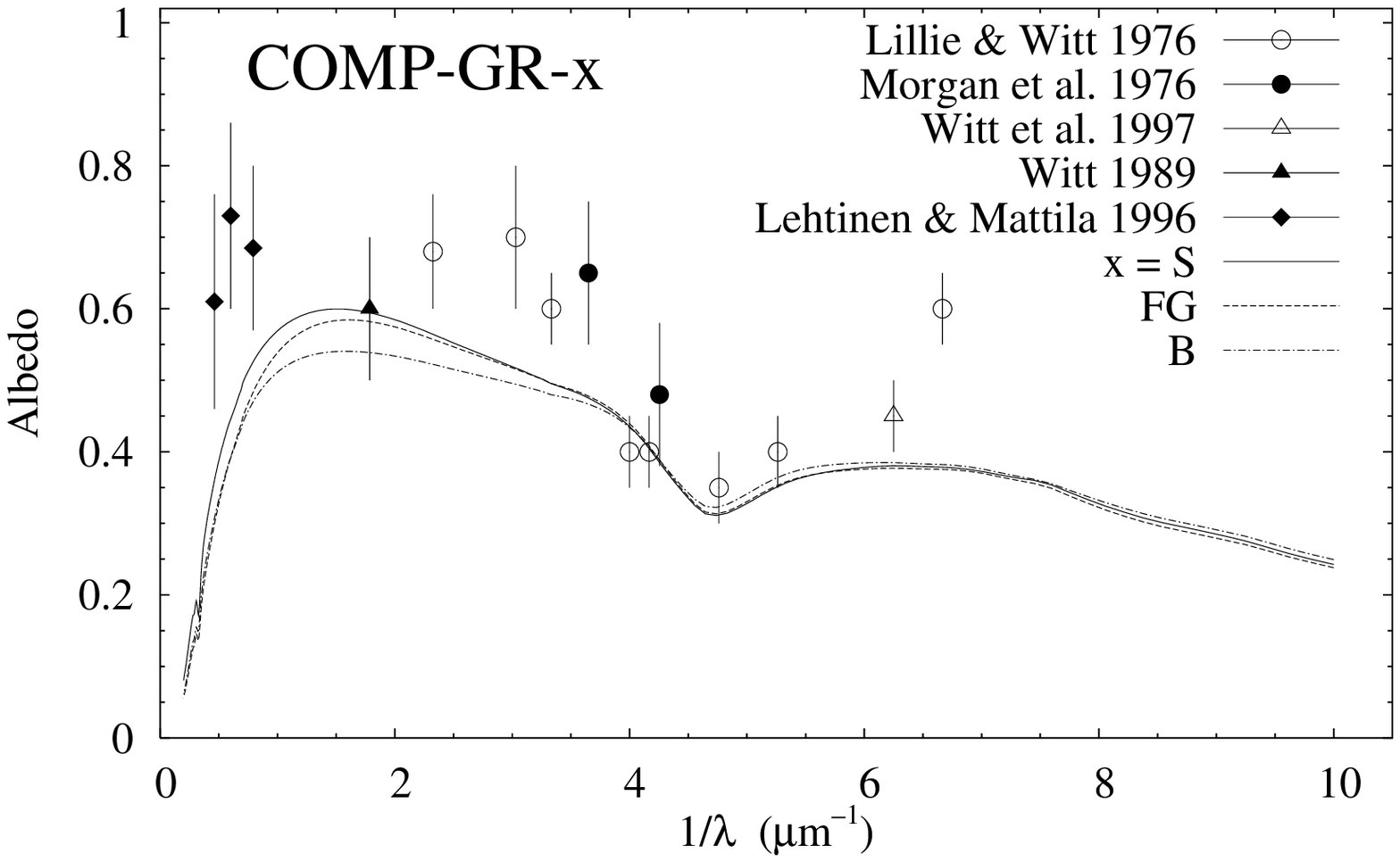}{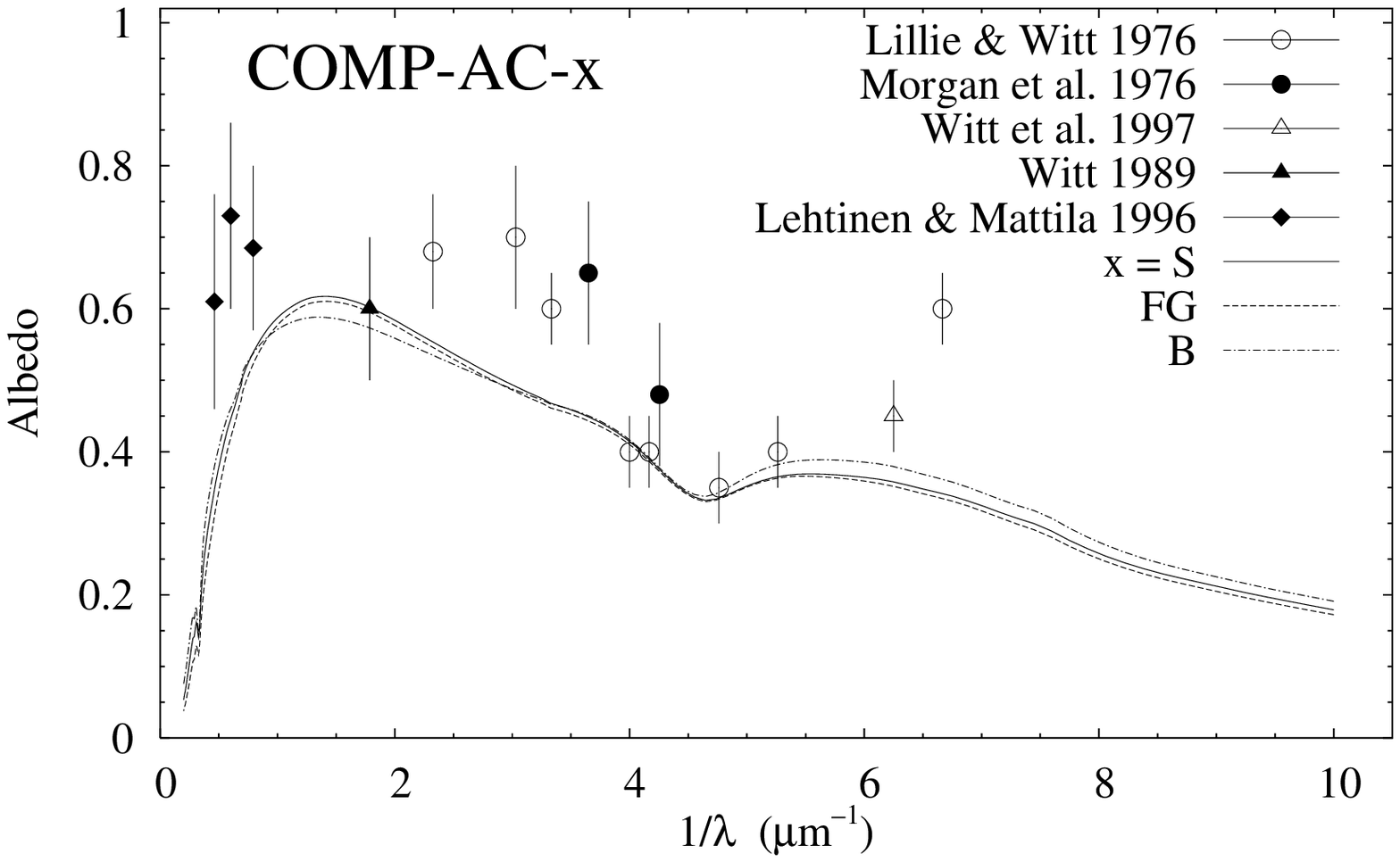}
\epsscale{2.45}
\plotone{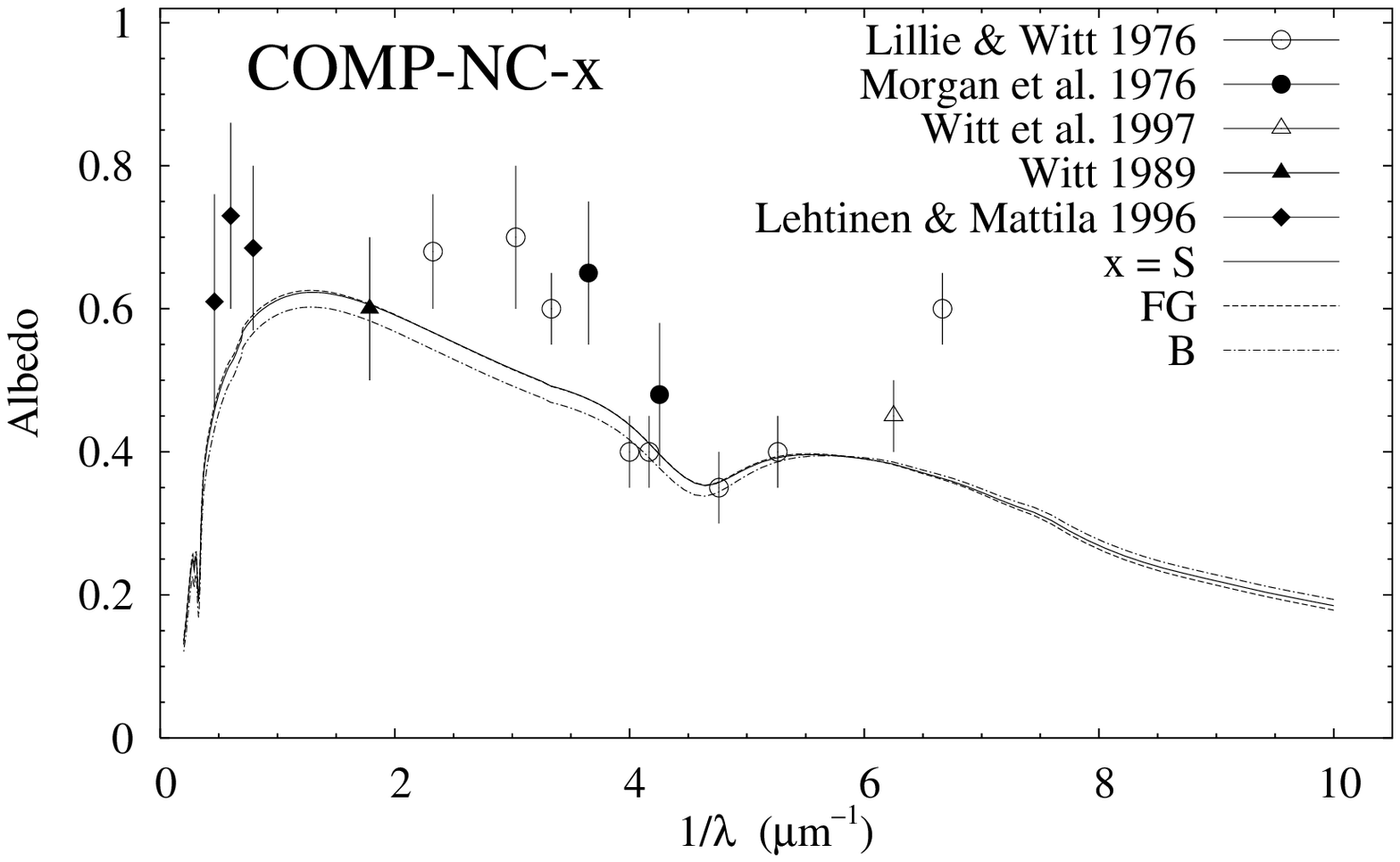}
\caption{Albedo for the dust models along with
   observational data for the diffuse Galactic light
   \citep{lw76,morgan76,witt97,witt89,lm96}.
}
   \label{fig:albedo}
\end{center}
\end{figure}

\begin{figure}
\begin{center}
\epsscale{1.1}
\plottwo{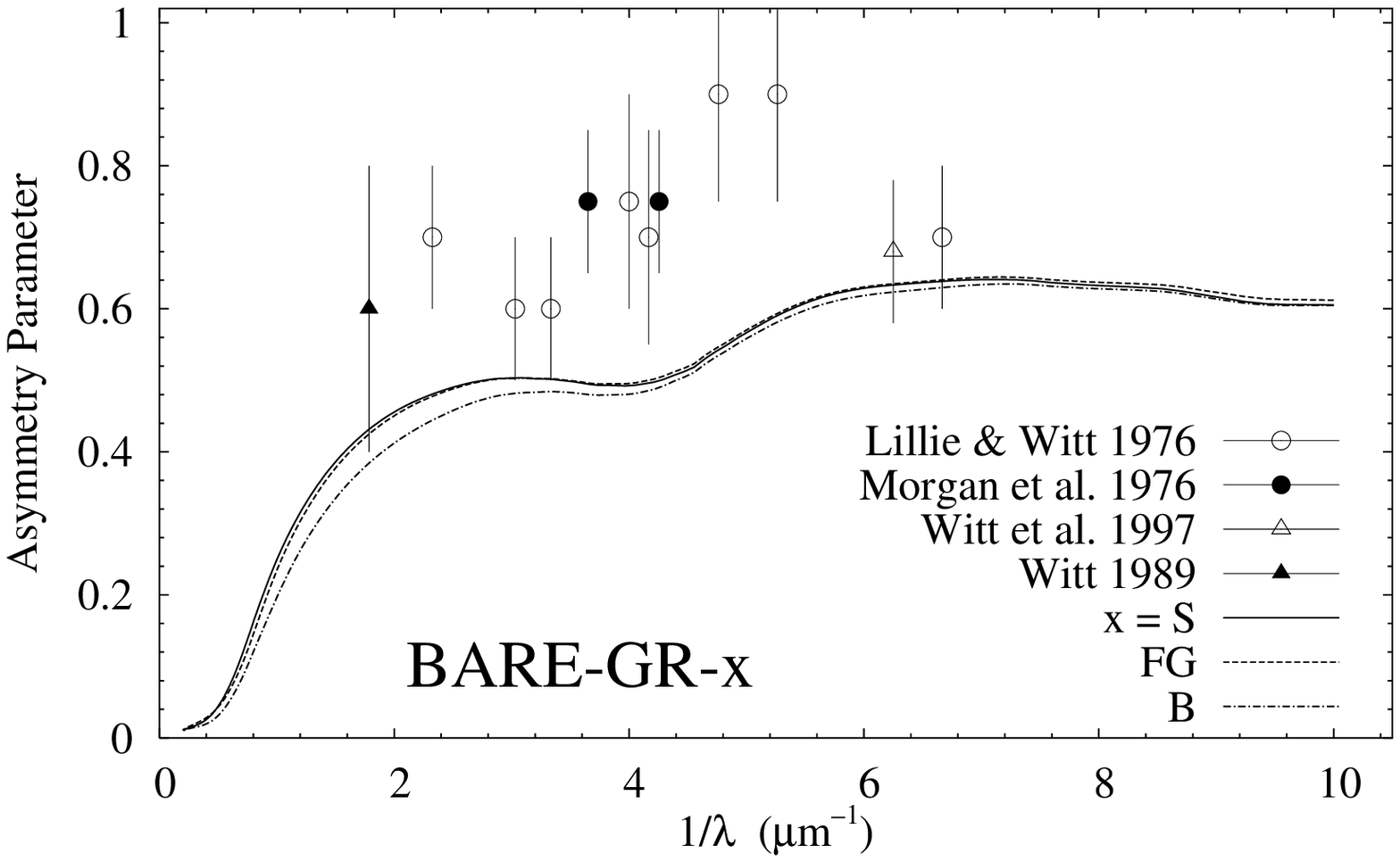}{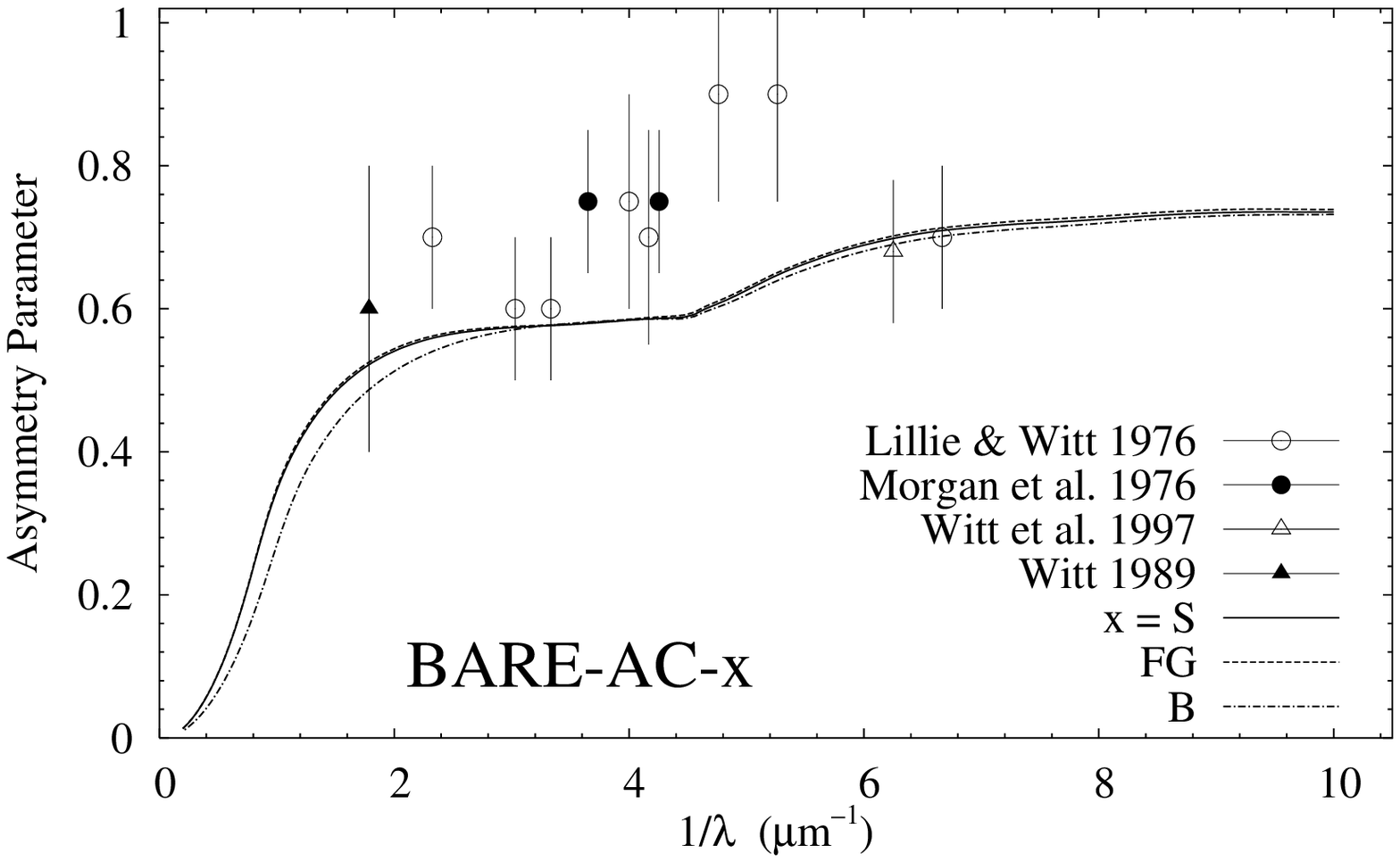}
\epsscale{2.45}
\plottwo{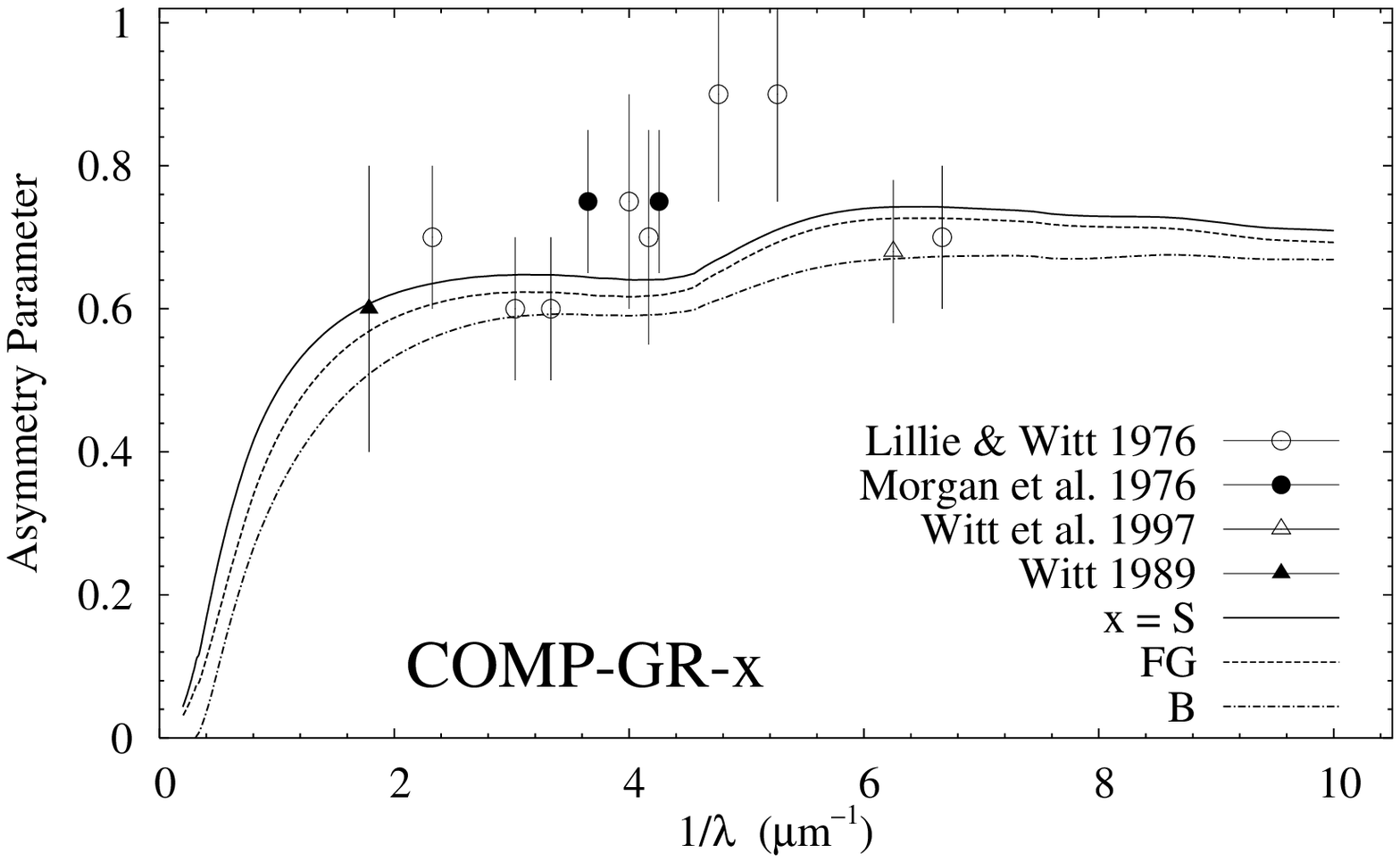}{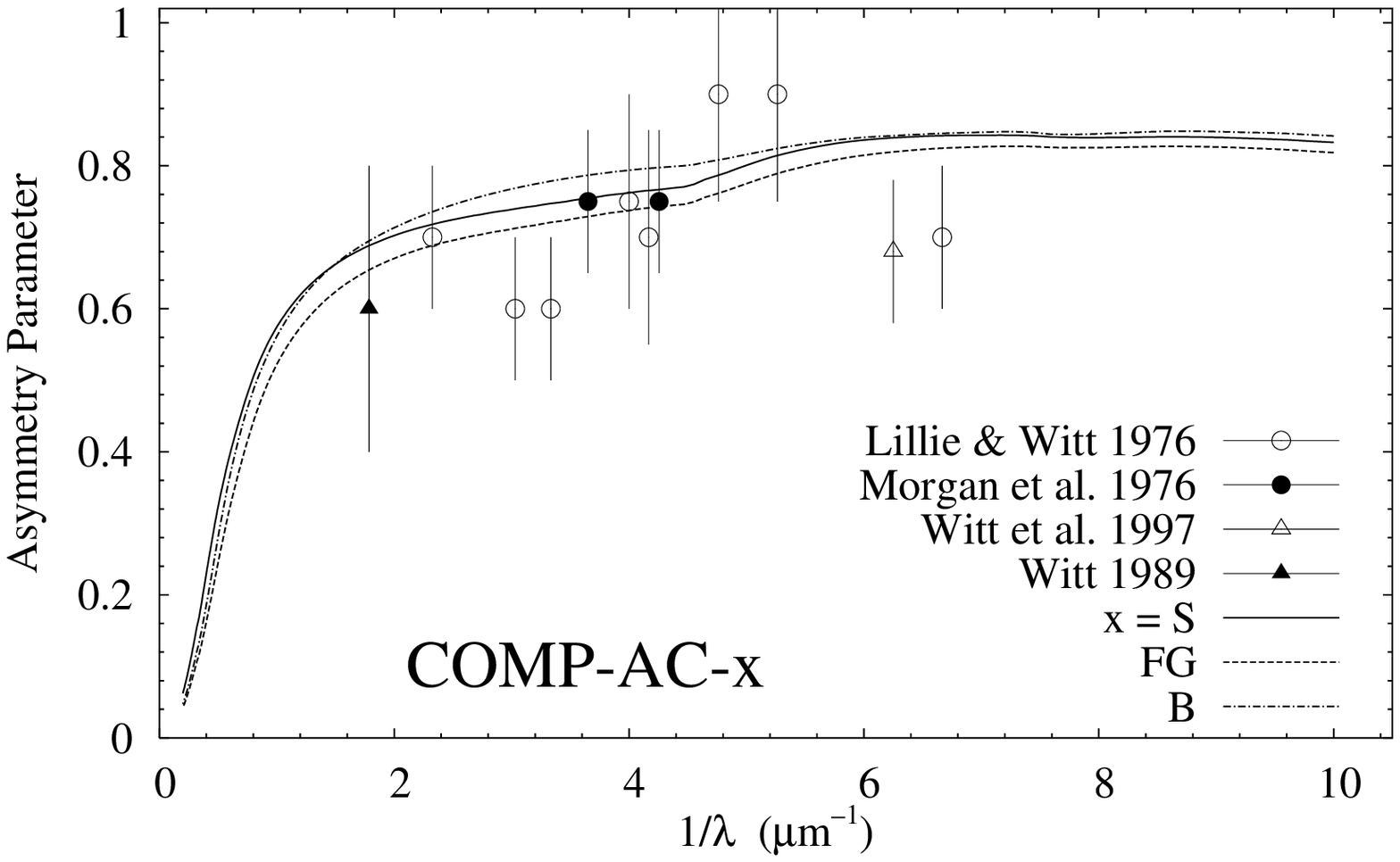}
\epsscale{2.45}
\plotone{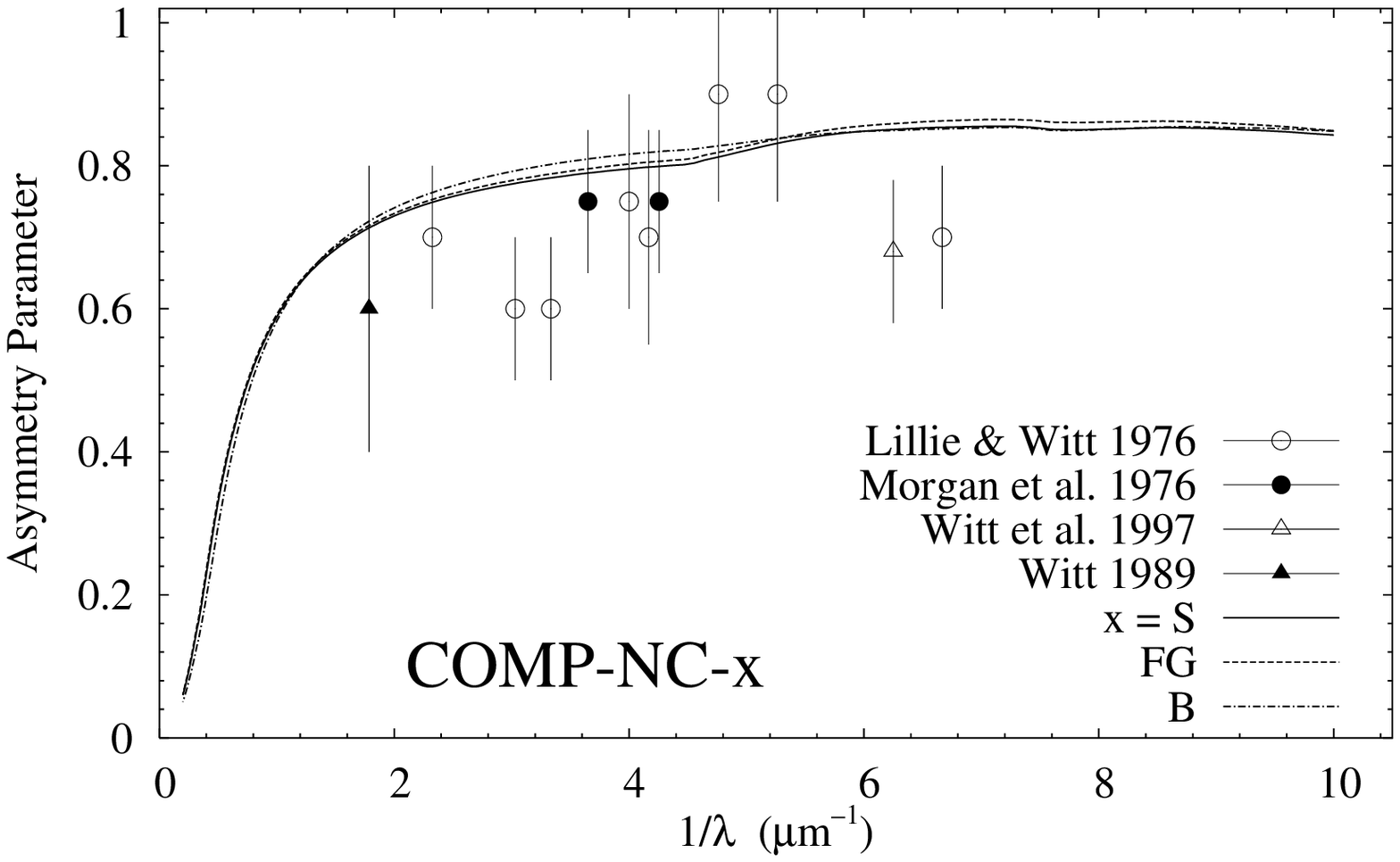}
\caption{ Asymmetry parameter for the dust models
   along with observational data for the diffuse
   Galactic light \citep{lw76,morgan76,witt97,witt89}.
}
   \label{fig:aspar}
\end{center}
\end{figure}

In Figures~\ref{fig:albedo}--\ref{fig:aspar}, we show the scattering
properties of our dust models: the albedo, that is the ratio of the mean
scattering cross section to the mean extinction cross section;
and phase function asymmetry parameter which is the mean value of
the cosine of the scattering angle. Generally, our models give
slightly lower values than the observed ones, but the COMP models
look preferable in comparison to the BARE models, especially for
the asymmetry parameter.

\section{ASTROPHYSICAL IMPLICATIONS}
      \label{section:implications}

An important result of our modeling is that the current extinction,
IR emission, and interstellar abundance constraints can be
simultaneously fit by several different classes of dust models. 

One class of models consist of the widely-used PAH, bare graphite
and silicate grains. With our addition of the abundance constraints
we derive a similar carbon abundance in PAHs and graphite as
\citet{ld01b} did, but with generally smaller graphite grains and
larger PAH molecules (see Fig. \ref{fig:size_ld01}). Our silicate
size distribution turns out to be similar to the MRN distribution,
with a shallower slope than the Li  \& Draine model.
The peak in our silicate grain size distribution is lower than that
in their model, which is why our Mg/H, Si/H, and Fe/H abundance
in the dust is only 33 ppm, compared to the $\approx$ 50 required
in the Li \& Draine model. 

A second class of dust models that fits the constraints contains
in addition to bare grains, composite particles, such as those
believed to have formed and observed to be present in dense
molecular clouds \citep[e.g.]{gl99,tielens96,schutte98,chiar00}. 
The sizes of the composite particles are generally larger than
those of the bare particles. This is consistent with the scenario
in which the ices and organic material of the composite grains
accumulated in molecular clouds, producing larger dust aggregates
than expected in the diffuse ISM. The presence of such particles
in the diffuse ISM would imply that their volatile components
survive the harsher radiation field in that environment.
This may be due to the fact that the volatile compounds may
have accreted in protected areas of the dust particles,
that the large size of the composite grains prevents surges in
their temperature which would evaporate the volatiles, and that
the volatiles may have been reprocessed by UV radiation into
a more refractory component. In our future works, we hope
to include other volatile materials, e.g. icy methane, ammonia,
carbon dioxide, and other oxides, when the respective optical
constants are measured in laboratory.

We also presented a class of BARE and COMP models in which graphite
is completely replaced by hydrogenated amorphous carbon. In these models,
PAHs are the only contributor to the UV extinction bump at 2175 {\AA}.
These models are motivated by the fact that the highly ordered structure
of graphite is unlikely to form in circumstellar environments,
and survive in the diffuse ISM, making amorphous carbon grains more
atractive alternative \citep{witt00}.  We note that the discrepancies
of models containing amorphous carbon are generally larger
than those containing graphite, which may be due to the fact that
their optical constants were not optimized to fit astronomical
observations. In contrast, the optical constants for ``astronomical''
graphite and ``astronomical'' silicate used in this study have been
basically derived by \citet{dl84} by fitting laboratory measurements
and interstellar extinction law, using the MRN model of graphite and
silicate grains with a power-law size distribution. 
So, it should not be surprising that dust models without graphite
will not provide as good a fit as those that do. 

A final class of models contains no carbon at all, except for that
present in PAHs, and in the organic constituent of the composite
particles. This model was motivated by the desire to fit the lowest
available carbon abundance in dust if the ISM had B star abundances.
The model, designated COMP-NC-B, consumes less carbon,
C/H $\approx$ 196 ppm, compared to models COMP-GR-B
(C/H $\approx$ 240 ppm) or COMP-AC-B (C/H $\approx$ 207 ppm)
(Table~\ref{tab_fin_abund}).
This class of models is made possible by the fact that the absorption
cross sections of a mixture of ``astronomical'' PAHs were adjusted
by \citet{ld01b} to fit the UV extinction bump at 2175 {\AA}.
These are idealized cross sections which ignore substructures in
the measured UV absorption of the PAH mixtures \citep{leger89,joblin92}.

Since several classes of dust models fit the observed extinction,
IR emission, and elemental abundances, any linear combination
of these models will provide equally good fits to these constraints.
This provides a wide range of latitude in the construction of
evolutionary models in which dust cycles between the various
ISM phases, allowing for spatial variations in dust composition
and size distribution in the diffuse ISM without significantly
affecting the average extinction.

\section{CONCLUSIONS}
      \label{section:conclusions}

We have taken an additional step in the characterization of
interstellar dust particles, by simultaneopusly fitting their
interstellar extinction, diffuse IR emission, and interstellar
abundance constraints. We used dust constituents with physical
optical constants and properties to characterize the components
of the dust model: PAHs, silicates, graphite, amorphous carbon,
and composite particles. Given the observational constraints
and the dust consituents and properties, we used the method
of regularization to solve for the optimal grain size
distribution of each dust component.

The principal results of this paper are as follows:
\begin{enumerate}
\item
   We found that bare grain (BARE) models (PAHs + graphite + silicate
   or PAHs + amorphous carbon + silicate) provide good simultaneous
   fits to the far-UV to near-IR extinction, thermal IR emission,
   and elemental abundance constraints.
\item
   Somewhat improved fits to the observational constraints are obtained
   by the composite grain (COMP) models, through the addition
   of composite grains (silicate + organic refractory + water ice
   + voids) to the bare grain model. 
\item
   The COMP-GR-FG model provides the best fit to the extinction and
   IR emission if the ISM has F and G stars abundances. The model also
   provides a good match to the observed infrared extinction and
   scattering properties: albedo and asymmetry parameter.
\item
   The results of our calculations show that it is harder to fit
   the B star abundances than the solar or F and G star abundances.
   Thus, we can conclude that solar or F and G star abundances for
   diffuse interstellar medium look preferable.
\item
   The results of our modeling demonstrate that there is no unique
   dust model that fits the extinction, IR emission, and elemental
   abundances constraints. We believe that adding more constraints,
   such as interstellar polarization could
   narrow the set of viable dust models.
\item
   Preliminary calculations show that haloes produced by small angle
   X-ray scattering can discriminate between the different dust models,
   favoring BARE over COMP models, which have a larger population
   of big particles \citep{dwek04}. 
\end{enumerate}

\acknowledgments

This work was supported by NASA's Astrophysics Theory Program NRA 99-OSS-01.
We thank the referee for his careful reading of the manuscript and his
insightful suggestions. We are grateful to Edward Fitzpatrick for
providing us with the Galactic extinction curve and estimates of its
uncertainty, and Bruce T. Draine for helpful discussions concerning
the IR emission from PAHs.

\clearpage


\begin{deluxetable}{llllllll}
\tabletypesize{\footnotesize}
\tablewidth{0pt}
\tablecaption{Inferred Dust Phase Abundances In the Diffuse ISM per 10$^6$ H Atoms.  \label{tab_dust_abund}}
\tablehead{
  \colhead{}   &
  \colhead{}   &
  \colhead{C}  &
  \colhead{O}  &
  \colhead{Si} &
  \colhead{Mg} &
  \colhead{Fe} &
  \colhead{N}
}
\startdata
Total ISM  & Solar\tablenotemark{a}        & 391$\pm$98 & 545$\pm$100 & 34.4$\pm$3.9  & 34.5$\pm$4.8  & 28.1$\pm$5.4  & 85.2$\pm$21.9\\
           & F \& G Stars\tablenotemark{b} & 358$\pm$82 & 445$\pm$156 & 39.9$\pm$13.1 & 42.7$\pm$17.2 & 27.9$\pm$7.7  & \nodata\\
           & B Stars\tablenotemark{b}      & 190$\pm$77 & 350$\pm$133 & 18.8$\pm$8.9  & 23.0$\pm$7.0  & 28.5$\pm$18.0 & 64.7$\pm$34.2\\
\tableline
Gas Phase\tablenotemark{c}  &              & 108$\pm$16 & 319$\pm$14  & $\approx$ 0   & $\approx$ 0   &  $\approx$ 0  & 75.0$\pm$4.0\\
\tableline
Dust Phase & Solar (S)                     & 283$\pm$99 & 226$\pm$101 & 34.4$\pm$3.9  & 34.5$\pm$4.8  & 28.1$\pm$5.4  & 10.2$\pm$22.3\\
           & F \& G Stars (FG)             & 250$\pm$84 & 126$\pm$157 & 39.9$\pm$13.1 & 42.7$\pm$17.2 & 27.9$\pm$7.7  & \nodata\\
           & B Stars (B)                   &  82$\pm$79 &  31$\pm$134 & 18.8$\pm$8.9  & 23.0$\pm$7.0  & 28.5$\pm$18.0 &  0.0$\pm$35\\
\enddata
\tablenotetext{a}{\citet{holweger01}}
\tablenotetext{b}{\citet{sofia01}}
\tablenotetext{c}{ The value for C is a straight average of two
  available estimates of 140$\pm$20 by \citet{cardelli96} and
  75$\pm$25 by \citet{dwek97}; the values for O and N
  are from \citet{meyer98} and \citet{meyer97}, respectively
}
\end{deluxetable}

\clearpage

\begin{deluxetable}{llcl}
\tablewidth{0pt}
\tablecaption{Properties of dust constituents.       \label{tab_dust_const}}
\tablehead{
  \multicolumn{1}{c}{Constituent}               &
  \multicolumn{1}{c}{Composition}               &
  \multicolumn{1}{c}{Mass density} &
  \multicolumn{1}{c}{Optical constants}\\
 & & \multicolumn{1}{c}{g cm$^{-3}$} & \multicolumn{1}{c}{References}
}
\startdata
PAHs               & C                      & 2.24       & Li \& Draine 2001b\\
Graphite           & C                      & 2.24       & Laor \& Draine 1993\\
Silicate           & MgFeSiO$_4$            & 3.5        & Weingartner \& Draine 2001\\
Amorphous Carbon   & C                      & 1.81--1.87 & Zubko et al. 1996\\
Organic Refractory & C$_{25}$H$_{25}$O$_5$N & 1.6        & Li \& Greenberg 1997\\
Water Ice          & H$_2$O                 & 0.92       & Warren 1984\\
\enddata
\end{deluxetable}

\clearpage

\begin{deluxetable}{lrrrcc}
\tabletypesize{\footnotesize}
\tablewidth{0pt}
\tablecaption{Discrepancies of the dust models.  \label{tab_discr}}
\tablehead{
  \colhead{Model}                        &
  \colhead{$\Phi_{\rm{d}}$}              &
  \colhead{$\Phi^{\rm{ext}}_{\rm{d}}$}   &
  \colhead{$\Phi^{\rm{emis}}_{\rm{d}}$}  &
  \colhead{$\Phi^{\rm{abund}}_{\rm{d}}$} &
  \colhead{$\alpha$}
}
\startdata
COMP-GR-FG        &  8.22 &  0.09 &  7.91 &  0.22  & 0.828\\
BARE-GR-FG        &  8.38 &  0.14 &  8.03 &  0.21  & 0.859\\
COMP-AC-FG        &  8.44 &  0.12 &  8.07 &  0.25  & 0.283\\
COMP-NC-FG        &  8.48 &  0.12 &  8.02 &  0.34  & 0.278\\
BARE-AC-FG        &  9.07 &  0.30 &  8.54 &  0.23  & 0.300\\[0.5cm]

COMP-GR-S         &  9.27 &  0.10 &  8.82 &  0.35  & 1.106\\
COMP-AC-S         &  9.46 &  0.14 &  8.98 &  0.34  & 0.325\\
BARE-GR-S         &  9.48 &  0.17 &  8.93 &  0.38  & 1.123\\
COMP-NC-S         &  9.51 &  0.14 &  8.99 &  0.38  & 0.322\\
BARE-AC-S         & 10.12 &  0.33 &  9.43 &  0.36  & 0.341\\[0.5cm]

COMP-GR-B         & 10.42 &  0.20 &  9.21 &  1.01  & 1.233\\
COMP-AC-B         & 10.57 &  0.14 &  9.62 &  0.81  & 0.346\\
COMP-NC-B         & 10.58 &  0.14 &  9.67 &  0.77  & 0.349\\
BARE-GR-B         & 10.76 &  0.27 &  9.33 &  1.16  & 1.290\\
BARE-AC-B         & 11.72 &  0.47 &  9.87 &  1.38  & 0.360\\
\enddata
\tablecomments{The table lists the total discrepancies for the models,
  $\Phi_{\rm{d}}$, together with the contributions from extinction,
  $\Phi^{\rm{ext}}_{\rm{d}}$, emission, $\Phi^{\rm{emis}}_{\rm{d}}$,
  and abundances, $\Phi^{\rm{abund}}_{\rm{d}}$. By definition:
  $\Phi_{\rm{d}} = \Phi^{\rm{ext}}_{\rm{d}} + \Phi^{\rm{emis}}_{\rm{d}}
  + \Phi^{\rm{abund}}_{\rm{d}}$. The last column contains
  the parameter of regularization, $\alpha$, in $\mu$m~H$^2$. 
}
\end{deluxetable}

\clearpage

\begin{deluxetable}{lccccc}
\tablewidth{0pt}
\tablecaption{Volume fractions of constituents and average mass densities
   of model composite particles.     \label{tab_vol_fr}}
\tablehead{
  \multicolumn{1}{c}{Model}        &
  \multicolumn{1}{c}{Voids}        &
  \multicolumn{1}{c}{Silicate}     &
  \multicolumn{1}{c}{Organic}      &
  \multicolumn{1}{c}{Water}        &
  \multicolumn{1}{c}{Density}\\
  & & & Refractory & Ice & (g cm$^{-3}$)\\
}
\startdata
COMP-GR-S  & 0.50 & 0.15 & 0.30 & 0.05 & 1.05 \\
COMP-AC-S  & 0.60 & 0.12 & 0.24 & 0.04 & 0.84 \\
COMP-NC-S  & 0.50 & 0.15 & 0.30 & 0.05 & 1.05 \\
COMP-GR-FG & 0.50 & 0.15 & 0.30 & 0.05 & 1.05 \\
COMP-AC-FG & 0.50 & 0.15 & 0.30 & 0.05 & 1.05 \\
COMP-NC-FG & 0.50 & 0.15 & 0.30 & 0.05 & 1.05 \\
COMP-GR-B  & 0.10 & 0.27 & 0.54 & 0.09 & 1.89 \\
COMP-AC-B  & 0.50 & 0.15 & 0.30 & 0.05 & 1.05 \\
COMP-NC-B  & 0.50 & 0.15 & 0.30 & 0.05 & 1.05 \\
\tableline
\enddata
\end{deluxetable}

\clearpage

\begin{deluxetable}{llrrrrrc}
\tabletypesize{\footnotesize}
\tablewidth{0pt}
\tablecaption{Atomic abundances in dust in atoms per 10$^6$ H Atoms for various models.   \label{tab_fin_abund}}
\tablehead{
 \multicolumn{1}{c}{Model}   & \multicolumn{1}{c}{Component}
    & \multicolumn{1}{c}{C}  & \multicolumn{1}{c}{O}  & \multicolumn{1}{c}{Si}
    & \multicolumn{1}{c}{Mg} & \multicolumn{1}{c}{Fe} & \multicolumn{1}{c}{N}
}
\startdata
BARE-GR-S   & {\em Total} & {\em 245.9}& {\em 133.2}& {\em 33.3}& {\em 33.3}& {\em 33.3}& {\em  0.0}\\
            & PAHs        &       33.0 &        0.0 &       0.0 &       0.0 &       0.0 &       0.0\\
            & Graphite    &      212.9 &        0.0 &       0.0 &       0.0 &       0.0 &       0.0\\
            & Silicate    &        0.0 &      133.2 &      33.3 &      33.3 &      33.3 &       0.0\\
\tableline
BARE-GR-FG  & {\em Total} & {\em 247.5}& {\em 132.4}& {\em 33.1}& {\em 33.1}& {\em 33.1}& {\em  0.0}\\
            & PAHs        &       35.2 &        0.0 &       0.0 &       0.0 &       0.0 &       0.0\\
            & Graphite    &      212.3 &        0.0 &       0.0 &       0.0 &       0.0 &       0.0\\
            & Silicate    &        0.0 &      132.4 &      33.1 &      33.1 &      33.1 &       0.0\\
\tableline
BARE-GR-B   & {\em Total} & {\em 254.4}& {\em 114.0}& {\em 28.5}& {\em 28.5}& {\em 28.5}& {\em  0.0}\\
            & PAHs        &       33.3 &        0.0 &       0.0 &       0.0 &       0.0 &       0.0\\
            & Graphite    &      221.1 &        0.0 &       0.0 &       0.0 &       0.0 &       0.0\\
            & Silicate    &        0.0 &      114.0 &      28.5 &      28.5 &      28.5 &       0.0\\
\tableline
BARE-AC-S   & {\em Total} & {\em 265.0}& {\em 134.0}& {\em 33.5}& {\em 33.5}& {\em 33.5}& {\em  0.0}\\
            & PAHs        &       51.4 &        0.0 &       0.0 &       0.0 &       0.0 &       0.0\\
            & ACH2        &      213.6 &        0.0 &       0.0 &       0.0 &       0.0 &       0.0\\
            & Silicate    &        0.0 &      134.0 &      33.5 &      33.5 &      33.5 &       0.0\\
\tableline
BARE-AC-FG  & {\em Total} & {\em 265.1}& {\em 136.8}& {\em 34.2}& {\em 34.2}& {\em 34.2}& {\em  0.0}\\
            & PAHs        &       52.4 &        0.0 &       0.0 &       0.0 &       0.0 &       0.0\\
            & ACH2        &      212.7 &        0.0 &       0.0 &       0.0 &       0.0 &       0.0\\
            & Silicate    &        0.0 &      136.8 &      34.2 &      34.2 &      34.2 &       0.0\\
\tableline
BARE-AC-B   & {\em Total} & {\em 275.3}& {\em 114.8}& {\em 28.7}& {\em 28.7}& {\em 28.7}& {\em  0.0}\\
            & PAHs        &       52.2 &        0.0 &       0.0 &       0.0 &       0.0 &       0.0\\
            & ACH2        &      223.1 &        0.0 &       0.0 &       0.0 &       0.0 &       0.0\\
            & Silicate    &        0.0 &      114.8 &      28.7 &      28.7 &      28.7 &       0.0\\
\tableline
COMP-GR-S   & {\em Total} & {\em 217.4}& {\em 153.6}& {\em 33.0}& {\em 33.0}& {\em 33.0}& {\em  3.0}\\
            & PAHs        &       33.5 &        0.0 &       0.0 &       0.0 &       0.0 &       0.0\\
            & Graphite    &      109.2 &        0.0 &       0.0 &       0.0 &       0.0 &       0.0\\
            & Silicate    &        0.0 &      100.0 &      25.0 &      25.0 &      25.0 &       0.0\\
            & Composites  &       74.7 &       53.6 &       8.0 &       8.0 &       8.0 &       3.0\\
\tableline
COMP-GR-FG  & {\em Total} & {\em 227.9}& {\em 146.7}& {\em 32.4}& {\em 32.4}& {\em 32.4}& {\em  2.4}\\
            & PAHs        &       35.8 &        0.0 &       0.0 &       0.0 &       0.0 &       0.0\\
            & Graphite    &      133.3 &        0.0 &       0.0 &       0.0 &       0.0 &       0.0\\
            & Silicate    &        0.0 &      104.4 &      26.1 &      26.1 &      26.1 &       0.0\\
            & Composites  &       58.8 &       42.3 &       6.3 &       6.3 &       6.3 &       2.4\\
\tableline
COMP-GR-B   & {\em Total} & {\em 239.8}& {\em 132.4}& {\em 27.8}& {\em 27.8}& {\em 27.8}& {\em  2.9}\\
            & PAHs        &       33.7 &        0.0 &       0.0 &       0.0 &       0.0 &       0.0\\
            & Graphite    &      133.0 &        0.0 &       0.0 &       0.0 &       0.0 &       0.0\\
            & Silicate    &        0.0 &       80.0 &      20.0 &      20.0 &      20.0 &       0.0\\
            & Composites  &       73.1 &       52.4 &       7.8 &       7.8 &       7.8 &       2.9\\
\tableline
COMP-AC-S   & {\em Total} & {\em 216.7}& {\em 159.7}& {\em 33.3}& {\em 33.3}& {\em 33.3}& {\em  3.6}\\
            & PAHs        &       50.6 &        0.0 &       0.0 &       0.0 &       0.0 &       0.0\\
            & ACH2        &       75.2 &        0.0 &       0.0 &       0.0 &       0.0 &       0.0\\
            & Silicate    &        0.0 &       94.8 &      23.7 &      23.7 &      23.7 &       0.0\\
            & Composites  &       90.9 &       64.9 &       9.6 &       9.6 &       9.6 &       3.6\\
\tableline
COMP-AC-FG  & {\em Total} & {\em 219.3}& {\em 160.0}& {\em 33.8}& {\em 33.8}& {\em 33.8}& {\em  3.5}\\
            & PAHs        &       51.7 &        0.0 &       0.0 &       0.0 &       0.0 &       0.0\\
            & ACH2        &       81.2 &        0.0 &       0.0 &       0.0 &       0.0 &       0.0\\
            & Silicate    &        0.0 &       98.0 &      24.5 &      24.5 &      24.5 &       0.0\\
            & Composites  &       86.4 &       62.0 &       9.2 &       9.2 &       9.2 &       3.5\\
\tableline
COMP-AC-B   & {\em Total} & {\em 206.9}& {\em 148.5}& {\em 27.9}& {\em 27.9}& {\em 27.9}& {\em  5.1}\\
            & PAHs        &       51.5 &        0.0 &       0.0 &       0.0 &       0.0 &       0.0\\
            & ACH2        &       28.1 &        0.0 &       0.0 &       0.0 &       0.0 &       0.0\\
            & Silicate    &        0.0 &       57.2 &      14.3 &      14.3 &      14.3 &       0.0\\
            & Composites  &      127.3 &       91.3 &      13.6 &      13.6 &      13.6 &       5.1\\
\tableline
COMP-NC-S   & {\em Total} & {\em 188.2}& {\em 173.6}& {\em 33.4}& {\em 33.4}& {\em 33.4}& {\em  5.5}\\
            & PAHs        &       50.0 &        0.0 &       0.0 &       0.0 &       0.0 &       0.0\\
            & Silicate    &        0.0 &       74.8 &      18.7 &      18.7 &      18.7 &       0.0\\
            & Composites  &      138.2 &       98.8 &      14.7 &      14.7 &      14.7 &       5.5\\
\tableline
COMP-NC-FG  & {\em Total} & {\em 189.9}& {\em 175.6}& {\em 33.9}& {\em 33.9}& {\em 33.9}& {\em  5.6}\\
            & PAHs        &       51.0 &        0.0 &       0.0 &       0.0 &       0.0 &       0.0\\
            & Silicate    &        0.0 &       76.4 &      19.1 &      19.1 &      19.1 &       0.0\\
            & Composites  &      138.9 &       99.2 &      14.8 &      14.8 &      14.8 &       5.6\\
\tableline
COMP-NC-B   & {\em Total} & {\em 196.1}& {\em 154.3}& {\em 28.1}& {\em 28.1}& {\em 28.1}& {\em  5.8}\\
            & PAHs        &       51.5 &        0.0 &       0.0 &       0.0 &       0.0 &       0.0\\
            & Silicate    &        0.0 &       50.8 &      12.7 &      12.7 &      12.7 &       0.0\\
            & Composites  &      144.6 &      103.5 &      15.4 &      15.4 &      15.4 &       5.8\\
\enddata
\tablecomments{ The first column is a name of the model; next column lists
   the components for each model; next six columns are the numbers of elements
   in dust components in ppm. The total abundances summed up over all
   components are in italics.
}
\end{deluxetable}

\clearpage

\begin{deluxetable}{lcccccccc}
\tabletypesize{\footnotesize}
\tablewidth{0pt}
\tablecaption{Mass in dust for various models.       \label{tab_mass}}
\tablehead{
  \colhead{Model}                       &
  \multicolumn{1}{c}{dust/gas} &
  \multicolumn{1}{c}{$M$ (g H$^{-1}$)}  &
  \multicolumn{1}{c}{PAHs}              &
  \multicolumn{1}{c}{Graphite}          &
  \multicolumn{1}{c}{ACH2}              &
  \multicolumn{1}{c}{Silicate}          &
  \multicolumn{1}{c}{Water Ice}         &
  \multicolumn{1}{c}{Organics}
}
\startdata
Li \& Draine 2001 & 0.00813 & 1.89$\times$10$^{-26}$ & \multicolumn{2}{c}{26.77} & \nodata &   73.23 & \nodata & \nodata \\
BARE-GR-S         & 0.00619 & 1.44$\times$10$^{-26}$ &            4.57 &   29.47 & \nodata &   65.96 & \nodata & \nodata \\
BARE-GR-FG        & 0.00618 & 1.43$\times$10$^{-26}$ &            4.88 &   29.44 & \nodata &   65.68 & \nodata & \nodata \\
BARE-GR-B         & 0.00568 & 1.32$\times$10$^{-26}$ &            5.02 &   33.38 & \nodata &   61.60 & \nodata & \nodata \\
BARE-AC-S         & 0.00639 & 1.49$\times$10$^{-26}$ &            6.89 & \nodata &   28.65 &   64.46 & \nodata & \nodata \\
BARE-AC-FG        & 0.00648 & 1.51$\times$10$^{-26}$ &            6.94 & \nodata &   28.15 &   64.91 & \nodata & \nodata \\
BARE-AC-B         & 0.00589 & 1.37$\times$10$^{-26}$ &            7.60 & \nodata &   32.47 &   59.93 & \nodata & \nodata \\
COMP-GR-S         & 0.00626 & 1.46$\times$10$^{-26}$ &            4.59 &   14.96 & \nodata &   64.78 &    1.37 &   14.30 \\
COMP-GR-FG        & 0.00620 & 1.44$\times$10$^{-26}$ &            4.94 &   18.43 & \nodata &   64.20 &    1.09 &   11.34 \\
COMP-GR-B         & 0.00580 & 1.35$\times$10$^{-26}$ &            4.98 &   19.64 & \nodata &   58.86 &    1.44 &   15.08 \\
COMP-AC-S         & 0.00637 & 1.48$\times$10$^{-26}$ &            6.81 & \nodata &   10.13 &   64.34 &    1.63 &   17.09 \\
COMP-AC-FG        & 0.00642 & 1.49$\times$10$^{-26}$ &            6.90 & \nodata &   10.85 &   64.60 &    1.54 &   16.11 \\
COMP-AC-B         & 0.00578 & 1.34$\times$10$^{-26}$ &            7.63 & \nodata &    4.16 &   59.31 &    2.53 &   26.37 \\
COMP-NC-S         & 0.00635 & 1.48$\times$10$^{-26}$ &            6.76 & \nodata & \nodata &   64.68 &    2.49 &   26.07 \\
COMP-NC-FG        & 0.00642 & 1.49$\times$10$^{-26}$ &            6.81 & \nodata & \nodata &   64.80 &    2.48 &   25.91 \\
COMP-NC-B         & 0.00579 & 1.35$\times$10$^{-26}$ &            7.62 & \nodata & \nodata &   59.64 &    2.86 &   29.88 \\
\enddata
\tablecomments{ The first column is name of the model; next two columns contain
   the dust-to-gas mass ratio and the mass in dust per one H atom; the rest
   of columns are the contributions of dust constituents in the dust mass in per cent.
}
\end{deluxetable}

\clearpage

\begin{deluxetable}{llll}
\tablewidth{0pt}
\tablecaption{Parameters of the analytical approximations
  to the size distributions for the BARE-GR-S model.       \label{f_bare-s-gr}}
\tablehead{
  \colhead{}                   &
  \multicolumn{1}{c}{PAHs}     &
  \multicolumn{1}{c}{Graphite} &
  \multicolumn{1}{c}{Silicate}
}
\startdata
$A$   & 2.227433$-$7 & 1.905816$-$7 & 1.471288$-$7 \\
$a_{\rm{min}}$
      & 0.00035    & 0.00035    & 0.00035    \\
$a_{\rm{max}}$
      & 0.005      & 0.33       & 0.37       \\
$\chi^2_{\rm{red}}$ &
        1.57548$-$4  & 8.60041$-$2  & 4.40336$-$2  \\
\tableline
$c_0$ & $-$8.02895   & $-$9.86000   & $-$8.47091   \\
$b_0$ & $-$3.45764   & $-$5.02082   & $-$3.68708   \\

$b_1$ & 1.18396+3  & 5.81215$-$3  & 2.37316$-$5  \\
$a_1$ & 1.0        & 0.415861   & 7.64943$-$3  \\
$m_1$ & $-$8.20551   & 4.63229    & 22.5489    \\

$b_2$ & \nodata    & \nodata    & \nodata    \\
$a_2$ & \nodata    & \nodata    & \nodata    \\
$m_2$ & \nodata    & \nodata    & \nodata    \\

$b_3$ & 1.0+24     & 1.12502+3  & 2.96128+3  \\
$a_3$ & $-$5.29496$-$3 & 0.160344   & 0.480229   \\
$m_3$ & 12.0146    & 3.69897    & 12.1717    \\

$b_4$ & \nodata    & 1.12602+3  & \nodata    \\
$a_4$ & \nodata    & 0.160501   & \nodata    \\
$m_4$ & \nodata    & 3.69967    & \nodata    \\

\enddata
\tablecomments{
The size distribution function $f(a)$ (in $\mu$m$^{-1}$~H$^{-1}$)
found in a numerical form is expressed as $f(a)=A \> g(a)$,
where $A$ is a normalization coefficient in H$^{-1}$,
$a$ is in $\mu$m, and function $g(a)$ (in $\mu$m$^{-1}$) has
a property $\int \! g(a) \> {\rm{d}}a = 1$.
To analytically approximate numerical values of $g(a)$,
we used a 14-parameter function $\overline g$:
$
\log \> {\overline g}(a) =
c_0 + b_0 \log(a) - b_1 \left|\log\left(a/a_1\right)\right|^{m_1}
                  - b_2 \left|\log\left(a/a_2\right)\right|^{m_2} -
b_3 |a - a_3|^{m_3} - b_4 |a - a_4|^{m_4}.
$
Parameters $a_{\rm{min}}$ and $a_{\rm{max}}$, both in {\micron}, define
the range of grain radii for which the approximations are valid.
$\chi^2_{\rm{red}}$ is the reduced $\chi^2$ for the fit of
the approximation to the model size distribution (not the fit
of the model to the constraints).
}
\end{deluxetable}

\clearpage

\begin{deluxetable}{llll}
\tablewidth{0pt}
\tablecaption{Parameters of the analytical approximations
  to the size distributions for the BARE-GR-FG model.       \label{f_bare-fg-gr}}
\tablehead{
  \colhead{}                   &
  \multicolumn{1}{c}{PAHs}     &
  \multicolumn{1}{c}{Graphite} &
  \multicolumn{1}{c}{Silicate}
}
\startdata
$A$   & 2.484404$-$7 & 1.901190$-$7 & 1.541199$-$7 \\
$a_{\rm{min}}$
      & 0.00035    & 0.00035    & 0.00035    \\
$a_{\rm{max}}$
      & 0.005      & 0.3        & 0.34       \\
$\chi^2_{\rm{red}}$ &
        2.18384$-$4  & 6.679$-$2    & 9.70614$-$2  \\
\tableline
$c_0$ & $-$8.54571   & $-$10.1149   & $-$8.53081   \\
$b_0$ & $-$3.60112   & $-$5.3308    & $-$3.70009   \\

$b_1$ & 1.86525+5  & 7.54276$-$2  & 3.96003$-$9  \\
$a_1$ & 1.0        & 8.08703$-$2  & 9.11246$-$3  \\
$m_1$ & $-$13.5755   & 3.37644    & 47.0606    \\

$b_2$ & \nodata    & \nodata    & \nodata    \\
$a_2$ & \nodata    & \nodata    & \nodata    \\
$m_2$ & \nodata    & \nodata    & \nodata    \\

$b_3$ & 1.0+24     & 1.12502+3  & 1.48+3     \\
$a_3$ & 1.98119$-$3  & 0.145378   & 0.484381   \\
$m_3$ & 9.25894    & 3.49042    & 12.3253    \\

$b_4$ & \nodata    & 1.12602+3  & 1.481+3    \\
$a_4$ & \nodata    & 0.169079   & 0.474035   \\
$m_4$ & \nodata    & 3.63654    & 12.0995    \\
\enddata
\end{deluxetable}

\clearpage

\begin{deluxetable}{llll}
\tablewidth{0pt}
\tablecaption{Parameters of the analytical approximations
  to the size distributions for the BARE-GR-B model.       \label{f_bare-b-gr}}
\tablehead{
  \colhead{}                   &
  \multicolumn{1}{c}{PAHs}     &
  \multicolumn{1}{c}{Graphite} &
  \multicolumn{1}{c}{Silicate}
}
\startdata
$A$   & 2.187355$-$7 & 1.879863$-$7 & 1.238052$-$7 \\
$a_{\rm{min}}$
      & 0.00035    & 0.00035    & 0.00035    \\
$a_{\rm{max}}$
      & 0.0055     & 0.32       & 0.32       \\
$\chi^2_{\rm{red}}$ &
        6.4061$-$4   & 0.165084   & 1.46712$-$2  \\
\tableline
$c_0$ & $-$8.84618   & $-$9.92887   & $-$8.53419   \\
$b_0$ & $-$3.69582   & $-$5.14159   & $-$3.7579    \\

$b_1$ & 1.23836+5  & 4.68489$-$3  & 3.89361$-$13 \\
$a_1$ & 1.0        & 0.450668   & 1.27635$-$3  \\
$m_1$ & $-$13.5577   & 4.85266    & 34.0815    \\

$b_2$ & \nodata    & \nodata    & \nodata    \\
$a_2$ & \nodata    & \nodata    & \nodata    \\
$m_2$ & \nodata    & \nodata    & \nodata    \\

$b_3$ & 1.0+24     & 1.12505+3  & 1.481+3    \\
$a_3$ & 2.3281$-$3   & 0.154046   & 0.268976   \\
$m_3$ & 9.36086    & 3.56481    & 13.3815    \\

$b_4$ & \nodata    & 1.12605+3  & 1.48003+3  \\
$a_4$ & \nodata    & 0.153688   & 0.836879   \\
$m_4$ & \nodata    & 3.56482    & 44.1634    \\
\enddata
\end{deluxetable}

\clearpage

\begin{deluxetable}{lllll}
\tablewidth{0pt}
\tablecaption{Parameters of the analytical approximations
  to the size distributions for the COMP-GR-S model.       \label{f_comp-s-gr}}
\tablehead{
  \colhead{}                     &
  \multicolumn{1}{c}{PAHs}       &
  \multicolumn{1}{c}{Graphite}   &
  \multicolumn{1}{c}{Silicate}   &
  \multicolumn{1}{c}{Composites}
}
\startdata
$A$   & 2.243145$-$7 & 1.965000$-$7 & 1.160677$-$7 & 6.975520$-$12 \\
$a_{\rm{min}}$                               
      & 0.00035    & 0.00035    & 0.00035    & 0.02        \\
$a_{\rm{max}}$                               
      & 0.0055     & 0.5        & 0.44       & 0.9         \\
$\chi^2_{\rm{red}}$ &
        8.24665$-$6  & 2.89125$-$3  & 2.70197$-$3  & 1.06335$-$3   \\
\tableline                                   
$c_0$ & $-$8.97672   & $-$10.4717   & $-$5.77068   & $-$3.90395    \\
$b_0$ & $-$3.73654   & $-$5.32268   & $-$3.82724   & $-$3.5354     \\

$b_1$ & 9.86507+10 & 5.63787$-$3  & 1.4815$-$7   & 9.85176$-$31  \\
$a_1$ & 1.0        & 7.75892$-$2  & 7.44945$-$3  & 2.30147$-$4   \\
$m_1$ & $-$14.8506   & 3.33491    & 12.3238    & 33.3071     \\

$b_2$ & \nodata    & \nodata    & \nodata    & \nodata     \\
$a_2$ & \nodata    & \nodata    & \nodata    & \nodata     \\
$m_2$ & \nodata    & \nodata    & \nodata    & \nodata     \\

$b_3$ & 1.0+24     & 1.12504+3  & 5.843      & \nodata     \\
$a_3$ & 2.34542$-$3  & 0.125304   & 0.398924   & \nodata     \\
$m_3$ & 9.33589    & 6.04033    & 0.561698   & \nodata     \\

$b_4$ & \nodata    & 1.12597+3  & \nodata    & \nodata     \\
$a_4$ & \nodata    & 0.271622   & \nodata    & \nodata     \\
$m_4$ & \nodata    & 4.67116    & \nodata    & \nodata     \\
\enddata
\end{deluxetable}

\clearpage

\begin{deluxetable}{lllll}
\tablewidth{0pt}
\tablecaption{Parameters of the analytical approximations
  to the size distributions for the COMP-GR-FG model.       \label{f_comp-fg-gr}}
\tablehead{
  \colhead{}                     &
  \multicolumn{1}{c}{PAHs}       &
  \multicolumn{1}{c}{Graphite}   &
  \multicolumn{1}{c}{Silicate}   &
  \multicolumn{1}{c}{Composites}
}
\startdata
$A$   & 2.520814$-$7 & 1.936847$-$7 & 1.309292$-$7 & 5.393662$-$12 \\
$a_{\rm{min}}$                               
      & 0.00035    & 0.00035    & 0.00035    & 0.02        \\
$a_{\rm{max}}$                               
      & 0.005      & 0.39       & 0.39       & 0.75        \\
$\chi^2_{\rm{red}}$ &
        1.10438$-$5  & 5.57211$-$3  & 1.74117$-$2  & 1.11274$-$3   \\
\tableline                                   
$c_0$ & $-$8.72489   & $-$11.1324   & $-$3.81346   & $-$3.82614    \\
$b_0$ & $-$3.65649   & $-$6.6148    & $-$3.76412   & $-$3.48373    \\

$b_1$ & 9.86507+10 & 3.66626$-$2  & 2.62792$-$9  & 9.86756$-$31  \\
$a_1$ & 1.0        & 0.144398   & 7.26393$-$3  & 4.13811$-$4   \\
$m_1$ & $-$14.6651   & 2.54938    & 15.5036    & 34.9122     \\

$b_2$ & \nodata    & \nodata    & \nodata    & \nodata     \\
$a_2$ & \nodata    & \nodata    & \nodata    & \nodata     \\
$m_2$ & \nodata    & \nodata    & \nodata    & \nodata     \\

$b_3$ & 1.0+24     & 1.12501+3  & 6.64727    & \nodata     \\
$a_3$ & 2.05181$-$3  & 0.166373   & 0.344185   & \nodata     \\
$m_3$ & 9.20391    & 4.58796    & 0.21785    & \nodata     \\

$b_4$ & \nodata    & 1.126+3    & \nodata    & \nodata     \\
$a_4$ & \nodata    & 0.400672   & \nodata    & \nodata     \\
$m_4$ & \nodata    & 6.14619    & \nodata    & \nodata     \\
\enddata
\end{deluxetable}

\clearpage

\begin{deluxetable}{lllll}
\tablewidth{0pt}
\tablecaption{Parameters of the analytical approximations
  to the size distributions for the COMP-GR-B model.       \label{f_comp-b-gr}}
\tablehead{
  \colhead{}                     &
  \multicolumn{1}{c}{PAHs}       &
  \multicolumn{1}{c}{Graphite}   &
  \multicolumn{1}{c}{Silicate}   &
  \multicolumn{1}{c}{Composites}
}
\startdata
$A$   & 2.216925$-$7 & 1.918716$-$7 & 1.082933$-$7 & 4.780856$-$12 \\
$a_{\rm{min}}$                               
      & 0.00035    & 0.00035    & 0.00035    & 0.02        \\
$a_{\rm{max}}$                               
      & 0.0055     & 0.52       & 0.33       & 0.45        \\
$\chi^2_{\rm{red}}$ &
        8.51192$-$6  & 5.39803$-$3  & 6.53544$-$4  & 2.64688$-$3   \\
\tableline                                   
$c_0$ & $-$9.04531   & $-$10.1159   & 1.39336+2  & $-$3.72463    \\
$b_0$ & $-$3.75834   & $-$5.45055   & $-$3.66338   & $-$3.4173     \\

$b_1$ & 9.86507+10 & 2.58749$-$3  & 2.85829$-$10 & 2.56334$-$26  \\
$a_1$ & 1.0        & 9.91702$-$2  & 5.26352$-$3  & 2.05195$-$4   \\
$m_1$ & $-$14.9148   & 3.71707    & 16.487     & 29.4592     \\

$b_2$ & \nodata    & \nodata    & \nodata    & \nodata     \\
$a_2$ & \nodata    & \nodata    & \nodata    & \nodata     \\
$m_2$ & \nodata    & \nodata    & \nodata    & \nodata     \\

$b_3$ & 1.0+24     & 1.0023+2   & 1.48931+2  & \nodata     \\
$a_3$ & 2.38145$-$3  & 0.200689   & 0.341914   & \nodata     \\
$m_3$ & 9.34323    & 3.52158    & 5.05577$-$3  & \nodata     \\

$b_4$ & \nodata    & 1.00027+2  & \nodata    & \nodata     \\
$a_4$ & \nodata    & 0.699922   & \nodata    & \nodata     \\
$m_4$ & \nodata    & 9.86403    & \nodata    & \nodata     \\
\enddata
\end{deluxetable}

\clearpage

\begin{deluxetable}{lllll}
\tablewidth{0pt}
\tablecaption{Parameters of the analytical approximations
  to the size distributions for the BARE-AC-S model.       \label{f_bare-s-ac}}
\tablehead{
  \colhead{}                     &
  \multicolumn{1}{c}{PAHs}       &
  \multicolumn{1}{c}{ACH2}       &
  \multicolumn{1}{c}{Silicate 1} &
  \multicolumn{1}{c}{Silicate 2}
}
\startdata
$A$   & 4.492237$-$7 & 8.185937$-$12 & 3.527574$-$7 & 6.134893$-$13 \\
$a_{\rm{min}}$                  
      & 0.00035    & 0.02        & 0.00035    & 0.0272      \\
$a_{\rm{max}}$                  
      & 0.0037     & 0.26        & 0.025      & 0.37        \\
$\chi^2_{\rm{red}}$ &
        7.73272$-$4  & 2.64754$-$3   & 5.19042$-$2  & 4.56595$-$2   \\
\tableline                      
$c_0$ & $-$9.05931   & $-$3.96337    & $-$8.88283   & 8.93254+3   \\
$b_0$ & $-$3.76458   & $-$3.57444    & $-$3.69508   & 5.76792+3   \\

$b_1$ & 6.28593+5  & 1.93427$-$18  & 3.03135$-$20 & 5.77029+3   \\
$a_1$ & 1.0        & 1.0046$-$4    & 3.00297$-$7  & 2.82861$-$2   \\
$m_1$ & $-$14.3443   & 33.923      & 28.9189    & 1.00027     \\

$b_2$ & \nodata    & \nodata     & \nodata    & 3.78160+2   \\
$a_2$ & \nodata    & \nodata     & \nodata    & 9.39447$-$2   \\
$m_2$ & \nodata    & \nodata     & \nodata    & 9.04197     \\

$b_3$ & 1.0+24     & \nodata     & \nodata    & \nodata     \\
$a_3$ & 1.69966$-$3  & \nodata     & \nodata    & \nodata     \\
$m_3$ & 8.8067     & \nodata     & \nodata    & \nodata     \\

$b_4$ & \nodata    & \nodata     & \nodata    & \nodata     \\
$a_4$ & \nodata    & \nodata     & \nodata    & \nodata     \\
$m_4$ & \nodata    & \nodata     & \nodata    & \nodata     \\
\enddata
\end{deluxetable}

\clearpage

\begin{deluxetable}{lllll}
\tablewidth{0pt}
\tablecaption{Parameters of the analytical approximations
  to the size distributions for the BARE-AC-FG model.       \label{f_bare-fg-ac}}
\tablehead{
  \colhead{}                     &
  \multicolumn{1}{c}{PAHs}       &
  \multicolumn{1}{c}{ACH2}       &
  \multicolumn{1}{c}{Silicate 1} &
  \multicolumn{1}{c}{Silicate 2}
}
\startdata
$A$   & 4.727727$-$7 & 7.862901$-$12 & 3.680573$-$7 & 6.218762$-$13 \\
$a_{\rm{min}}$                  
      & 0.00035    & 0.02        & 0.00035    & 0.026       \\
$a_{\rm{max}}$                  
      & 0.0036     & 0.28        & 0.024      & 0.37        \\
$\chi^2_{\rm{red}}$ &
        6.30466$-$4  & 2.69604$-$3   & 4.25529$-$2  & 6.01087$-$2   \\
\tableline                      
$c_0$ & $-$8.91244   & $-$3.92513    & $-$8.88283   & 9.04443+3   \\
$b_0$ & $-$3.72015   & $-$3.54913    & $-$3.69508   & 5.7679+3    \\

$b_1$ & 6.78215+5  & 2.13708$-$17  & 2.17105$-$20 & 5.77024+3   \\
$a_1$ & 1.0        & 2.03908$-$4   & 3.0$-$7      & 2.7051$-$2    \\
$m_1$ & $-$14.2532   & 34.7835     & 29.2       & 1.00024     \\

$b_2$ & \nodata    & \nodata     & \nodata    & 3.82848+2   \\
$a_2$ & \nodata    & \nodata     & \nodata    & 9.39615$-$2   \\
$m_2$ & \nodata    & \nodata     & \nodata    & 8.94494     \\

$b_3$ & 1.0+24     & \nodata     & \nodata    & \nodata     \\
$a_3$ & 1.58225$-$3  & \nodata     & \nodata    & \nodata     \\
$m_3$ & 8.71891    & \nodata     & \nodata    & \nodata     \\

$b_4$ & \nodata    & \nodata     & \nodata    & \nodata     \\
$a_4$ & \nodata    & \nodata     & \nodata    & \nodata     \\
$m_4$ & \nodata    & \nodata     & \nodata    & \nodata     \\
\enddata
\end{deluxetable}

\clearpage

\begin{deluxetable}{lllll}
\tablewidth{0pt}
\tablecaption{Parameters of the analytical approximations
  to the size distributions for the BARE-AC-B model.       \label{f_bare-b-ac}}
\tablehead{
  \colhead{}                     &
  \multicolumn{1}{c}{PAHs}       &
  \multicolumn{1}{c}{ACH2}       &
  \multicolumn{1}{c}{Silicate 1} &
  \multicolumn{1}{c}{Silicate 2}
}
\startdata
$A$   & 4.47594$-$7  & 7.813825$-$12 & 3.157791$-$7 & 8.130755$-$13 \\
$a_{\rm{min}}$                  
      & 0.00035    & 0.02        & 0.00035    & 0.021       \\
$a_{\rm{max}}$                  
      & 0.0037     & 0.25        & 0.03       & 0.33        \\
$\chi^2_{\rm{red}}$ &
        8.9447$-$4   & 1.12192$-$2   & 9.47682$-$2  & 2.2829$-$2    \\
\tableline                      
$c_0$ & $-$9.1238    & $-$3.79483    & $-$8.95818   & 2.38667+4   \\
$b_0$ & $-$3.78439   & $-$3.46375    & $-$3.71928   & 1.4412+4    \\

$b_1$ & 6.02165+5  & 3.41188$-$17  & 4.01642$-$14 & 1.44139+4   \\
$a_1$ & 1.0        & 1.94208$-$4   & 2.45908$-$6  & 2.20822$-$2   \\
$m_1$ & $-$14.3958   & 34.2285     & 22.8203    & 1.00016     \\

$b_2$ & \nodata    & \nodata     & \nodata    & 5.12554+2   \\
$a_2$ & \nodata    & \nodata     & \nodata    & 7.85859$-$2   \\
$m_2$ & \nodata    & \nodata     & \nodata    & 10.2469     \\

$b_3$ & 1.0+24     & \nodata     & \nodata    & \nodata     \\
$a_3$ & 1.76358$-$3  & \nodata     & \nodata    & \nodata     \\
$m_3$ & 8.85147    & \nodata     & \nodata    & \nodata     \\

$b_4$ & \nodata    & \nodata     & \nodata    & \nodata     \\
$a_4$ & \nodata    & \nodata     & \nodata    & \nodata     \\
$m_4$ & \nodata    & \nodata     & \nodata    & \nodata     \\
\enddata
\end{deluxetable}

\clearpage

\begin{deluxetable}{llllll}
\tablewidth{0pt}
\tablecaption{Parameters of the analytical approximations
  to the size distributions for the COMP-AC-S model.       \label{f_comp-s-ac}}
\tablehead{
  \colhead{}                     &
  \multicolumn{1}{c}{PAHs}       &
  \multicolumn{1}{c}{ACH2}       &
  \multicolumn{1}{c}{Silicate 1} &
  \multicolumn{1}{c}{Silicate 2} &
  \multicolumn{1}{c}{Composites}
}
\startdata
$A$   & 4.480636$-$7 & 2.760470$-$12 & 3.758253$-$7 & 1.347406$-$13 & 2.485468$-$12 \\
$a_{\rm{min}}$                                              
      & 0.00035    & 0.02        & 0.00035    & 0.045       & 0.02        \\
$a_{\rm{max}}$                                              
      & 0.0038     & 0.25        & 0.02       & 0.4         & 0.91        \\
$\chi^2_{\rm{red}}$ &
        8.34817$-$5  & 1.55806$-$3   & 4.91354$-$4  & 1.22049$-$3   & 1.73353$-$2   \\
\tableline                                                  
$c_0$ & $-$9.14864   & $-$3.88834    & $-$8.83908   & 7.60836+3   & $-$2.61479    \\
$b_0$ & $-$3.79425   & $-$3.52976    & $-$3.68116   & 5.77288+3   & $-$2.68416    \\

$b_1$ & 9.86507+10 & 5.2005$-$24   & 1.11155$-$13 & 2.5083+3    & 9.97435$-$31  \\
$a_1$ & 1.0        & 8.805$-$4     & 3.68131$-$5  & 4.81218$-$2   & 2.53994$-$4   \\
$m_1$ & $-$14.3335   & 31.7703     & 16.6919    & 1.00005     & 33.679      \\

$b_2$ & \nodata    & \nodata     & \nodata    & 1.27299     & \nodata    \\
$a_2$ & \nodata    & \nodata     & \nodata    & 0.12723     & \nodata    \\
$m_2$ & \nodata    & \nodata     & \nodata    & 6.80997     & \nodata    \\

$b_3$ & 1.0+24     & \nodata     & \nodata    & \nodata     & \nodata    \\
$a_3$ & 1.65191$-$3  & \nodata     & \nodata    & \nodata     & \nodata    \\
$m_3$ & 8.71635    & \nodata     & \nodata    & \nodata     & \nodata    \\

$b_4$ & \nodata    & \nodata     & \nodata    & \nodata     & \nodata    \\
$a_4$ & \nodata    & \nodata     & \nodata    & \nodata     & \nodata    \\
$m_4$ & \nodata    & \nodata     & \nodata    & \nodata     & \nodata    \\
\enddata
\end{deluxetable}

\clearpage

\begin{deluxetable}{llllll}
\tablewidth{0pt}
\tablecaption{Parameters of the analytical approximations
  to the size distributions for the COMP-AC-FG model.       \label{f_comp-fg-ac}}
\tablehead{
  \colhead{}                     &
  \multicolumn{1}{c}{PAHs}       &
  \multicolumn{1}{c}{ACH2}       &
  \multicolumn{1}{c}{Silicate 1} &
  \multicolumn{1}{c}{Silicate 2} &
  \multicolumn{1}{c}{Composites}
}
\startdata
$A$   & 4.725913$-$7 & 2.868609$-$12 & 3.926154$-$7 & 1.821520$-$13 & 1.320657$-$12 \\
$a_{\rm{min}}$                                              
      & 0.00035    & 0.02        & 0.00035    & 0.036       & 0.02        \\
$a_{\rm{max}}$                                              
      & 0.0035     & 0.25        & 0.02       & 0.4         & 0.66        \\
$\chi^2_{\rm{red}}$ &
        1.05143$-$5  & 1.93094$-$3   & 5.43861$-$4  & 6.74454$-$3   & 8.39873$-$4   \\
\tableline                                                  
$c_0$ & $-$9.38386   & $-$3.91187    & $-$8.82601   & 1.03553+5   & $-$0.951599   \\
$b_0$ & $-$3.86273   & $-$3.54546    & $-$3.67646   & 5.76819+3   & $-$1.81202    \\

$b_1$ & 1.37152+13 & 9.86048$-$31  & 3.93265$-$14 & 2.50619+3   & 5.27084$-$14  \\
$a_1$ & 1.0        & 4.65102$-$4   & 3.36955$-$5  & 1.13857$-$18  & 2.72956$-$6   \\
$m_1$ & $-$17.038    & 38.1558     & 17.1386    & 1.00002     & 12.5346     \\

$b_2$ & \nodata    & \nodata     & \nodata    & 1.0829      & \nodata     \\
$a_2$ & \nodata    & \nodata     & \nodata    & 0.120892    & \nodata     \\
$m_2$ & \nodata    & \nodata     & \nodata    & 8.01741     & \nodata     \\

$b_3$ & 1.0+24     & \nodata     & \nodata    & \nodata     & 1.00014+3   \\
$a_3$ & 1.60933$-$3  & \nodata     & \nodata    & \nodata     & 0.298067    \\
$m_3$ & 8.66626    & \nodata     & \nodata    & \nodata     & 5.93761     \\

$b_4$ & \nodata    & \nodata     & \nodata    & \nodata     & \nodata     \\
$a_4$ & \nodata    & \nodata     & \nodata    & \nodata     & \nodata     \\
$m_4$ & \nodata    & \nodata     & \nodata    & \nodata     & \nodata     \\
\enddata
\end{deluxetable}

\clearpage

\begin{deluxetable}{llllll}
\tablewidth{0pt}
\tablecaption{Parameters of the analytical approximations
  to the size distributions for the COMP-AC-B model.       \label{f_comp-b-ac}}
\tablehead{
  \colhead{}                     &
  \multicolumn{1}{c}{PAHs}       &
  \multicolumn{1}{c}{ACH2}       &
  \multicolumn{1}{c}{Silicate 1} &
  \multicolumn{1}{c}{Silicate 2} &
  \multicolumn{1}{c}{Composites}
}
\startdata
$A$   & 4.470182$-$7 & 2.068952$-$13 & 3.383484$-$7 & 7.022871$-$14 & 4.516269$-$12 \\
$a_{\rm{min}}$                                              
      & 0.00035    & 0.022       & 0.00035    & 0.046       & 0.02        \\
$a_{\rm{max}}$                                              
      & 0.0039     & 0.21        & 0.02       & 0.25        & 0.7         \\
$\chi^2_{\rm{red}}$ &
        1.36914$-$5  & 3.5196$-$5    & 4.2771$-$4   & 2.93878$-$5   & 2.0045$-$3    \\
\tableline                                                  
$c_0$ & $-$9.6296    & $-$3.92565    & $-$8.82349   & $-$6.03894    & $-$2.70635    \\
$b_0$ & $-$3.94002   & $-$4.49943    & $-$3.67603   & $-$7.42085    & $-$2.66966    \\

$b_1$ & 1.37152+13 & 0.64013     & 8.30679$-$29 & 10.7406     & 1.60242$-$19  \\
$a_1$ & 1.0        & 6.90874$-$2   & 6.23404$-$7  & 0.106099    & 2.60559$-$4   \\
$m_1$ & $-$17.2214   & 9.27863     & 28.0109    & 9.89337     & 21.4322     \\

$b_2$ & \nodata    & 0.94057     & \nodata    & 0.698926    & \nodata     \\
$a_2$ & \nodata    & 0.115074    & \nodata    & 0.231661    & \nodata     \\
$m_2$ & \nodata    & 1.85136     & \nodata    & 3.02112     & \nodata     \\

$b_3$ & 1.0+24     & \nodata     & \nodata    & \nodata     & \nodata     \\
$a_3$ & 1.81574$-$3  & \nodata     & \nodata    & \nodata     & \nodata     \\
$m_3$ & 8.82131    & \nodata     & \nodata    & \nodata     & \nodata     \\

$b_4$ & \nodata    & \nodata     & \nodata    & \nodata     & \nodata     \\
$a_4$ & \nodata    & \nodata     & \nodata    & \nodata     & \nodata     \\
$m_4$ & \nodata    & \nodata     & \nodata    & \nodata     & \nodata     \\
\enddata
\end{deluxetable}

\clearpage

\begin{deluxetable}{lllll}
\tablewidth{0pt}
\tablecaption{Parameters of the analytical approximations
  to the size distributions for the COMP-NC-S model.       \label{f_comp-s-nc}}
\tablehead{
  \colhead{}                     &
  \multicolumn{1}{c}{PAHs}       &
  \multicolumn{1}{c}{Silicate 1} &
  \multicolumn{1}{c}{Silicate 2} &
  \multicolumn{1}{c}{Composites}
}
\startdata
$A$   & 4.457245$-$7  & 3.894306$-$7 & 8.424593$-$14 & 3.936063$-$12 \\
$a_{\rm{min}}$                                 
      & 0.00035     & 0.00035    & 0.048       & 0.02        \\
$a_{\rm{max}}$                                 
      & 0.0036      & 0.02       & 0.34        & 0.8         \\
$\chi^2_{\rm{red}}$ &
        9.09294$-$6   & 5.38936$-$4  & 1.45554$-$3   & 3.85852$-$3   \\
\tableline                                     
$c_0$ & $-$9.36312    & $-$8.82792   & 7.40265+3   & $-$2.48599    \\
$b_0$ & $-$3.85681    & $-$3.67806   & 5.76918+3   & $-$2.58455    \\

$b_1$ & 2.68843+12  & 2.89768$-$20 & 2.50652+3   & 9.87203$-$31  \\
$a_1$ & 1.0         & 7.88794$-$6  & 5.21321$-$2   & 6.35373$-$6   \\
$m_1$ & $-$16.1999    & 22.3415    & 1.0003      & 28.5023     \\

$b_2$ & \nodata     & \nodata    & 2.32676     & \nodata     \\
$a_2$ & \nodata     & \nodata    & 0.127735    & \nodata     \\
$m_2$ & \nodata     & \nodata    & 7.0254      & \nodata     \\

$b_3$ & 1.0+24      & \nodata    & \nodata     & \nodata     \\
$a_3$ & 1.68273$-$3   & \nodata    & \nodata     & \nodata     \\
$m_3$ & 8.74084     & \nodata    & \nodata     & \nodata     \\

$b_4$ & \nodata     & \nodata    & \nodata     & \nodata     \\
$a_4$ & \nodata     & \nodata    & \nodata     & \nodata     \\
$m_4$ & \nodata     & \nodata    & \nodata     & \nodata     \\
\enddata
\end{deluxetable}

\clearpage\begin{deluxetable}{lllll}

\tablewidth{0pt}
\tablecaption{Parameters of the analytical approximations
  to the size distributions for the COMP-NC-FG model.       \label{f_comp-fg-nc}}
\tablehead{
  \colhead{}                     &
  \multicolumn{1}{c}{PAHs}       &
  \multicolumn{1}{c}{Silicate 1} &
  \multicolumn{1}{c}{Silicate 2} &
  \multicolumn{1}{c}{Composites}
}
\startdata
$A$   & 4.703264$-$7  & 4.093516$-$7 & 8.907133$-$14 & 2.552983$-$12 \\
$a_{\rm{min}}$                                 
      & 0.00035     & 0.00035    & 0.0489      & 0.019       \\
$a_{\rm{max}}$                                 
      & 0.0035      & 0.02       & 0.36        & 0.85        \\
$\chi^2_{\rm{red}}$ &
        7.51744$-$6   & 8.68976$-$4  & 8.60985$-$3   & 3.52374$-$3   \\
\tableline                                     
$c_0$ & $-$9.23696    & $-$8.78955   & 7.40319+3   & $-$1.64917    \\
$b_0$ & $-$3.81837    & $-$3.66574   & 5.76731+3   & $-$2.01144    \\

$b_1$ & 2.68843+12  & 9.91223$-$31 & 2.50684+3   & 1.7781$-$28   \\
$a_1$ & 1.0         & 7.68829$-$7  & 5.20919$-$2   & 3.42324$-$7   \\
$m_1$ & $-$16.0932    & 30.2697    & 0.999787    & 24.2268     \\

$b_2$ & \nodata     & \nodata    & 2.95527     & \nodata     \\
$a_2$ & \nodata     & \nodata    & 0.12907     & \nodata     \\
$m_2$ & \nodata     & \nodata    & 5.4247      & \nodata     \\

$b_3$ & 1.0+24      & \nodata    & \nodata     & \nodata     \\
$a_3$ & 1.55501$-$3   & \nodata    & \nodata     & \nodata     \\
$m_3$ & 8.62883     & \nodata    & \nodata     & \nodata     \\

$b_4$ & \nodata     & \nodata    & \nodata     & \nodata     \\
$a_4$ & \nodata     & \nodata    & \nodata     & \nodata     \\
$m_4$ & \nodata     & \nodata    & \nodata     & \nodata     \\
\enddata
\end{deluxetable}

\clearpage

\begin{deluxetable}{lllll}
\tablewidth{0pt}
\tablecaption{Parameters of the analytical approximations
  to the size distributions for the COMP-NC-B model.       \label{f_comp-b-nc}}
\tablehead{
  \colhead{}                     &
  \multicolumn{1}{c}{PAHs}       &
  \multicolumn{1}{c}{Silicate 1} &
  \multicolumn{1}{c}{Silicate 2} &
  \multicolumn{1}{c}{Composites}
}
\startdata
$A$   & 4.473155$-$7  & 3.368198$-$7 & 3.037852$-$14 & 5.103841$-$12 \\
$a_{\rm{min}}$                                 
      & 0.00035     & 0.00035    & 0.0579      & 0.02        \\
$a_{\rm{max}}$                                 
      & 0.0038      & 0.02       & 0.18        & 0.8         \\
$\chi^2_{\rm{red}}$ &
        1.14807$-$5   & 6.28528$-$4  & 3.43037$-$5   & 9.61637$-$3   \\
\tableline                                     
$c_0$ & $-$9.45486    & $-$8.81100   & 7.03663+3   & $-$2.86045    \\
$b_0$ & $-$3.8855     & $-$3.67229   & 5.77097+3   & $-$2.84458    \\

$b_1$ & 2.68843+12  & 9.83854$-$31 & 2.50716+3   & 9.97738$-$31  \\
$a_1$ & 1.0         & 4.07425$-$7  & 6.0368$-$2    & 2.03298$-$4   \\
$m_1$ & $-$16.2777    & 29.4547    & 1.00102     & 33.5112     \\

$b_2$ & \nodata     & \nodata    & 26.3599     & \nodata     \\
$a_2$ & \nodata     & \nodata    & 9.2944$-$2    & \nodata     \\
$m_2$ & \nodata     & \nodata    & 5.99341     & \nodata     \\

$b_3$ & 1.0+24      & \nodata    & \nodata     & \nodata     \\
$a_3$ & 1.7925$-$3    & \nodata    & \nodata     & \nodata     \\
$m_3$ & 8.82824     & \nodata    & \nodata     & \nodata     \\

$b_4$ & \nodata     & \nodata    & \nodata     & \nodata     \\
$a_4$ & \nodata     & \nodata    & \nodata     & \nodata     \\
$m_4$ & \nodata     & \nodata    & \nodata     & \nodata     \\
\enddata
\end{deluxetable}

\clearpage


\clearpage

\clearpage

\clearpage

\clearpage

\clearpage

\clearpage

\clearpage

\clearpage

\clearpage

\clearpage

\clearpage

\clearpage

\clearpage

\clearpage

\clearpage

\clearpage

\clearpage

\clearpage

\clearpage

\clearpage

\clearpage

\clearpage

\clearpage


\begin{thebibliography}{}

\bibitem[Allamandola et al.(1985)]{atb85} Allamandola, L.J.,
  Tielens, A.G.G.M., \& Barker, J.R. 1985, \apj, 290, L25

\bibitem[Arendt et al.(1998)]{arendt98} Arendt, R.G., et al. 1998,
  \apj, 508, 74

\bibitem[Backus \& Gilbert(1970)]{bg70} Backus, G.E., \& Gilbert, F. 1970,
  Geophysical Journal of the Royal Astronomical Society, 16, 169

\bibitem[Bohlin et al.(1978)]{bohlin79} Bohlin, F.C., Savage, B.D.,
  \& Drake, J.F. 1978, \apj, 224, 132

\bibitem[Bohren \& Huffman(1983)]{bh83} Bohren, C.F., \& Huffman, D.R.
  1983, Absorption and Scattering of Light by Small Particles
  (New York: Wiley-Interscience)

\bibitem[Cardelli et al.(1989)]{ccm89} Cardelli, J.A., Clayton, G.C.,
  Mathis, J.S. 1989, \apj, 345, 245

\bibitem[Cardelli et al.(1996)]{cardelli96} Cardelli, J.A., Meyer,
  D.M., Jura, M., \& Savage, B.D. 1996, \apj, 467, 334

\bibitem[Chiar et al.(2000)]{chiar00} Chiar, J.E.,
  Tielens, A.G.G.M., Whittet, D.C.B., Schutte, W.A.,
  Boogert, A.C.A., Lutz, D., van Dishoeck, E.F., \&
  Bernstein, M.P. 2000, \apj, 537, 749

\bibitem[Colangeli et al.(1995)]{col95} Colangeli, L., Mennella, V.,
  Palumbo, P., Rotundi, A., Bussoletti, E. 1995, \aaps, 113, 561

\bibitem[D\'esert et al.(1990)]{dbp90} D\'esert, F.X.,
  Boulanger, F., \& Puget, J.L. 1990, \aap, 237, 215

\bibitem[Draine(1989)]{draine89} Draine, B.T. 1989,
  in Infrared Spetroscopy in Astronomy, ed. B.H. Kaldeich
  (Paris: ESA), 93

\bibitem[Draine \& Lee(1984)]{dl84} Draine, B.T., \& Lee, H.M. 1984,
  \apj, 285, 89

\bibitem[Draine \& Anderson(1985)]{da85} Draine, B.T., \& Anderson, N. 1985,
  \apj, 292, 494

\bibitem[Draine \& Li(2001)]{dl01} Draine, B.T., \& Li, A. 2001,
  \apj, 551, 807

\bibitem[Dwek et al.(1997)]{dwek97} Dwek, E., et al. 1997, \apj,
  475, 565
  
\bibitem[Dwek et al.(2004)]{dwek04} Dwek, E., Zubko, V., Arendt, R. G.,
  \& Smith, R. K. 2004, in ASP Conference Series Vol. 000,
  eds. A.N. Witt, B.T. Draine, \& G.C. Clayton (San Francisco: ASP),
  page 000

\bibitem[Fitzpatrick(1999)]{fitzpatrick99} Fitzpatrick, E.L. 1999, \pasp,
  111, 63

\bibitem[Greenberg(1968)]{greenberg68} Greenberg, J.M. 1968,
  in Stars and Stellar Systems, Vol. 7, ed. B.M. Middlehurst \&
  L.H. Aller (Chicago: Univ. of Chicago Press), 221

\bibitem[Greenberg \& Li(1999)]{gl99} Greenberg, J.M., \&
  Li, A. 1999, Adv. Space Res., 24, 497

\bibitem[Guhathakurta \& Draine(1989)]{gd89} Guhathakurta, P. \&
  Draine, B.T. 1989, \apj, 345, 230

\bibitem[Holweger(2001)]{holweger01} Holweger, H. 2001,
  in Joint SOHO/ACE workshop ``Solar and Galactic Composition'',
  ed. R.F. Wimmer-Schweingruber, American Institute
  of Physics Conference proceedings, vol. 598, p.23 

\bibitem[Joblin, L\'eger, \& Martin(1992)]{joblin92} Joblin, C., L\'eger, A.,
  \& Martin, P. 1992, \apj, 393, L79

\bibitem[Kim \& Martin(1994)]{km94} Kim, S.-H., \& Martin, P.G.,
  1994, \apj, 431, 783

\bibitem[Kim \& Martin(1995)]{km95} Kim, S.-H., \& Martin, P.G.,
  1995, \apj, 444, 293

\bibitem[Kim \& Martin(1996)]{km96} Kim, S.-H., \& Martin, P.G.,
  1996, \apj, 462, 296

\bibitem[Kim et al.(1994)]{kmh94} Kim, S.-H., Martin, P.G.,
  \& Hendry, P.D. 1994, \apj, 422, 164

\bibitem[Laor \& Draine(1993)]{ld93} Laor, A., \& Draine, B.T.
  1993, \apj, 402, 441

\bibitem[L\'eger \& Puget(1984)]{lp84} L\'eger, A., \& Puget, J.L.
  1984, \aap, 137, 5L

\bibitem[L\'eger et al.(1989)]{leger89} L\'eger, A., Verstraete, L.,
  d'Hendecourt, L., D\'efourneau, D., Dutuit, O., Schmidt, W.,
  \& Lauer, J. 1989, in IAU Symp. 135, Interstellar Dust,
  ed. L.J. Allamandola \& A.G.G.M. Tielens (Dordrecht: Kluwer), 173

\bibitem[Lehtinen \& Mattila(1996)]{lm96} Lehtinen, K., \& Mattila, K.
  1996, \aap, 309, 570

\bibitem[Li \& Draine(2001a)]{ld01a} Li, A., \& Draine, B.T. 2001, \apj, 550, L213

\bibitem[Li \& Draine(2001b)]{ld01b} Li, A., \& Draine, B.T. 2001, \apj, 554, 778

\bibitem[Li \& Greenberg(1997)]{lg97} Li, A., \& Greenberg, J.M. 1997,
  \aap, 323, 566

\bibitem[Lillie \& Witt(1976)]{lw76} Lillie, C.F., \& Witt, A.N. 1976,
  \apj, 208, 64

\bibitem[Lutz et al.(1996)]{lutz96} Lutz, D., et al. 1996, \aap, 315, L269

\bibitem[Mather et al.(1994)]{mather94} Mather, J.C., et al. 1994,
  \apj, 420, 439

\bibitem[Mathis et al.(1977)]{mrn77} Mathis, J.S., Rumpl, W.,
  \& Nordsieck, K.H. 1977, \apj, 217, 425 (MRN)

\bibitem[Mathis et al.(1983)]{mathis83} Mathis, J.S., Mezger, P.G.,
  \& Panagia, N. 1983, \aap, 128, 212

\bibitem[Meyer et al.(1997)]{meyer97} Meyer, D.M., Cardelli,
  J.A., \& Sofia, U.J. 1997, \apj, 490, L103

\bibitem[Meyer et al.(1998)]{meyer98} Meyer, D.M., Jura, M.,
  \& Cardelli, J.A. 1998, \apj, 493, 222

\bibitem[Morgan et al.(1976)]{morgan76} Morgan, D.H., Nandy, K., \&
  Thompson, G.I. 1976, \mnras, 177, 531

\bibitem[Narayan \& Nityananda(1986)]{narayan86} Narayan, R., \&
  Nityananda, R. 1986, \araa, 24, 127

\bibitem[Ossenkopf(1991)]{ossenkopf91} Ossenkopf, V. 1991,
  \aap, 251, 210

\bibitem[Press et al.(1992)]{numerical_recipes} Press, W.H.,
  Teukolsky, S.A., Vetterling, W.T., \& Flannery, B.P. 1992,
  Numerical Recipes in C: The Art of Scientific Computing.
  Second Edition (Cambridge University Press)

\bibitem[Sasseen et al.(2002)]{sasseen01} Sasseen, T.P., Hurwitz, M.,
  Dixon, W.V., \& Airieau, S. 2002, \apj, 566, 267

\bibitem[Savage \& Mathis(1979)]{sm79} Savage, B.D., \& Mathis, J.S. 1979,
  \araa, 17, 73

\bibitem[Savage et al.(1985)]{savage85} Savage, B.D., Massa, D.L.,
  Meade, M., \& Wesselius, P.R., 1985, \apjs, 59, 397

\bibitem[Schutte et al.(1998)]{schutte98} Schutte, W.A., et al. 1998,
  \aap, 337, 261

\bibitem[Snow \& Witt(1996)]{sw96} Snow, T.P., \& Witt, A.N.
  1996, \apj, 468, L65

\bibitem[Sofia \& Meyer(2001)]{sofia01} Sofia, U.J., \& Meyer, D.M.
  2001, \apj, 554, L221

\bibitem[Stognienko et al.(1995)]{sho95} Stognienko, R., Henning, Th.,
  \& Ossenkopf, V. 1995, \aap, 296, 797

\bibitem[Tielens et al.(1996)]{tielens96} Tielens, A.G.G.M.,
  Wooden, D.H., Allamandola, L.J., Brigman, J., \&
  Witteborn, F.C. 1996, \apj, 461, 210

\bibitem[Tikhonov et al.(1995)]{tikhonov95} Tikhonov, A.N.,
  Goncharsky, A.V., Stepanov, V.V., \& Yagola, A. 1995,
  Numerical Methods for the Solution of Ill-Posed Problems
  (Dordrecht, the Netherlands: Kluwer Academic Publishers)

\bibitem[Tokunaga(1997)]{tokunaga97} Tokunaga, A. T. 1997,
  in "Diffuse Infrared radiation and the IRTS", eds. H. Okuda,
  T. Matsumoto, and T. L. Roellig (A.S.P.; san Francisco), p. 149 

\bibitem[Warren(1984)]{warren84} Warren, S.G. 1984, \ao, 23, 1206

\bibitem[Weingartner \& Draine(2001)]{wd01} Weingartner, J.C.,
  \& Draine, B.T. 2001, \apj, 548, 296

\bibitem[Whittet et al.(1997)]{whittet97} Whittet, D.C.B., et al.
  1997, \apj, 490, 729

\bibitem[Witt(1989)]{witt89} Witt, A.N. 1989,
  in IAU Symp. 135, Interstellar Dust, ed. L.J. Allamandola \&
  A.G.G.M. Tielens (Dordrecht: Reidel), 87

\bibitem[Witt(2000)]{witt00} Witt, A. N. 2000, in IAU Symp. 197,
  Astrochemistry: From Molecular Clouds to Planetary Systems,
  ed. Y.C. Minh \& E.F. van Dishoeck (San Francisco: ASP), 317

\bibitem[Witt et al.(1997)]{witt97} Witt, A.N., Friedmann, B.C., \&
  Sasseen, T.P. 1997, \apj, 481, 809

\bibitem[Zubko(1997)]{z97} Zubko, V.G. 1997, \mnras, 289, 305

\bibitem[Zubko et al.(1996a)]{zkw96} Zubko, V.G., Kre{\l}owski, J.,
  \& Wegner, W. 1996, \mnras, 283, 577

\bibitem[Zubko et al.(1998)]{zkw98} Zubko, V.G., Kre{\l}owski, J.,
  \& Wegner, W. 1998, \mnras, 294, 548

\bibitem[Zubko et al.(1996b)]{zubko96} Zubko, V.G., Mennella, V.,
  Colangeli, L., \& Bussoletti, E. 1996, \mnras, 282, 1321
\end{thebibliography}
\end{document}